\documentclass[pdflatex,11pt]{article}

\usepackage{jheppub}
\usepackage{comment}
\usepackage{xcolor}
\usepackage{graphicx}
\usepackage{wrapfig}
\usepackage{multirow}
\usepackage{parskip}
\usepackage{tikz-cd}
\usepackage{chngcntr}
\counterwithout{figure}{section} 

\newcommand{\ali}[1]{\begin{align} #1 \end{align}}
\newcommand{\p}{\partial}

\newcommand{\ra}{\rightarrow}

\newcommand{\vev}[1]{\langle #1 \rangle} 
\newcommand{\mn}{{\mu\nu}}

\makeatletter
\gdef\@fpheader{}
\makeatother

\begin{document}
	\thispagestyle{empty}

	\title{Brane effective actions and their island rule from $T\bar T$ flows} 
	
	\author{Nele Callebaut}\author{and Matteo Selle}
	\affiliation{Institute for Theoretical Physics, University of Cologne, Z\"{u}lpicher Stra\ss e 77, 50937 K\"{o}ln, Germany} 
	
	\abstract{We derive a $2d$ effective action for semi-classical AdS$_3$ gravity with an EOW brane of arbitrary tension by solving the radial Hamiltonian flow in a derivative expansion. Interpreting the radial flow as an RG flow, the effective theory is identified with a $T\bar T$-like deformed CFT. This leads us to propose that the holographic dual for semi-classical AdS$_3$ gravity with Neumann or conformal boundary conditions is obtained by setting the $T\bar T$-like deformed CFT free. We verify our proposal explicitly for saddles corresponding to $(A)dS_2$ branes in empty AdS$_3$. We discuss three useful ways to decompose the effective brane theory into a matter sector ($T\bar T$-deformed, $T\bar T$-like deformed, and undeformed CFT, respectively) and a corresponding braneworld gravity sector. In particular, the split into $T\bar T$-deformed CFT coupled to $T\bar T$-like deformed Liouville gravity directly extends the Liouville description of the integrated Weyl anomaly into the bulk. The effective action is then used to perform an island-rule calculation for a non-asymptotic brane in an AdS/bCFT set-up, finding exact agreement with the Ryu-Takayanagi result. This extends previous holographic checks on the island rule to non-asymptotic regimes of application, as well as provides a Wald entropy interpretation for $T\bar T$ entanglement. Finally, we relate our effective theory and its higher derivative corrections to AdS$_3$ gravity with a fluctuating radial  cutoff by explicit calculation of the on-shell action.
}
	
\maketitle

\section{Introduction and overview of results}

Since the island papers  \cite{Almheiri:2019hni,Penington:2019npb,Almheiri:2019qdq,Penington:2019kki} 
proposing a resolution of the black hole information paradox, there has been a revival of braneworld holography   \cite{Chen:2020uac,Chen:2020hmv,Geng:2022slq,Geng:2022tfc,Neuenfeld:2024gta} 
(among many others). 
The typical set-up for island rule applications involves a two-dimensional brane plus bath configuration, where CFT matter extends over the whole system but is coupled to gravity in the brane region. In toy models, one typically uses Jackiw-Teitelboim (JT) gravity as the two-dimensional theory of gravity on the brane \cite{Almheiri:2019hni,Almheiri:2019yqk,Almheiri:2019psf}. The island rule was originally holographically inspired \cite{Almheiri:2019hni}, before being derived from a two-dimensional replica trick calculation \cite{Almheiri:2019qdq,Penington:2019kki}, and has been holographically matched to Ryu-Takayanagi entanglement calculations in the bulk for certain $3D/2d$ holographic set-ups \cite{Almheiri:2019hni,Suzuki:2022xwv,Chen:2020hmv,Chen:2020uac}. 
However, these matches 
were restricted to asymptotic branes \cite{Suzuki:2022xwv} or involved placing JT gravity on the brane by hand \cite{Chen:2020uac,Chen:2020hmv}, whereas it is debatable if the effective brane gravity theory induced by the three-dimensional bulk gravity is in fact JT and if so in which regimes  \cite{Geng:2022slq,Geng:2022tfc,Wang:2025bcx,Neuenfeld:2024gta,Callebaut:2025thw}. 
Another ambiguity in the use of the island rule is the treatment of the UV cutoff of the matter sector  implied by the use of the terminology `cutoff CFT' \cite{Gubser:1999vj} in the context of braneworld holography \cite{Geng:2026asi,DeVuyst:2022bua,Geng:2020qvw}. 
Related to this comment, 
island braneworld literature is focused typically on branes that are close to the asymptotic boundary of AdS. From an original braneworld perspective, however, one can consider placing branes at any finite cutoff location in the bulk \cite{Randall:1999ee,Randall:1999vf,Karch:2000ct}.  
The cutoff issues play a central role in the ongoing debate on the interpretation of islands \cite{Geng:2026asi,Antonini:2025sur}.

In \cite{Callebaut:2025thw}, we set out to systematically study braneworld holography, allowing for branes that are non-asymptotic, and focused on the questions: what is the induced two-dimensional gravity theory on the brane (is it JT?) and how can we concretely interpret `cutoff CFT'? The latter question is answered by using modern holographic $T\bar T$ language \cite{McGough:2016lol} applied to the original work by Gubser \cite{Gubser:1999vj}, which allows to interpret his `cutoff CFT' as $T\bar T$-deformed CFT. This led us to propose the construction of a braneworld holographic dictionary as $T \bar T$ set free, in the terminology of \cite{Compere:2008us} and at the semi-classical level. It involves first imposing Dirichlet boundary conditions at the brane before allowing the two-dimensional geometry to fluctuate. We will consider Dirichlet, Neumann and conformal boundary conditions imposed at the end-of-the-world (EOW) brane in this work. While in \cite{Callebaut:2025thw} we restricted to branes of zero curvature or unit tension (in units of the AdS scale $\ell$),  in this paper we extend the discussion to non-zero curvature or non-unit tension $T_0$. 

In this work, we are able to derive the effective gravity action on the brane by applying the $T\bar T$ set free proposal. 
The first main result is given in \eqref{Weff} and repeated here. 
We find that the induced effective theory on the 
EOW brane with two-dimensional induced metric $g = e^\Phi \gamma$ 
is given, at lowest order in derivatives of $\Phi$, by 
\begin{align}
	W[\gamma,\Phi]
	= \frac{c}{48\pi}\int d^2x \sqrt{-\gamma}&\Bigg[  2 \log \left( \frac{ e^{\Phi/2} + \sqrt{e^\Phi - \lambda}}{2-\lambda^2} \right) R_{\gamma}  + 4e^{\Phi}\left(\sqrt{1-\lambda e^{-\Phi}}-T_0\right) \Bigg] 
\end{align} 
in terms of the Weyl factor $\Phi$, and $\gamma$ an $AdS_2$, $dS_2$ or $Mink_2$ reference metric of unit radius, for values of the sign parameter $\lambda$ respectively equal  to $1,-1$ or $0$ (directly related to the tension by \eqref{lambdaofT0}). 
We propose this defines the holographic dual to the bulk theory with Neumann boundary conditions (NBC) or conformal boundary conditions (CBC) at the brane as respectively 
\ali{
	Z_{NBC} = \int D \gamma \int D \Phi e^{i W[\gamma,\Phi]} 
} 
and 
\ali{
	Z_{CBC}(\gamma) = \int D \Phi e^{i W[\gamma,\Phi]}. 
}
The result for $W[\gamma,\Phi]$ is obtained in section \ref{sec:braneeffective} by solving the radial Hamilton-Jacobi evolution equation into the bulk, for the boundary data dependence of the bulk on-shell action $S_{tot}(\gamma,\Phi)$. This can be equivalently interpreted as solving a $T\bar T$-like trace flow equation in the boundary, 
for the $T\bar T$-like generating functional $W[\gamma,\Phi]$. The flow that $W[\gamma,\Phi]$ satisfies is given explicitly in \eqref{TTbarlikeTraceflow}. 
In practice, we solve it in a derivative expansion making use of the ansatz  \eqref{Wansatz}. 
We think of this as extending the holographic RG methods of \cite{deBoer:1999tgo} deeper into the bulk. 
The above result for $W[\gamma,\Phi]$ is given to lowest derivative order in $\Phi$, 
but does not require an asymptotic expansion: both the potential and curvature coupling are obtained in a closed form that is valid at any finite EOW location in the bulk. In $Z_{NBC}$, with both $\gamma$ and $\Phi$ dynamical, $W[\gamma,\Phi]$ is a dilaton gravity theory. Its higher-order derivative contributions and non-local contributions are discussed to some extent, in particular in Appendices \ref{app:nonlocal} and \ref{app:higherderivatives}.   

The proposed holographic dictionaries above state that AdS$_3$ gravity with NBC or CBC is dual to a $T\bar T$-like deformed CFT set free. 
We proceed in section \ref{sec:identifyingmatter} to split the effective brane action into a matter sector and a gravity sector. By putting forward which flow equation it obeys separately, the choice of the matter sector allows to decouple the full flow equation and solve it for the remaining gravity sector. 
We discuss three useful such decompositions: 
\ali{
	W &= W_{T\bar T-like} + \tilde S \\ 
	  &= W_{T\bar T} + \hat S \\ 
	  &= W_{CFT} + \bar S  
}
with the respective gravitational sectors on the brane given explicitly, again to lowest 
orders in a derivative expansion, in \eqref{Stildae}, \eqref{STT} and \eqref{Sbar} (and the corresponding matter sectors in \eqref{Wmatfctional}, \eqref{Wttbarghatsplit} and \eqref{WCFTSdiv}). They are all dilaton gravity theories of the form of the original ansatz for $W$ in \eqref{Wansatz}. Asymptotic limits can be considered using \eqref{sigmadef1} and expanding for small $\epsilon$. 

The first choice, with a $T\bar T$-like matter sector, 
leads to the timelike Liouville action $\tilde S$ that was proposed by Allameh and Shaghoulian in \cite{Allameh:2025gsa} in the context of the CBC problem, at least to lowest order in derivatives of $\Phi$. Our procedure 
thus provides a viable strategy to derive their proposal and identify   
the type of trace flow equation that the corresponding matter sector obeys.

The second choice, with a $T\bar T$-deformed CFT matter sector, provides the direct generalization to non-zero brane 
intrinsic curvature of our previous results in \cite{Callebaut:2025thw}, with $W_{T\bar T}$ in the role of `cutoff CFT'. The corresponding gravitational sector $\hat S$ is a $T\bar{T}$-like deformed Liouville action, and plays the role of the into-the-bulk extended 
timelike Liouville action describing the Weyl anomaly. 

The third choice, with a CFT matter sector, still appears technically possible, also for a non-asymptotic brane. It allows to consider an island calculation set-up inspired by \cite{Suzuki:2022xwv} where we can compare the Ryu-Takayanagi result for the radiation entropy to the result from an island rule in the effective brane theory. We find exact agreement 
\ali{
	S_{island} = S_{RT} , 
} in a set-up where the effective brane gravity is derived (from both bulk and boundary perspectives) rather than ad hoc imposed. It is of the form of CFT matter coupled to the induced gravity $\bar S$ in \eqref{Sbar}, which is a dilaton gravity theory but not JT. The match between the island rule result and holographic entanglement works for branes at any finite location in the bulk. The island set-up and calculation is discussed in section \ref{subsec:islandrule}. In the preceding section \ref{sect:TTEE} we discuss a related holographic entanglement application of our effective $W[\gamma,\Phi]$ theory:  
it provides a Wald entropy interpretation of $T\bar T$ entanglement entropy or a finite RT length 
\ali{
	S_{T\bar{T}} = S_{Wald} = S_{RT}  
}
for $T\bar T$ on an $(A)dS_2$ background and a half-space entangling region. This extends into the bulk the three-fold duality between boundary-brane-bulk perspectives on entanglement entropy discussed by Hawking, Maldacena and Strominger in \cite{Hawking:2000da}.  

It is sufficient for the 
entanglement applications in section \ref{sect:EEapps} to consider $W[\gamma,\Phi]$ in its lowest-derivatives form. The 
derivation of the higher-derivative and non-local contributions quickly becomes  cumbersome. In section \ref{sec:fluctuating}, we consider the bulk perspective more typically used in holographic braneworld where the bulk on-shell action is calculated up to a fluctuating brane location. This allows to systematically discuss derivative contributions. 
We discuss the matching to our $T\bar T$-flow-derived results in sections \ref{sect:bulk} and \ref{holomatch}, including the $T\bar T$ and CFT matter splits, in sections \ref{subsLiou} and \ref{sectannular}.  
We compare to literature results on derived JT on the brane (at linear order in the brane fluctuation) in \eqref{Sphi}. The discussion in section  \ref{subsLiou} most closely follows the strategy of \cite{Callebaut:2025thw} and focuses on the equivalence between 
a fluctuating brane in a chosen (typically empty AdS$_3$) bulk and the change of the bulk as it reacts 
to Dirichlet boundary conditions that differ by a Weyl factor 
(Figure \ref{holoLiouville4}). In terms of the modes we introduced, it focuses on the relation between the brane fluctuation and the Weyl mode on the brane. In the asymptotic limit, the brane gravity role is played by the undeformed timelike Liouville theory for the integrated Weyl anomaly, which includes (kinetic) derivative terms and of which we give a detailed holographic derivation in Appendix \ref{AppLiou}, again focusing on the relation between the brane fluctuation and asymptotic Weyl mode (which labels the relevant finite Brown-Henneaux diffeomorphisms). We include in Appendix \ref{app:curvedbanados} the derivation of a curved-sliced Banados solution \eqref{curvedBanados+-1} and of the finite Brown-Henneaux diffeomorphisms  
between Poincar\'e AdS$_3$ 
and the Banados geometry  
with asymptotic $(A)dS_2$ slicing. This extends the known diffeomorphisms  from Poincar\'e AdS$_3$ to flat-sliced Banados \cite{Roberts:2012aq}.

\section{Brane effective action from integration of $T\bar T$-like flow} \label{sec:braneeffective}

\begin{figure}[t]
	\centering
	\includegraphics[scale=0.42]{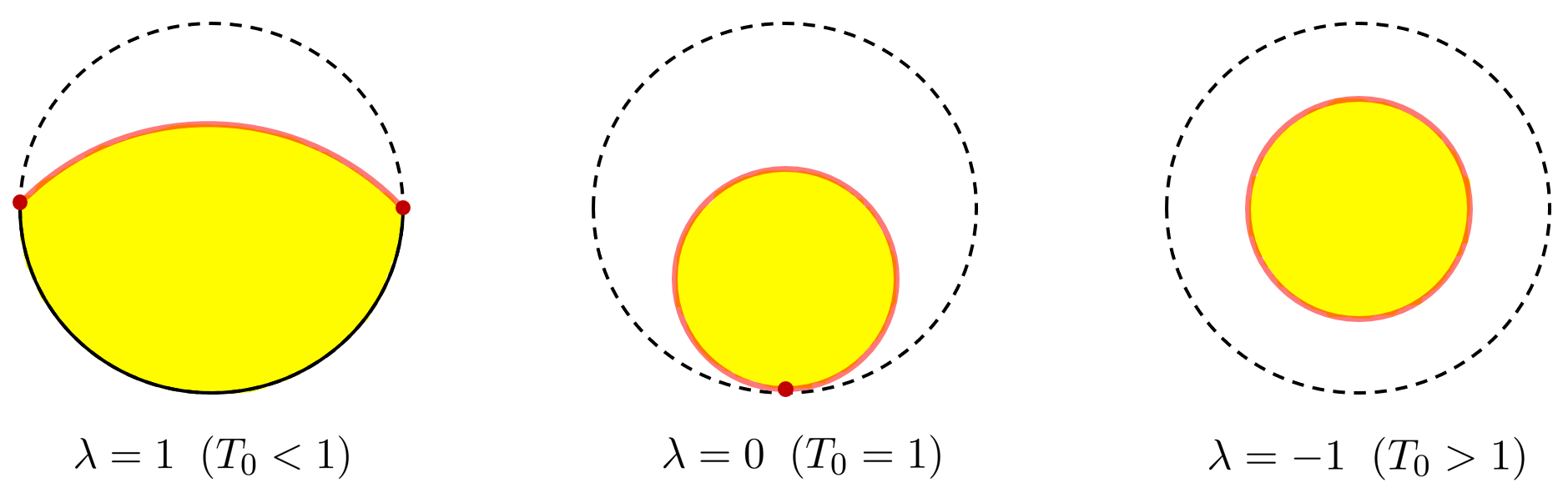}
	\caption{Two-dimensional end-of-the-world branes of constant curvature, shown as solid red lines, embedded in a time-reflection-symmetric slice of empty AdS$_3$ (the disk, whose dashed circumference denotes the asymptotic boundary). The white/yellow region indicates the portion of AdS$_3$ that is excised/retained. The parameter $\lambda$ characterizes the brane type: $\lambda = 1,0,-1$ correspond respectively to subcritical, critical, and supercritical tension ($T_0<1$, $T_0=1$, $T_0>1$), giving rise to $AdS_2$, $Mink_2$, and $dS_2$ brane geometries with constant negative, zero, and positive intrinsic curvature. In the $\lambda=1$ case, the brane intersects the asymptotic boundary at two points, leaving a finite portion of the boundary (solid black arc) intact. For vanishing tension, $T_0=0$, the $AdS_2$ brane divides the bulk into two symmetric halves. In the $\lambda=0$ case, the flat brane touches the asymptotic boundary at a single point.} \label{3typesofbranes}
\end{figure}

We study asymptotically-AdS$_3$ geometries $\mathcal{M}$ with a single timelike end-of-the-world brane (EOW) which constitutes (a portion or the entirety of) its outer boundary EOW $\subseteq \partial \mathcal{M}$, 
as in the examples illustrated in Figure \ref{3typesofbranes}. 
We define the total action as 
\begin{equation} \label{Stot}
	S_{tot}[G]=S_{EH}[G]+S_{GH}[g]+S_{T_0}[g].
\end{equation}
It consists of an Einstein-Hilbert term with negative cosmological constant 
\begin{equation}
	S_{EH}[G]=\frac{1}{2\kappa}\int_{\mathcal{M}} d^3X  \sqrt{-G}\left(R_G+2\right), \label{SEH}
\end{equation}
a Gibbons-Hawking boundary term on the EOW brane
\begin{equation}
	S_{GH}[g]=\frac{1}{\kappa}\int_{EOW} d^2x \sqrt{-g} \, K, \label{SGH}
\end{equation}
and an intrinsic tension term for the brane
\begin{equation} \label{STconstant}
	S_{T_0}[g]=-\frac{1}{\kappa}\int_{EOW} d^2x \sqrt{-g} \, T_0.
\end{equation}
The variation of the total action (\ref{Stot}) reads
\ali{
	\delta S_{tot} = \int_{\mathcal M} d^3 X (EOM)\delta G - \frac{1}{2 \kappa} \int_{\p \mathcal M} d^2 x \sqrt{-g} \, \left(K^\mn - K g^\mn + T_0 g^\mn \right) \delta g_\mn . \label{deltaStot}
}
Here and in the following, we set the bulk curvature radius to unity ($\ell=1$). We denote the induced metric at the brane and its Ricci scalar with $g_{\mu\nu}$ and $R$ respectively, while we use $R_{[.]}$ for all other metrics $[.]$, e.g. $R_G$ for the Ricci scalar of the bulk metric $G_{MN}$. $T_0$ is the brane's tension, $\kappa=8\pi G$ the three-dimensional gravitational constant, and  $K=g^{\mu\nu}K_{\mu\nu}$ is the trace of the extrinsic curvature on the brane. $X^M$ (with capital Latin indices) and $x^{\mu}$ (with Greek indices) denote the bulk and the brane coordinates respectively. Throughout, we disregard boundary contributions away from the brane, including one-dimensional corner terms associated with potential intersections of the brane with the asymptotic boundary.

\par If the tension is set to unity, the tension term coincides with the holographic counterterm of Balasubramanian-Kraus \cite{Balasubramanian:1999re}. 
When the total action is evaluated with unit tension, such that it is asymptotically free of power-law divergences when evaluated on-shell, we refer to it as the gravitational action
\begin{equation}
	S_{grav}[G]=S_{tot}[G]\big|_{T_0=1}. 
\end{equation}
According to the proposal of \cite{McGough:2016lol}, the $T\bar{T}$-deformation of a holographic CFT$_2$ provides a  
definition of the gravitational path integral for AdS$_3$ gravity with Dirichlet boundary conditions (DBC) at finite cutoff
\begin{equation} \label{TTbardictionary}
	Z_{T\bar{T}}(g)=\int_{G_{\partial}=g} \mathcal{D}G e^{iS_{grav}[G]}, \qquad \text{Dirichlet bc:} \qquad \delta g=0.
\end{equation}
The $T\bar{T}$-deformation is defined via the following flow equation for the field theory action $S_{T\bar{T}}[g]$ with respect to the deformation parameter $t$ \cite{Zamolodchikov:2004ce}
\begin{equation}
	\frac{d}{dt} S_{T\bar{T}}[g]=\frac{1}{4\pi}\int d^2 x \sqrt{-g} \, \mathcal{O}_{T\bar{T}},
\end{equation}
with the irrelevant $T\bar{T}$ operator given by $\mathcal{O}_{T\bar{T}}=\left(\mathcal T_{\mu\nu}\mathcal T_{\alpha\beta}g^{\mu\alpha}g^{\nu\beta}-(\mathcal T_{\mu\nu}g^{\mu\nu})^2\right)$ in terms of the field theory stress tensor $\mathcal T_{\mu\nu}=4\pi \delta S_{T\bar{T}}[g]/(\delta g^{\mu\nu}\sqrt{-g})$. When the seed theory is taken to be a holographic CFT$_2$ with large central charge $c=3/2G$, under suitable assumptions the $T\bar{T}$ operator factorizes and the expectation value of the stress tensor $\langle \mathcal T_{\mu\nu} \rangle = 4\pi \delta W_{T\bar{T}}[g]/(\delta g^{\mu\nu}\sqrt{-g})$, with $Z_{T\bar{T}}(g)=e^{iW_{T\bar{T}}[g]}$, obeys the trace flow equation $\langle \mathcal T_{\mu\nu}g^{\mu\nu}\rangle=-\frac{c}{12}R+t\langle O_{T\bar{T}} \rangle$, which is dual 
to the radial Hamiltonian constraint for AdS$_3$ gravity with negative cosmological constant, i.e. $R+K_{\mu\nu}K^{\mu\nu}-K^2+2=0$,  under the 
identification $t=-2G$ between the deformation parameter and the bulk parameters \cite{Marolf18}. 
We will be building on this holographic dictionary and propose a duality for semi-classical AdS$_3$ gravity with Neumann or conformal boundary conditions \cite{Witten:2022xxp,Anninos:2023epi,Coleman:2020jte,Chu:2021mvq,Anninos:2024wpy} at finite cutoff (NBC and CBC respectively), following our earlier proposal \cite{Callebaut:2025thw} for braneworld holography but generalizing it to include EOW branes with non-vanishing intrinsic curvature and CBC.

Typically one imposes Neumann boundary conditions $K^\mn - K g^\mn + T_0 g^\mn = 0$ at the EOW brane to make the boundary variation part of $\delta S_{tot}$ in \eqref{deltaStot} vanish. Instead, we will start our discussion by imposing Dirichlet boundary conditions $\delta g_\mn = 0$ at the EOW brane and treat it initially just as part of the fixed boundary. (This is an intermediate step in our proposal and we will proceed to impose Neumann and conformal boundary conditions by a setting-free procedure in section \ref{sec:settingfree}.)

The gravitational path integral with Dirichlet boundary conditions weighted by the total action \eqref{Stot}
\begin{equation} 
	Z_{tot}(g)=e^{iW[g]}=\int_{G_{\partial}=g} \mathcal{D}G e^{iS_{tot}[G]}   \label{Ztot}
\end{equation} 
defines an associated dual boundary theory with generating functional (of connected correlators) given by $W[g]$ \footnote{
	We will consistently denote the field dependencies of generating functionals with square brackets, a notation we otherwise reserve for dynamical field dependence (while round brackets are reserved for parametric dependence). The reason for this is that we will later promote the dependence on boundary field distributions to dynamical field dependence and thus $W$ to an effective action, by setting the boundary theory free (in section \ref{sec:settingfree}). 
}. In the semi-classical limit of small $\kappa$, the generating functional can be equated to the on-shell total action $S_{tot}[G^{\star}]\equiv S_{tot}(g)$
\begin{equation} \label{WStotsemicl}
	\lim\limits_{\kappa\rightarrow0}W[g]=S_{tot}(g).
\end{equation} 
Throughout the paper, we will restrict to the semi-classical limit and often omit the $\kappa \rightarrow 0$ limit and the star superscript on the bulk metric $G$ to denote a solution of the bulk Einstein equations.

\subsection{From Hamilton-Jacobi to a $T\bar{T}$-like flow} \label{subsec:fromhamilton}

The Brown-York stress tensor at the brane, including the tension contribution, can be read off from the total variation \eqref{deltaStot}
\begin{equation}
	T_{\mu\nu}^{BY}=-\frac{4\pi}{\sqrt{-g}}\frac{\delta S_{tot}}{\delta g^{\mu\nu}}=-\frac{1}{4 G}\left(K_{\mu\nu}-K g_{\mu\nu}+T_0g_{\mu\nu}\right).
\end{equation}
This allows us to rewrite the radial Hamiltonian constraint for AdS$_3$ gravity
\begin{equation}
	R+K_{\mu\nu}K^{\mu\nu}-K^2+2=0
\end{equation}
as the following Hamilton-Jacobi equation for the on-shell total action
\begin{equation}
	T_0 \, T^{\mu\nu}_{BY}g_{\mu\nu}= \frac{R}{8G}-\frac{(T^2_0-1)}{4G}+2 G \left(T^{\mu\nu}_{BY}T^{\alpha\beta}_{BY} g_{\alpha\mu}g_{\beta \nu} -(T^{\mu\nu}_{BY}g_{\mu\nu})^2\right).
\end{equation} 
In its interpretation as a Hamilton-Jacobi equation, it can be solved for the dependence of the bulk on-shell total action $S_{tot}(g)$ on the fixed boundary data $g$.  
We use the semi-classical holographic dictionary \eqref{WStotsemicl} to identify the on-shell total action $S_{tot}(g)$ with the generating functional of a dual boundary theory $W[g]$ with stress-tensor expectation value $\langle \mathcal T_{\mu\nu}\rangle=-T_{\mu\nu}^{BY}$, i.e.
\begin{equation}
	\langle \mathcal T_{\mu\nu} \rangle= \frac{4\pi}{\sqrt{-g}}\frac{\delta W[g]}{\delta g^{\mu\nu}}.
\end{equation}
Thus, we can express the radial Hamiltonian constraint as the following trace-flow equation for the dual boundary theory 
\begin{equation}\label{traceflowtot}
	T_0 \langle \mathcal T^{\mu}_{\mu}\rangle=-\frac{c}{12}R+\frac{c}{6}(T^2_0-1)+t\langle \mathcal{O}_{T\bar{{T}}} \rangle  ,
\end{equation}
with $c=3/2G$, $\mathcal T^{\mu}_{\mu}=\mathcal T^{\mu\nu}g_{\mu\nu}$, $\mathcal{O}_{T\bar{{T}}}=\mathcal T^{\mu\nu}\mathcal T^{\alpha\beta}g_{\mu\alpha}g_{\nu\beta}-(\mathcal T^{\mu\nu}g_{\mu\nu})^2$ and coupling parameter $t$ for the $\mathcal{O}_{T\bar{{T}}}$ operator
\begin{equation} \label{TTbarcoupling}
	t=-2G=-\frac{3}{c}.
\end{equation} 
When the tension has the critical value $T_0=1$, the tension term coincides with the holographic counterterm \cite{Balasubramanian:1999re} and the boundary trace flow equation reduces to the trace flow equation for the $T\bar{T}$-deformation of the dual CFT$_2$, i.e. $\langle \mathcal T^{\mu}_{\mu}\rangle=-\frac{c}{12}R+t\langle \mathcal{O}_{ T\bar{{T}}}\rangle$. In the undeformed limit one recovers the Weyl anomaly of the dual CFT$_2$: $\lim_{t \rightarrow 0}\langle \mathcal T^{\mu}_{\mu}\rangle=-\frac{c}{12}R$. As first conjectured by \cite{McGough:2016lol}, the dual boundary theory is thus a $T\bar{T}$-deformed CFT$_2$: $W[g]\equiv W_{T\bar{T}}[g]$. Allowing for a general tension $T_0$, we may then call Eq.~(\ref{traceflowtot}) a generalized $T\bar{T}$-trace flow equation, because the trace term is now multiplied by $T_0$ and an additional constant term $\sim (T_0^2-1)$ is produced. These features are not entirely new in the holographic $T\bar{T}$ literature. For example, an overall factor multiplying the trace of the stress tensor, as on the left-hand-side of equation \eqref{traceflowtot}, can also be produced by the need for bulk matter counterterms, as found in \cite{Callebaut:2025uye}. Moreover, additional constant terms as the one on the right-hand-side of the trace flow \eqref{traceflowtot} appear for example in the context of $T\bar{T}+\Lambda$ deformations \cite{Gorbenko:2018oov}, although there they are related to the curvature of the bulk rather than that of the brane. Here, the modifications to the standard $T\bar{T}$-trace flow equation in \eqref{traceflowtot} will correspond to non-vanishing curvature of the brane, as illustrated in Fig.~\ref{3typesofbranes}, when the brane geometry is treated as dynamical in section \ref{sec:settingfree} (imposing the NBC or CBC at the brane).

We write the induced metric on the EOW brane as a Weyl factor times a reference metric  
\begin{equation} \label{gPhigamma}
	g_{\mu\nu}(x)=e^{\Phi(x)}\gamma_{\mu\nu}(x),
\end{equation}
with $\gamma_{\mu\nu}(x)$ the metric on $dS_2$ ($\lambda=-1$), $Mink_2$ ($\lambda=0$) or $AdS_2$ ($\lambda=+1$) with unit radius of curvature. The curvature of the reference metric is thus $R_\gamma = -2\lambda$. The sign parameter $\lambda$ used to distinguish between the three cases is defined in terms of the tension as follows
\begin{equation}
	\lambda=\text{sgn}(1-T_0^2). \label{lambdaofT0}
\end{equation} 
The trace flow equation (\ref{traceflowtot}) for the dual boundary theory with generating functional $W[g]$ 
now reads 
\ali{
	T_0 \left\langle T_{\mu\nu}\gamma^{\mu\nu} \right\rangle=-\frac{c}{12}\left(R_{\gamma}-\Box\Phi\right) +\frac{c}{6}e^{\Phi}(T^2_0-1) +te^{-\Phi}\left\langle T_{\mu\nu}T_{\alpha\beta}\gamma^{\mu\alpha}\gamma^{\nu\beta}-\left(T_{\mu\nu}\gamma^{\mu\nu}\right)^2 \right\rangle  \label{TTbarlikeTraceflow}
} 
where we used  $\sqrt{-g}R=\sqrt{-\gamma}\left(R_{\gamma}-\Box \Phi\right)$ and 
\ali{ 
	\langle T_{\mu\nu} \rangle =\frac{4\pi}{\sqrt{-\gamma}}\frac{\delta W}{\delta \gamma^{\mu\nu}}
}
is the stress tensor whose indices are raised and lowered with the reference metric\footnote{
	Since the total effective theory, 
	including a possible non-local sector, can be written as a functional of $g_{\mu\nu}$ only, we must have 
	\begin{align}
		\delta W[g]&=\frac{1}{4\pi}\int d^2x \sqrt{-g}\langle \mathcal T_{\mu\nu}\rangle \delta g^{\mu\nu}=\frac{1}{4\pi}\int d^2x \sqrt{-\gamma}e^{\Phi}\langle \mathcal T_{\mu\nu}\rangle (e^{-\Phi}\delta\gamma^{\mu\nu}-e^{-\Phi}\gamma^{\mu\nu}\delta \Phi)
		\\
		&=\frac{1}{4\pi}\int d^2x \sqrt{-\gamma}(\langle \mathcal T_{\mu\nu}\rangle\delta \gamma^{\mu\nu}-\langle \mathcal T_{\mu\nu}\gamma^{\mu\nu}\rangle \delta \Phi). \label{delWdelPhi}
	\end{align}
	Hence it follows that $\langle \mathcal T_{\mu\nu}\rangle=\frac{4\pi}{\sqrt{-g}}\frac{\delta W[g]}{\delta g^{\mu\nu}}=\frac{4\pi}{\sqrt{-\gamma}}\frac{\delta W[\gamma,\Phi]}{\delta \gamma^{\mu\nu}} = \langle T_{\mu\nu}\rangle$ and $\langle T_{\mu\nu}\gamma^{\mu\nu}\rangle=-\frac{4\pi}{\sqrt{-\gamma}}\frac{\delta W[\gamma,\Phi]}{\delta \Phi}$. 
}. The box operator is $\Box=\gamma^{\mu\nu}\nabla_{\mu}\nabla_{\nu}$, with $\nabla_{\mu}$ denoting covariant differentiation with respect to $\gamma_{\mu\nu}$.  We call \eqref{TTbarlikeTraceflow} the generalized trace flow equation of a ``$T\bar{T}$--like" deformed theory, to emphasize that the coupling $t$ for the $T\bar{T}$ operator gets rescaled by the conformal factor as $t\rightarrow te^{-\Phi}$. 

The asymptotic Weyl anomaly $\lim_{t \rightarrow 0}\langle \mathcal T^{\mu}_{\mu}\rangle=-\frac{c}{12 T_0}R + \frac{c}{6T_0} (T_0^2-1)$ or $\frac{\delta W[g]}{\delta \Phi} = -\frac{1}{4\pi} \sqrt{-g} \vev{\mathcal T_\mu^\mu}$ is integrated straightforwardly to the timelike Liouville description, with non-unit tension corresponding to a non-zero 
$\sim (T_0^2-1) e^\Phi$ Liouville potential term. 
To solve the full flow, we proceed with an ansatz for $W[g]$.

\subsection{Solving the flow} \label{subsec:solvingtheflow}

We make the following ansatz for the local part of the bulk on-shell total action or generating functional of the dual boundary theory $W[g]\equiv W[\gamma,\Phi]$ 
\begin{equation} 
	W[\gamma,\Phi]=\frac{c}{48\pi}\int d^2x \sqrt{-\gamma}\left(D(\Phi) R_{\gamma}+\frac{1}{2}E(\Phi)(\nabla\Phi)^2+ V(\Phi)\right)+\text{higher}\;\text{derivatives},  \label{Wansatz}
\end{equation}
with $(\nabla \Phi)^2=\gamma^{\mu\nu}\nabla_{\mu}\Phi\nabla_{\nu}\Phi$. 
We will refer to $W[\gamma,\Phi]$ as the brane effective action. 
We are using a derivative expansion to solve the $T\bar T$-like flow \eqref{TTbarlikeTraceflow} for the dependence of the functional $W[\gamma,\Phi]$ on fixed boundary data $(\gamma,\Phi)$, with the potential $V(\Phi)$ being the leading zero-derivative term. Both the curvature coupling term $D(\Phi)R_{\gamma}$ and the kinetic term $\frac{1}{2}E(\Phi)(\nabla \Phi)^2$ contain two derivatives. This type of ansatz is standard in the context of solving the holographic renormalization group \cite{deBoer:1999tgo, Verlinde:1999xm}. In our equations, we use the notation  ``+ derivatives" to denote additional terms that contain derivatives of the conformal factor specifically, e.g. $\nabla_{\mu}\Phi$, while we use ``+ higher derivatives" to denote terms containing more than two derivatives. 
The leading behaviour of the stress tensor corresponding to our choice of ansatz is thus
\begin{equation}
	\langle T_{\mu\nu}\rangle =-\frac{c}{24}V(\Phi)\gamma_{\mu\nu}+\text{derivatives}.  \label{formTmn}
\end{equation}
Plugging this into the trace flow equation \eqref{TTbarlikeTraceflow} with $R_{\gamma}=-2\lambda$ (with $\lambda$ just a sign) and $t=-\frac{3}{c}$, we obtain
\begin{equation}
	-\frac{c}{12}T_0 V(\Phi)=\frac{c}{6}\lambda +\frac{c}{6}e^{\Phi}(T_0^2-1)+\frac{c}{96}e^{-\Phi}(V(\Phi))^2+\text{derivatives}. \label{quadraticeqV}
\end{equation} 
A solution to the non-derivative sector, which is a simple quadratic equation, is given by  	
\begin{align} \label{Vsol}
	V(\Phi)=4 e^{\Phi}\left(\sqrt{1-\lambda e^{-\Phi}}-T_0\right),
\end{align}
where we picked the branch yielding $V(\Phi)=0$ for $\lambda=0$. 
Including terms containing up to two derivatives, the stress tensor is
\begin{align}
	\langle T_{\mu\nu}\rangle &=\frac{c}{12}\left(\frac{1}{2}E(\Phi)\nabla_{\mu}\Phi\nabla_{\nu}\Phi-\frac{1}{2}\gamma_{\mu\nu}\left(\frac{1}{2}E(\Phi)(\nabla\Phi)^2+V(\Phi)\right)+\gamma_{\mu\nu}\Box D(\Phi)-\nabla_{\mu}\nabla_{\nu}D(\Phi)\right) \nonumber
	\\
	& \qquad \qquad \qquad +\text{higher}\;\text{derivatives}. \label{stressCarroll}
\end{align}
Plugging this into the trace flow equation \eqref{TTbarlikeTraceflow} once again together with the solution (\ref{Vsol}) for the potential, we arrive at the differential equation 
\begin{equation} \label{DprimeDseconddiffeq}
	\Box \Phi = \sqrt{1-\lambda e^{-\Phi}}\, \Box D(\Phi) +\text{higher}\;\text{derivatives}. 
\end{equation}
Taking care of the $\sim D''(\Phi)(\nabla \Phi)^2$ contribution from $\Box D(\Phi) = D'(\Phi)\Box \Phi+D''(\Phi)(\nabla \Phi)^2$ requires the addition of some non-local term to the effective action (see Appendix \ref{app:nonlocal} for the form of such term). Demanding the $\sim \Box \Phi$ terms to match on both sides of equation \eqref{DprimeDseconddiffeq}  imposes
\begin{equation} 
	D'(\Phi)=\frac{1}{\sqrt{1-\lambda e^{-\Phi}}}.
\end{equation} 
This differential equation is solved by
\begin{equation} \label{DPhisol}
	D(\Phi)=\Phi+2\log\left(1+\sqrt{1-\lambda  e^{-\Phi}}\right)+D_0,  
\end{equation}
with $D_0$ an  
integration constant (independent of $\Phi$) that we fix to 
\begin{equation} \label{D0}
	D_0=-2\log(2-\lambda^2),
\end{equation}
i.e.
\begin{align}
	&\lambda=\pm1: \qquad D_0=0, 
	\\
	&\lambda=0: \;\;\; \qquad D_0=-2\log 2.
\end{align}
This choice of integration constant ensures a direct match 
with the on-shell evaluation of the bulk total action in section \ref{sect:bulk} and is related to the choice of renormalization scheme for the asymptotic CFT. 
With this $D_0$,  
we can write out $D(\Phi)$ alternatively as 
\begin{align}
	&\lambda=1: \qquad D(\Phi)=2\cosh^{-1}\left( e^{\Phi/2}\right), \label{DPhicases0}
	\\
	&\lambda=0:  \qquad D(\Phi)=\Phi, 
	\\
	&\lambda=-1: \;\;\;\; D(\Phi)=2\sinh^{-1}\left( e^{\Phi/2}\right). \label{DPhicases0f}
\end{align}

Plugging our solutions into the ansatz \eqref{Wansatz}, we find 
\begin{align}
	W[\gamma,\Phi]
	= \frac{c}{48\pi}\int d^2x \sqrt{-\gamma}&\Bigg[  2 \log \left( \frac{ e^{\Phi/2} + \sqrt{e^\Phi - \lambda}}{2-\lambda^2} \right) R_{\gamma} \nonumber
	\\
	& \qquad \qquad + 4e^{\Phi}\left(\sqrt{1-\lambda e^{-\Phi}}-T_0\right) \Bigg] +\;\text{derivatives}. \label{Weff}
\end{align}
This is one of our main results. The non-derivative (in $\Phi$) sector of $W[\gamma,\Phi]$ is obtained in a closed form to all orders in an asymptotic expansion, and therefore it will be able to function in the following sections as brane effective action for branes at any location in the bulk. 
The derivative terms include, at lowest order in derivatives, the kinetic term for $\Phi$, which depends on the so far undetermined function $E(\Phi)$. By demanding consistency with the undeformed or asymptotic limit $\Phi \rightarrow \infty$, we will see that we must require the asymptotic behaviour
\begin{equation} \label{EPhiasymptotic}
	E(\Phi)=1+\mathcal{O}(e^{-\Phi}).
\end{equation}
Moreover, we will restrict to values for the tension with the following asymptotics
\begin{equation} \label{T0asymptotic}
	T_0=1+\mathcal{O}(e^{-\Phi}).
\end{equation}
This ensures that the undeformed CFT limit is free of power law divergences. Under the condition $T_0=K/2$ imposed by both the NBC and CBC, we will see that this is naturally required by the asymptotic structure of AdS$_3$ geometries\footnote{
	See the argument in Appendix \ref{AppLiou} for \eqref{tensionKarg}. 
}.  
With the assumptions \eqref{EPhiasymptotic}-\eqref{T0asymptotic}, we
may determine the higher derivative terms within the local sector \eqref{Wansatz} to leading order in an asymptotic expansion. The result \eqref{higherderivativesolutionappendix}, derived in Appendix \ref{app:higherderivatives}, is 
\begin{align} \label{higherderivatives}
	\text{higher}\;\text{derivatives}\;\text{in}\; W[\gamma,\Phi]=-\frac{c}{48\pi}\int d^2x\sqrt{-\gamma}&\frac{e^{-\Phi}}{32}\Big((\nabla \Phi)^4-4\Box\Phi (\nabla \Phi)^2   \nonumber \\ 
	&   - 8 R_\mn[\gamma]\nabla^\mu \Phi \nabla^\nu \Phi  
	\Big)+\mathcal{O}(e^{-2\Phi}), 
\end{align} 
with $R_\mn[\gamma]$ the Ricci tensor of $\gamma$. 

We observe that the effective action that solves the Hamilton-Jacobi equation can be naturally organized in an asymptotic expansion in powers of $e^{-\Phi}$, with the asymptotic limit given by $\Phi \rightarrow \infty$ \cite{Witten:2022xxp}. The solutions that we obtained for the potential $V(\Phi)$ (zero-derivative term) and curvature coupling $D(\Phi)$ (two-derivative term) are already closed-form expressions, i.e.~resummed in the expansion parameter. We only managed to treat the terms involving derivatives of the conformal factor perturbatively, as they become increasingly cumbersome. 
The structure of the deformation flow implies that at 
order $\mathcal{O}(e^{-(n+1)\Phi})$ the local effective action requires
terms containing at most $4+2n$ derivatives. 
We do not address in this paper how to solve for the function $E(\Phi)$
within the present construction (notice that it doesn't enter the trace of the stress tensor in \eqref{stressCarroll}).   
However, the asymptotic description of the dual CFT's integrated Weyl anomaly in terms of a timelike Liouville requires it to behave asymptotically as $E(\Phi)=1+\mathcal{O}(e^{-\Phi})$. We will postpone to section \ref{subsec:TTbarmatter} a more precise discussion of the asymptotic limit, which captures the integrated Weyl anomaly and whose  
holographic derivation is detailed in Appendix \ref{AppLiou}. Finally, note that the identity $\frac{\delta W[\gamma,\Phi]}{\delta \Phi}=-\gamma^{\mu\nu}\frac{\delta W[\gamma,\Phi]}{\delta \gamma^{\mu\nu}}$, derived in equation \eqref{delWdelPhi}, is not a condition to be imposed on the local sector of the effective action alone, but rather on the combined local and non-local sectors.

\section{Setting $T\bar{T}$ free for holography with NBC and CBC} \label{sec:settingfree}
We formally define the partition function $Z_{bw}$ describing the braneworld system \cite{Randall:1999ee,Randall:1999vf,Karch:2000ct,Porrati:2001gx,Perez-Victoria:2001lex,Arkani-Hamed:2000ijo} as the gravitational path integral $Z_{NBC}$ with Neumann boundary condition (NBC) weighted by the total action \cite{Gubser:1999vj,Giddings:2000mu,Verlinde:1999fy}
\ali{
	Z_{bw} \equiv Z_{NBC}= \int_{NBC} \mathcal{D}G \, e^{i S_{tot}[G]}, \qquad \text{Neumann bc: } \quad K_\mn - K g_\mn + T_0 g_\mn = 0 .  \label{Neumann}
}
Writing the brane's metric in conformal gauge
\begin{equation}
	g=e^{\Phi}\gamma,
\end{equation}
its variation decomposes as $\delta g=e^{\Phi}\delta \gamma+g\delta \Phi$. Therefore, the boundary variation of the total action, given by equation \eqref{deltaStot}, splits as follows
\begin{align}
	\delta S_{tot}=\int_{\mathcal{M}}d^3 X(EOM)\delta G&-\frac{1}{2\kappa}\int_{\partial \mathcal{M}}d^2x \sqrt{-g}e^{\Phi}(K^{\mu\nu}-Kg^{\mu\nu}+T_0g^{\mu\nu})\delta \gamma_{\mu\nu} \nonumber
	\\
	&+\frac{1}{2\kappa}\int_{\partial \mathcal{M}}d^2x \sqrt{-g}(K-2T_0)\delta \Phi. \label{Stotvariationsplit}
\end{align} 
 We define the gravitational path integral $Z_{CBC}(\gamma)$ with conformal boundary conditions (CBC) \cite{Witten:2022xxp} as
\begin{equation}
	Z_{CBC}(\gamma)=\int_{CBC} \mathcal{D}G e^{iS_{tot}[G]}, \qquad \text{Conformal bc:} \qquad K=2T_0, \qquad \delta \gamma=0.
\end{equation}
In the semi-classical limit $c\rightarrow\infty$ ($\kappa\rightarrow 0$), we propose that
the gravitational path integral with Neumann boundary conditions (NBC) and conformal boundary conditions (CBC) 
can then be obtained by ``setting the boundary theory free" \cite{Callebaut:2025thw}, where we borrowed the terminology of \cite{Compere:2008us}, i.e.\footnote{
	We don't explicitly include in the notation of $Z_{NBC}$ and $Z_{CBC}(\gamma)$ their dependence on $T_0$ and $K$ respectively. 
}
\begin{equation} \label{ZNBCsetfree}
	Z_{NBC}=\int \mathcal{D}\gamma \int \mathcal{D}\Phi Z_{tot}(\gamma,\Phi),  
\end{equation}
and, for $T_0=K/2$ in $S_{tot}$,
\begin{equation} \label{ZCBCsetfree}
	Z_{CBC}(\gamma)=\int \mathcal{D}\Phi Z_{tot}(\gamma,\Phi). 
\end{equation}
The boundary theory to be set free is defined by the gravitational path integral with Dirichlet boundary conditions, given in equation \eqref{Ztot}. The set-free proposal is motivated by \cite{Gubser:1999vj,Giddings:2000mu,Compere:2008us}: since $W[\gamma,\Phi]$ is of leading order $\mathcal{O}(c=\frac{1}{G})$, on the semi-classical saddle of the $\mathcal{D}\gamma$ path integral the NBC condition $\delta S_{tot}/\delta \gamma_\mn$ is imposed as an equation of motion. Similarly, on the saddle of the $\mathcal{D}\Phi$ path integral, the trace of the NBC condition $\delta S_{tot}/\delta \Phi$ is imposed as an equation of motion\footnote{ The equations of motion conjugate to $\delta \Phi$ and $\delta \gamma$ are thus not independent, with the former being identical to the trace of the latter. This can also be seen from the variation (\ref{Stotvariationsplit}) of the total action, with the momentum conjugate to $\Phi$ proportional to the trace of the momentum conjugate  to $\gamma$, or from the observation that the CBC condition $K=2T_0$ is equivalent to the trace of the NBC.}. We set both $\gamma$ and $\Phi$ free in \eqref{ZNBCsetfree} for braneworld, rather than only $\Phi$ as we previously argued in \cite{Callebaut:2025thw}. The reason for this is that in \cite{Callebaut:2025thw}, the emphasis was on setting the boundary theory free $\int \mathcal{D}g Z_{tot}(g)$. It follows from known results of two-dimensional quantum gravity that the path integral $\mathcal{D}g$ reduces to a path integral over its conformal factor $\Phi$ only, plus a ghost-action contribution which is subleading in the semi-classical expansion \cite{ZZ_Liouville_gravity}. However, here setting the boundary free $\int \mathcal{D}g Z_{tot}(g)$ comes with the additional constraint that the NBC is satisfied. Only by setting $\gamma$ free is the full NBC (rather than just its trace) semi-classically imposed. We then further require integration over $\Phi$ such that $Z_{NBC}$ is independent of both $\gamma$ and $\Phi$.

For CBC, the choice is less ambiguous. Conformal boundary conditions consist of keeping the conformal class on the boundary fixed as well as the trace of the extrinsic curvature. This could be 
achieved by setting only the Weyl mode $\Phi$ free in \eqref{ZCBCsetfree}, as the saddle condition $\delta S_{tot}/\delta \Phi$ sets $K$ equal to 
the constant $2T_0$ (and keeps $\gamma$ fixed as it was, in \eqref{Ztot}).    
In our notation aimed primarily at the NBC discussion, it is precisely the condition $T_0=K/2$ that reduces the used total action $S_{tot}$ to the one standardly used for CBC, namely the Einstein-Hilbert term \eqref{SEH} plus half of the Gibbons-Hawking term \eqref{SGH}. Nonetheless, when considering the CBC, we will take $T_0=K/2$ to be kept fixed 
from the outset in $S_{tot}$ (before making $\Phi$ dynamical). Once again, we still have to integrate over $\Phi$ 
such that $Z_{CBC}(\gamma)$ is independent of $\Phi$.

We emphasize that in going from the $Z_{tot}$ theory in \eqref{Ztot} to the $Z_{NBC}$ and $Z_{CBC}$ theories in \eqref{ZNBCsetfree}-\eqref{ZCBCsetfree}, the background fields $\gamma$ and $\Phi$ are made dynamical or `set free' by hand. 
To avoid introducing too much notation, we  
write $W[\gamma,\Phi]$ in all these expressions, but it takes on different interpretations depending on the boundary conditions: in \eqref{Ztot}, $W[\gamma,\Phi]$ is the dual theory on a fixed background with $\gamma$ and $\Phi$ fixed by the DBC. In \eqref{ZNBCsetfree}, $W[\gamma,\Phi]$ takes on the interpretation of effective dilaton gravity action on the brane, with both $\gamma$ and $\Phi$ dynamical gravitational fields. In \eqref{ZCBCsetfree}, finally, it gives the effective action on the boundary of fixed $\gamma$, with $\Phi$ dynamical. In all those cases, we keep using the square bracket notation that we otherwise reserve for denoting dependence on dynamical fields (as opposed to parametric dependence by round brackets). The interpretation should be clear from context, with the NBC application (both $\gamma$ and $\Phi$ dynamical) the most central in our paper.

Given the results of section \ref{sec:braneeffective}, in which we identify the theory dual to DBC AdS$_3$ gravity with a $T\bar{T}$-like deformed CFT$_2$ obeying the generalized trace flow equation \eqref{TTbarlikeTraceflow} (or a $T\bar{T}$-deformed CFT$_2$ obeying the trace flow \eqref{traceflowtot}) 
we could summarize our proposal for the holographic dual for semi-classical AdS$_3$ gravity with NBC/CBC at finite cutoff as a $T\bar{T}$-like deformed CFT$_2$ that is set free,  
\begin{equation}
	\text{AdS$_3$ gravity with NBC/CBC}=\text{$T\bar{T}$-like deformed CFT$_2$ set free}.
\end{equation}
Our obtained $T\bar{T}$-like deformed theory $W[\gamma,\Phi]$ provides an expression for $Z_{NBC}$ and $Z_{CBC}$, using \eqref{Ztot} in \eqref{ZNBCsetfree} and \eqref{ZCBCsetfree}, and takes on the interpretation of effective action. 
We are thus ready to use our effective action \eqref{Weff} to compute $Z_{NBC}$ and $Z_{CBC}$. Following the above arguments, in the saddle-point approximation of the boundary path integrals, equations \eqref{ZNBCsetfree}-\eqref{ZCBCsetfree} reduce to
\begin{equation}
	Z_{CBC}(\gamma)=e^{iW[\gamma,\Phi^{\star}]}, \qquad Z_{NBC}=e^{iW[\gamma^{\star},\Phi^{\star}]},
\end{equation}
with $\Phi^{\star}$ and $\gamma^{\star}$ solutions to the equations of motion   
\begin{equation}
	\frac{\delta W[\gamma,\Phi]}{\delta \Phi}
	=0, \qquad \frac{\delta W[\gamma,\Phi]}{\delta \gamma^{\mu\nu}}
	=0.
\end{equation}
The equation of motion conjugate to $\delta\Phi$ is
\begin{equation} \label{deltaWdeltatilderho}
	\left(\frac{R_{\gamma}+2\lambda}{\sqrt{1-\lambda e^{-\Phi}}}+4e^{\Phi}\left(\sqrt{1-\lambda e^{-\Phi}}-T_0\right)+\text{derivatives}\right)
	=0
\end{equation}
and the one conjugate to $\delta \gamma$ is
\begin{equation}
	\left(4e^{\Phi}\left(\sqrt{1-\lambda e^{-\Phi}}-T_0\right)+\text{derivatives}\right)
	=0.
\end{equation} 
A constant solution to these equations of motion is provided by any constant conformal factor for $\lambda=0$ and by the constant conformal factor
\begin{equation} \label{Phistarsol}
	\Phi^{\star}=\log\left(\frac{\lambda}{1-T_0^2}\right)
\end{equation}
for $\lambda=\pm1$, together with the constant curvature metric $R_{\gamma^{\star}}=-2\lambda$, i.e. $\gamma^{\star}=\gamma$ the reference $(A)dS_2$ or $Mink_2$ metric as introduced in section \ref{sec:braneeffective} Eq.~\eqref{gPhigamma}.   
Note of course that the derivative terms as well as the non-local terms vanish on such constant $\Phi^{\star}$ solutions and therefore do not affect them. 
Thus, for the chosen solution \eqref{Phistarsol}, we find 
\ali{
	D(\Phi^\star) = \log\left(\lambda\frac{1+T_0}{1-T_0}\right) 
} 
and $V(\Phi^\star) = 0$, so that 
\begin{align} \label{ZNBCT0new}
	\log Z_{NBC}&=-i\frac{\lambda c}{24\pi} V_2 \,  \log\left(\lambda\frac{1+T_0}{1-T_0}\right), 
\end{align}
and
\begin{align} \label{ZCBCKnew}
	\log Z_{CBC}&=-i\frac{\lambda c}{24\pi} V_2 \,  \log\left(\lambda\frac{2+K}{2-K}\right),
\end{align}
where we introduced $V_2 = \int d^2 x \sqrt{-\gamma}$ as a volume factor.

\subsection{Bulk geometries and on-shell actions} \label{sect:bulk}
Having discussed the saddles of the EOW theory, let us now discuss the corresponding bulk on-shell geometries. 
They are of the form 
\begin{equation} \label{bulkmetricansatz}  
	ds^2=G_{MN}(X)dX^{M}dX^{N}=d\rho^2+e^{2 A(\rho)}\gamma_{\mu\nu}(x)dx^{\mu}dx^{\nu},
\end{equation}
where $G_{MN}$ and $X^{M}:(\rho,x^{\mu})$ denote the bulk metric and coordinates respectively, $\rho$ is a radial coordinate such that the asymptotic boundary is reached as $\rho\rightarrow \infty$, $x^{\mu}$ are coordinates spanning over radial slices, $A(\rho)$ is the warping factor and $\gamma_{\mu\nu}(x)$ is the 2d metric. We restrict to the following three choices of warping factors, corresponding respectively to $AdS_2$, flat and $dS_2$ radial slicing of empty AdS$_3$\footnote{The range of the $\rho$ coordinate is $\rho\in(-\infty,+\infty)$ for $\lambda=1,0$, while $\rho\in[0,+\infty)$ for $\lambda=-1$. Note that these coordinate ranges refer to the corresponding slicing coordinates and do not necessarily constitute global coordinate charts on AdS$_3$. In particular, for $\lambda=0$, $\rho\to-\infty$ corresponds to the Poincar\'e horizon. }
\begin{align} \label{warpingfactors}
	\begin{split} 
	&AdS_2 \; (\lambda=1): \; \; \; \; \;\; \; \; \; \; e^{2A(\rho)}=(\cosh\rho)^2, \; \; \; \; \;\; \; \; \; \; 
	\\
	&Mink_2 \; (\lambda=0):\; \; \; \; \; \; \; \,  e^{2A(\rho)}= e^{2\rho}, \; \; \; \; \;\; \; \; \; \; \; \; \; \;\; \; \; \;  \;  
	\\
	&dS_2 \; (\lambda=-1): \; \; \; \; \;\; \; \; \; \;  e^{2A(\rho)}= (\sinh\rho)^2, \; \; \; \; \;\; \; \; \; \;  \, 
	\end{split} 
\end{align}
For example, in planar coordinates, $\gamma_\mn dx^\mu dx^\nu = \frac{-dt^2 + dZ^2}{Z^2}$ for $\lambda=1$,  $\gamma_\mn dx^\mu dx^\nu =-dt^2 + dZ^2$ for $\lambda=0$ and $\gamma_\mn dx^\mu dx^\nu = \frac{-dt^2 + dZ^2}{t^2}$ for $\lambda=-1$.  
Using $e^{2A(\rho)}= \left(\frac{e^{\rho}+\lambda e^{-\rho}}{1+\lambda^2}\right)^2$, these geometries may be written in Fefferman-Graham (FG) gauge 
 \begin{equation} \label{FG}
	G_{MN}dX^MdX^N=d\rho^2+e^{2\rho}\left(g^{(0)}_{\mu\nu}(x)+ e^{-2\rho}g^{(2)}_{\mu\nu}(x)+e^{-4\rho}g^{(4)}_{\mu\nu}(x)\right)dx^{\mu}dx^{\nu},
\end{equation}
with
\begin{equation}
	g_{(0)}=\frac{1}{(1+\lambda^2)^2}\gamma, \qquad g_{(2)}=\frac{\lambda}{2}\gamma, \qquad g_{(4)}=\frac{\lambda^2}{(1+\lambda^2)^2}\gamma.
\end{equation} 
Inserting an EOW brane at some constant radius
\begin{equation}
	EOW: \; \; \; \; \; \rho=\bar{\rho},
\end{equation}
we consider cutting off the portion of the asymptotic region of the bulk geometry contained between the brane and the asymptotic boundary. The induced metric on the EOW brane is
\begin{equation}
	g_{\mu\nu}\big|_{\rho=\bar{\rho}}\equiv \hat{g}_{\mu\nu}=e^{2 A(\bar{\rho})}\gamma_{\mu\nu}.  
\end{equation}
The extrinsic curvature and its trace on the brane are given by $K_{\mu\nu}\big|_{\rho=\bar{\rho}}= \dot{A}g_{\mu\nu}\big|_{\rho=\bar{\rho}}$ and $K\big|_{\rho=\bar{\rho}}=2 \dot{A}\big|_{\rho=\bar{\rho}}$, where $\dot{A}\equiv \partial_{\rho}A(\rho)$. For the warping factors in \eqref{warpingfactors}, for which
$\ddot A+\dot A^2=1$, the Ricci scalar of the bulk metric $G_{MN}$ and of the metric $\gamma_{\mu\nu}$ are related as follows
\begin{equation} \label{riccirelations}
	R_G=e^{-2A}R_{\gamma}-2 \ddot{A} - 4 \dot{A}^2-2,
\end{equation}
where $\ddot{A}=\partial_{\rho}^2 A(\rho)$. Using the above expressions, we find that the on-shell total action induces an effective action on the brane of the form 
\begin{equation} \label{bulkactionconstantgamma}
	S_{tot}[G(\rho<\bar{\rho},x)]=S_0[\gamma]+\frac{1}{8\pi G}\int d^2 x \sqrt{-\gamma}\,  e^{2A}\left(\dot{A}-T_0\right)\Big|_{\rho=\bar{\rho}},
\end{equation}
where
\begin{equation} \label{S_0}
	S_0[\gamma]=\frac{\bar{\rho}}{16\pi G}\int d^2 x \sqrt{-\gamma} R_{\gamma}
\end{equation}
is an entirely topological term: its variation vanishes, as it is proportional to the Einstein-tensor of the 2d metric $\gamma_{\mu\nu}$. Note that, here and in the following, we discard any contribution from the lower bound of radial integration in the Einstein-Hilbert term.
The equation of motion conjugate to $\delta \gamma$ variations derived from Eq.~(\ref{bulkactionconstantgamma}) imposes a constraint on the warping factor
\begin{equation}
	\frac{\delta 	S_{tot}[G(\rho<\bar{\rho},x)]}{\delta \gamma^{\mu\nu}}=0 \implies T_0= \dot{A}\big|_{\rho=\bar{\rho}},
\end{equation}
which is equivalently enforced by the conformal boundary condition (CBC)
\begin{equation} \label{cbckt0}
	CBC: \; \; \; \; K\big|_{\rho=\bar{\rho}}=2T_0,
\end{equation}
or by the trace of the Neumann boundary condition (NBC)
\begin{equation}
	NBC: \; \; \; \;  \left[K_{\mu\nu}-(K-T_0)g_{\mu\nu}\right]\Big|_{\rho=\bar{\rho}}=0.
\end{equation}
This establishes a relation between the tension (or the trace of the extrinsic curvature on the brane for CBC using Eq.~(\ref{cbckt0})) and the cutoff radius
\begin{align} 
	\begin{split} 
	&AdS_2 \; (\lambda=1): \; \; \; \; \;\; \; \; \; \; T_0=\tanh\bar{\rho}<1, 
	\\
	&Mink_2 \; (\lambda=0):\; \; \; \; \; \; \; \,  T_0=1, 
	\\
	&dS_2 \; (\lambda=-1): \; \; \; \; \;\; \; \; \; \; T_0=\coth\bar{\rho}>1  
	\end{split} \label{tensions}
\end{align}
or more succinctly 
\ali{
	\bar \rho = \frac{1}{2}\log\left(\lambda\frac{1+T_0}{1-T_0}\right). 
}
It is also useful to introduce an alternative notation $\epsilon$ for the constant $\bar \rho$ brane location  
\ali{
	\epsilon = e^{-A(\bar \rho)} \label{defepsilon}
}
in terms of which the tension is given by\footnote{For notational simplicity, we restrict here to a brane located at $\rho=\bar{\rho}\geq 0$ and thus to non-negative tensions $T_0\ge0$. This entails no loss of generality for the $\lambda=1$ case, since the geometry is invariant under the $Z_2$ reflection $\rho\to-\rho$, owing to the evenness of $\cosh\rho$.} 
\ali{
	T_0 = \sqrt{1 - \lambda \epsilon^2}. \label{T0epsilon}
}
In this notation, $\epsilon$ is any finite parameter but becomes small in the asymptotic brane limit 
$\bar \rho \ra \infty$, which is useful to consider for comparison to previous work.  
The on-shell total action evaluated with the CBC or NBC can be expressed in terms of the cutoff radius satisfying \eqref{tensions}, as
\begin{equation} \label{StotNBC}
	S_{tot}[G(\rho<\bar{\rho},x)]=S_0[\gamma]=\frac{\bar{\rho}}{16\pi G}\int d^2 x \sqrt{-\gamma}R_{\gamma}\; . 
\end{equation} 
In the semi-classical approximation of the holographic dictionary, we may then write the logarithm of the partition functions for $\lambda=\pm 1 $ in terms of the tension  for NBC
\begin{equation} \label{ZNBCT0}
	\log Z_{NBC}=-i\frac{\lambda c}{24\pi} \, V_2 
	\log\left(\lambda\frac{1+T_0}{1-T_0}\right), 
\end{equation}
or in terms of the trace of the extrinsic curvature for CBC
\begin{equation} \label{ZCBCK}
	\log Z_{CBC}=-i\frac{\lambda c}{24\pi} \, V_2 
	\log\left(\lambda\frac{2+K}{2-K}\right). 
\end{equation}
This matches the results of the previous section, with
\ali{
	\Phi^{\star}=-2\log\epsilon, \qquad D(\Phi^\star) = 2  \bar \rho. \label{Dtobarrho} 
} 
Note that while in section \ref{sec:braneeffective} we solved the radial Hamiltonian flow or $T\bar T$-like flow for the  
bulk on-shell action, 
in this section we have calculated the bulk on-shell action directly.  
As postulated by the holographic dictionary, the on-shell total action with the DBC $g=e^{\Phi^{\star}}\gamma=\hat{g}(\gamma)=\gamma/\epsilon^2$ matches the effective action \eqref{Weff}
	\begin{align}
		W[\gamma,\Phi^{\star}]&=	S_{tot}[G(\rho<\bar{\rho},x)]=S_0[\gamma]+\frac{1}{8\pi G}\int d^2 x \sqrt{-\gamma}\,  e^{2A}\left(\dot{A}-T_0\right)\Big|_{\rho=\bar{\rho}} \nonumber
		\\
		&=\frac{c}{48\pi}\int d^2x \sqrt{-\gamma}\Big[ \left(\Phi^{\star}+2\log\left(1+\sqrt{1-\lambda  e^{-\Phi^{\star}}}\right)-2\log(2-\lambda^2)\right)R_{\gamma} \nonumber
		\\
		& \qquad \qquad \qquad \qquad \qquad \qquad \qquad  + 4e^{\Phi^{\star}}\left(\sqrt{1-\lambda e^{-\Phi^{\star}}}-T_0\right)\Big]. 
		\label{Wbulkviolet}
	\end{align}
Notice once again that the derivative terms in \eqref{Weff} (including the non-local sector) vanish for constant $\Phi^{\star}$. 
The match justifies the choice of integration constant in equation \eqref{D0}. 
In the FG expansion \eqref{FG}, there is a factor of 4 difference between the conformal boundary $g_{(0)}$ in FG coordinates and the reference metric $\gamma$ in the $\lambda=0$ versus the $\lambda = \pm 1$ case. It is this factor 4 difference that is responsible for the $\lambda$-dependency of $D_0$ in \eqref{D0}.

Similarly, from the semi-classical limit of the holographic $T\bar{T}$ dictionary \eqref{TTbardictionary}, we can compute the generating functional of the $T\bar{T}$-deformed CFT$_2$ dual to our bulk geometry with DBC
	\begin{align}
		W_{T\bar{T}}[\hat{g}(\gamma)]&=	S_{grav}[G(\rho<\bar{\rho},x)]=S_0[\gamma]+\frac{1}{8\pi G}\int d^2 x \sqrt{-\gamma}\,  e^{2A}\left(\dot{A}-1\right)\Big|_{\rho=\bar{\rho}} \nonumber 
		\\
		&=\frac{c}{48\pi}\int d^2x \sqrt{-\gamma}\Big[ \left(2\log\left(\frac{1}{\epsilon}+\sqrt{\frac{1}{\epsilon^2}-\lambda  }\right)-2\log(2-\lambda^2)\right)R_{\gamma} \nonumber
		\\
		& \qquad \qquad \qquad \qquad \qquad \qquad \qquad  + \frac{4}{\epsilon^2}\left(\sqrt{1-\lambda \epsilon^2}-1\right)\Big]. \label{Wttbarghatbulk}		 
	\end{align}
We will use this result in the next section, where we will show how our effective action can be naturally split into a matter sector and a gravitational sector. In section \ref{sec:fluctuating}, we will consider evaluating the on-shell total action with a position-dependent cutoff radius $\rho=\tilde{\rho}(x)$, rather than the constant $\rho=\bar{\rho}$. This will allow us to compare also the derivative terms in the effective action with a direct bulk action calculation and, by studying the asymptotic limit, will provide an adequate framework to holographically derive 
the timelike Liouville description of the integrated Weyl anomaly.

\section{Identifying matter and gravitational sectors} \label{sec:identifyingmatter}
Motivated by  
standard strategies in braneworld holography \cite{Gubser:1999vj,Giddings:2000mu,Verlinde:1999xm}, we consider decomposing the full effective brane theory $W[\gamma,\Phi]$ into a matter sector and a gravitational sector
\ali{
	W = W_{mat} + S.  \label{generalsplit}
} 
The stress tensor $\langle T_{\mu\nu} \rangle=\frac{4\pi}{\sqrt{-\gamma}}\frac{\delta W}{\delta \gamma^{\mu\nu}}$ correspondingly splits into the sum of a matter stress tensor and a geometrical part 
\ali{
	\vev{T_\mn} = \vev{\hat T_\mn} + t_\mn , 
}
with $\langle \hat{T}_{\mu\nu} \rangle=\frac{4\pi}{\sqrt{-\gamma}}\frac{\delta W_{mat}}{\delta \gamma^{\mu\nu}}$ and $t_{\mu\nu} =\frac{4\pi}{\sqrt{-\gamma}}\frac{\delta S}{\delta \gamma^{\mu\nu}}$. 
There is freedom in how to identify or separate out a matter sector.
We discuss three different choices that each have their use. 

The first, 
\ali{
W_{mat} = W_{T\bar T-like}[\gamma,\Phi], \qquad S=\tilde{S}[\gamma,\Phi],
}
discussed in section \ref{subsec:TTbarlikematter}, is the one that makes contact with 
the 
proposal of \cite{Allameh:2025gsa}. 
In \cite{Allameh:2025gsa}, the particular case of interest is the CBC theory, 
but our discussion applies to NBC and CBC (according to \eqref{ZNBCsetfree} and \eqref{ZCBCsetfree}). 

The second, 
\ali{
	W_{mat} = W_{T\bar T}[\hat g(\gamma),\sigma], \qquad  S=\hat{S}[\gamma,\sigma],
}
is discussed in section \ref{subsec:TTbarmatter} and gives the direct generalization to non-unit tension of our previous work \cite{Callebaut:2025thw}. It naturally extends the timelike Liouville description of the integrated Weyl anomaly further into the bulk. 

The third, 
\ali{
	W_{mat} = W_{CFT}[\gamma],  \qquad S=\bar{S}[\gamma,\Phi],
}
provides the most useful description to apply in an island rule set-up. 
This is discussed in section \ref{subsec:undeformedmatter}, with an application to the island rule away from the asymptotic brane regime presented in section \ref{subsec:islandrule}.

\subsection{$T\bar T$-like deformed conformal matter} \label{subsec:TTbarlikematter}

In \cite{Callebaut:2025thw}, we introduced the term ``$T\bar T$-like deformed'' to describe a theory that satisfies a trace flow of the form \eqref{TTbarlikeTraceflow}, including in particular a $T\bar T$ deformation $t e^{-\Phi} \vev{\mathcal O_{T\bar T}}$ that comes with a Weyl rescaled $T\bar T$ coupling, $t \ra t e^{-\Phi}$. The specific case considered in \cite{Callebaut:2025thw} was that of $R_\gamma = 0$ and $T_0=1$, whereas the discussion in section \ref{sec:braneeffective} is for constant
reference curvature $R_\gamma = -2\lambda$ (for $\lambda=0,\pm 1$) and arbitrary constant tension $T_0$. The whole effective theory $W[\gamma,\Phi]$ in \eqref{Weff} was obtained from integrating the trace flow \eqref{TTbarlikeTraceflow} and is thus a $T\bar T$-like deformed CFT$_2$. For constant $\Phi$, the full $T\bar T$-like deformed functional \eqref{Weff} can be written as 
\ali{
	W &= -\frac{V_2}{4\pi} \left[ \lambda \frac{c}{3} \log \left(\frac{ \sqrt{\frac{3}{-c t} e^\Phi} + \sqrt{\frac{3}{-c t} e^\Phi - \lambda}}{2-\lambda^2} \right) + \frac{1}{t} e^\Phi \left(  \sqrt{1 + \lambda \frac{c t}{3} e^{-\Phi}} - T_0  \right) \right] \label{WDSTTlike}
}    
in a direct generalization of the $T\bar T$ procedures in \cite{Donnelly:2018bef,Deng:2023pjs}, which we summarize in Appendices \ref{Appendix:B.1}-\ref{Appendix:B.2}. By construction, it satisfies the flow\footnote{On constant $\Phi$ and for $T_0=1$, the $T\bar T$-like $W$ reduces back to the $T\bar T$ theory $W$ of \eqref{normZS2} or \eqref{normZH2}, with $e^{\Phi/2}=r$ the radius of the boundary geometry in the notation of Appendix \ref{appendix:B} and \cite{Donnelly:2018bef}.} 
\ali{
	t \frac{dW}{dt} 
	= -\frac{\delta W}{\delta \Phi} = \frac{1}{4\pi}\int d^2x\sqrt{-\gamma}\langle \gamma^{\mu\nu}T_{\mu\nu}\rangle=
	\frac{V_2}{4\pi} (2\alpha),  \label{consistentflows}
}  
consistent with scaling arguments based on the dimension of $t$  
\cite{Marolf18}. 
Here $\alpha$ refers to the trace of the stress tensor, whose leading behavior (up to derivatives of $\Phi$ which vanish for constant $\Phi$) is given by $\vev{T_\mn} = \alpha \gamma_\mn$. It is given explicitly by 
\ali{
	\alpha = \frac{1}{2t} e^{\Phi} \left( \sqrt{1 + \lambda \frac{c t}{3} e^{-\Phi}} - T_0 \right)  \label{alphaTTbarlike}
}
in order to satisfy the zeroth order trace flow equation  $2 T_0 \alpha = \frac{c}{6} \lambda + \frac{c}{6}e^\Phi (T_0^2-1) - 2 t e^{-\Phi} \alpha^2$.
Indeed these equations reduce to 
equations \eqref{formTmn}, \eqref{quadraticeqV} and \eqref{Vsol} of section \ref{subsec:solvingtheflow}.    
 
Now let us consider writing the following decomposition of the effective theory
\ali{ 
	W[g]=W_{T\bar{T}-like}[\gamma,\Phi]+\tilde{S}[\gamma,\Phi],  
}
as a matter sector consisting of $T\bar T$-like deformed matter with stress tensor $\vev{\hat T_\mn} = \frac{4\pi}{\sqrt{-\gamma}} \frac{\delta W_{mat}}{\delta \gamma^\mn}$, coupled to the effective gravity action $\tilde{S}$, with stress tensor $t_\mn = \frac{4\pi}{\sqrt{-\gamma}} \frac{\delta \tilde S}{\delta \gamma^\mn}$. 
In particular, we want to consider a matter sector consisting of a $T\bar T$-like deformed CFT$_2$ that satisfies a trace flow equation of the form 
\ali{
	 W_{mat}=W_{T\bar T-like}[\gamma,\Phi]: \qquad \vev{\hat T_\mn \gamma^\mn} = -\frac{c}{12} R_\gamma + t e^{-\Phi} \vev{\mathcal{\hat O}_{T\bar T}} \, + \, \text{derivatives}, \label{TTbarlikeflow}  
}
containing a deformation $t e^{-\Phi}  \langle \hat{\mathcal{O}}_{T\bar{T}} \rangle$ with $\hat{\mathcal{O}}_{T\bar{T}}=\hat{T}_{\mu\nu}\hat{T}_{\alpha\beta}\gamma^{\alpha\mu}\gamma^{\beta\nu}-(\hat{T}_{\mu\nu}\gamma^{\mu\nu})^2$, but also derivative terms of $\Phi$, captured here again in `derivatives'. Since the trace flow equation \eqref{TTbarlikeflow} for the above chosen matter sector coincides with the trace flow equation \eqref{TTbarlikeTraceflow} for the total effective action when $T_0=1$, the stress tensor for $T\bar T$-like matter must also take the form 
\ali{
	\vev{\hat T_\mn} = \alpha \gamma_\mn \, + \, \text{derivatives} \label{Tmnlike}
}
with $\alpha$ given by \eqref{alphaTTbarlike} with $T_0$ set to one. 
Given that $W$ itself is a $T\bar T$-like deformed theory, imposing the matter sector to satisfy the simplest version of a $T\bar T$-like flow \eqref{TTbarlikeflow}, 
namely the same form of the trace flow equation as 
\eqref{TTbarlikeTraceflow} up to the potential term $\frac{c}{6} e^{\Phi}(T_0^2-1)$ and the factor $T_0$ multiplying the trace, 
simply results in a gravitational part $\tilde S$ containing only a potential contribution correcting for the cases in which $T_0 \neq 1$, i.e.~$W[\gamma,\Phi]=W_{T\bar{T}-like}[\gamma,\Phi]+\frac{1}{8\pi G}\int d^2x \sqrt{-\gamma}e^{\Phi}(1-T_0)+\text{derivatives}$, for $\langle \hat{T}_{\mu\nu}\gamma^{\mu\nu}\rangle=-\frac{c}{12}(R_{\gamma}-\Box \Phi)+te^{-\Phi}\langle \hat{O}_{T\bar{T}}\rangle$. 
However, integrating the flow of the matter generating functional 
\ali{
	\frac{dW_{T\bar{T}-like}[\gamma,\Phi]}{dt} = \frac{1}{4\pi} \int d^2 x \sqrt{-\gamma} \left(-\frac{c}{12 t} R_\gamma +  e^{-\Phi} \vev{\mathcal{\hat O}_{T\bar T}} \, + \, \text{derivatives}\right)   \label{dtWmat}
} 
corresponding to the trace flow equation \eqref{TTbarlikeflow}, one can also choose the different normalization 
\ali{
	W_{T\bar{T}-like} [\gamma,\Phi]
	&= -\frac{V_2}{4\pi} \left[ - \Phi R_\gamma + \lambda \frac{c}{3} \log \left(\frac{ \sqrt{\frac{3}{-c t} e^\Phi} + \sqrt{\frac{3}{-c t} e^\Phi - \lambda}}{2-\lambda^2} \right)
	\right. \nonumber \\ 
	&\qquad \qquad \left. + \frac{1}{t} e^\Phi \left(  \sqrt{1 + \lambda \frac{c t}{3} e^{-\Phi}} - 1 \right) \right] \, + \, \text{derivatives}. \label{WTTbarlikedifferentnormalization}
}
The $(\Phi R_\gamma)$ term can be thought of as a different choice of ($t$-independent) integration constant \cite{Allameh:2025gsa}, compared to the integration of \eqref{consistentflows} to \eqref{WDSTTlike} with $T_0$ set to one.  
Restoring the full functional form in equation \eqref{WTTbarlikedifferentnormalization}, we can write
\ali{
	W_{T\bar{T}-like}[\gamma,\Phi] &= \frac{c}{48\pi} \int d^2 x \sqrt{-\gamma} \left( (D(\Phi) - \Phi) R_\gamma - 4 e^{\Phi} (1 - \sqrt{1 - \lambda e^{-\Phi}}) \right) \nonumber \\ 
	&\qquad \qquad \qquad \qquad \qquad  \qquad \qquad \qquad +\text{derivatives} \label{Wmatfctional}
}
with $D(\Phi)$ defined in \eqref{DPhisol}. 
We see from the curvature coupling terms in \eqref{Wmatfctional} that indeed one must include derivative terms in the stress tensor ansatz \eqref{Tmnlike} for self-consistency. Consistency with the general solution strategy of section \ref{subsec:solvingtheflow} reveals 
that the choice of normalization in \eqref{WTTbarlikedifferentnormalization} in particular corresponds to a $T\bar T$-like matter sector satisfying the trace flow 
\ali{
	\vev{\hat T_\mn \gamma^\mn} = -\frac{c}{12} R_\gamma + t e^{-\Phi} \vev{\mathcal{\hat O}_{T\bar T}} \, + \, \frac{c}{12} ( 1 - \sqrt{1 - \lambda e^{-\Phi}}) \Box \Phi \, + \, \text{higher derivatives}.   \label{TTlikeflow}
}
By simple subtraction from the full effective theory $W$ given in \eqref{Weff}, the corresponding gravitational sector is then determined to be 
\begin{equation} \label{Stildae}
	\tilde{S}[\gamma,\Phi]=\frac{1}{32\pi G}\int d^2x \sqrt{-\gamma}\left[\Phi R_{\gamma}+4(1-T_0)e^{\Phi}\right]+\text{derivatives} . 
\end{equation}
A kinetic term of the form $\frac{1}{2}\tilde{E}(\Phi)(\nabla \Phi)^2$, with $\tilde{E}(\Phi)$ so far unconstrained, is allowed within the unspecified derivative terms. For the CBC case, with $T_0=K/2$, equation \eqref{Stildae} agrees with the non-derivative sector of the timelike Liouville action in the proposal of \cite{Allameh:2025gsa}.
 
\subsection{$T\bar T$-deformed conformal matter} \label{subsec:TTbarmatter}

The Weyl mode $\Phi$ introduced in \eqref{gPhigamma} contains the information on the 
location of the brane in the bulk. We have referred to  
this location in section \ref{sect:bulk} using $\bar \rho$ or $\epsilon$ defined in \eqref{defepsilon}. 
We can extrapolate the constant (in general finite) radial cutoff $\epsilon$ by introducing a new Weyl mode 
\ali{
	\sigma(x) = \Phi(x) + 2 \log \epsilon, \label{sigmadef1}  
}
such that the induced metric on the brane takes the form 
\ali{
	g_\mn(x) = e^{\sigma(x)}\hat g_\mn(x) = \frac{e^{\sigma(x)}}{\epsilon^2} \gamma_\mn(x). \label{sigmadef2}
} 
The $(\sigma, \hat g)$ notation directly matches that of the flat brane discussion in \cite{Callebaut:2025thw}, which we extend in this section to include curved branes. As we will see, the extrapolation of a constant cutoff $\epsilon$ from the Weyl factor $\Phi$, together with the introduction of the new conformal mode $\sigma$, is particularly useful in order to study the asymptotic limit $\epsilon \rightarrow 0$ of the brane theory without killing its dynamical content, which would occur if one takes  the $\Phi \rightarrow \infty$ limit directly in \eqref{Weff}. It is also convenient to introduce the rescaled deformation coupling
\begin{equation}
	\tilde{t}=t\epsilon^2=-\frac{3\epsilon^2}{c}.
\end{equation} 
The asymptotic limit $\epsilon \ra 0$ or $\bar{\rho}\rightarrow \infty$ then corresponds to the undeformed limit $\tilde{t}\rightarrow 0$.

We  employ the following decomposition for the boundary theory's generating functional 
\begin{equation}
	W[g]=W_{T\bar{T}}[\hat{g}(\gamma)]+\hat{S}[\gamma,\sigma], \label{splitghat}
\end{equation} 
and for the corresponding dual stress tensor
\begin{equation} \label{stresstendeco}
	\langle \mathcal T_{\mu\nu}\rangle=\langle \hat{T}_{\mu\nu}\rangle +t_{\mu\nu},
\end{equation}
with $\langle \hat{T}_{\mu\nu}\rangle=\frac{4\pi}{\sqrt{-\gamma}}\frac{\delta W_{T\bar{T}}[\hat{g}(\gamma)]}{\delta \gamma^{\mu\nu}}=\frac{4\pi}{\sqrt{-\hat{g}}}\frac{\delta W_{T\bar{T}}[\hat{g}(\gamma)]}{\delta \hat{g}^{\mu\nu}}$ and $t_{\mu\nu}=\frac{4\pi}{\sqrt{-\gamma}}\frac{\delta \hat{S}[\gamma,\sigma]}{\delta \gamma^{\mu\nu}}$. Plugging the stress tensor decomposition \eqref{stresstendeco} into the trace flow equation \eqref{traceflowtot} and using $\sqrt{-g}R=\sqrt{-\hat{g}}\left(R_{\hat{g}}-\hat{\Box}\sigma\right)$, we obtain 
\begin{align} \label{tracefloweqT0hatTt}
	&T_0 \left\langle \left(\hat{T}_{\mu\nu}+t_{\mu\nu}\right)\hat{g}^{\mu\nu} \right\rangle=-\frac{c}{12}\left(R_{\hat{g}}-\hat{\Box}\sigma\right) +\frac{c}{6}e^{\sigma}(T^2_0-1) \nonumber
	\\
	&\qquad +te^{-\sigma}\left\langle \left(\hat{T}_{\mu\nu}+t_{\mu\nu}\right)\left(\hat{T}_{\alpha\beta}+t_{\alpha\beta}\right)\hat{g}^{\mu\alpha}\hat{g}^{\nu\beta}-\left(\left(\hat{T}_{\mu\nu}+t_{\mu\nu}\right)\hat{g}^{\mu\nu}\right)^2 \right\rangle.  
\end{align}

We can then proceed to separate out a $T\bar T$ matter sector that lives on the $\hat g_\mn$ geometry. That is, we impose that the matter stress tensor obeys the $T\bar T$ trace flow equation  
\begin{align} \label{ttbartraceflowreference}
	W_{mat}=W_{T\bar T}[\hat{g}(\gamma)]: \qquad&\left\langle \hat{T}_{\mu\nu}\hat{g}^{\mu\nu} \right\rangle=-\frac{c}{12}R_{\hat{g}} +t\left\langle \hat{T}_{\mu\nu}\hat{T}_{\alpha\beta}\hat{g}^{\mu\alpha}\hat{g}^{\nu\beta}-\left(\hat{T}_{\mu\nu}\hat{g}^{\mu\nu}\right)^2 \right\rangle.  
\end{align}  
As is standard in $T\bar T$, it is immediate from this equation that the $T\bar T$ theory on $\hat g_\mn$ with deformation parameter $t$ is equivalent to a $T\bar T$ theory on $\gamma_\mn$ with deformation parameter $\tilde t$. 
The stress tensor will be proportional to the background metric \cite{Donnelly:2018bef} 
\ali{ 
	\vev{\hat{T}_{\mu\nu}}=\epsilon^2\alpha\hat{g}_{\mu\nu}=\alpha \gamma_{\mu\nu}, \qquad \alpha=\frac{1}{2\tilde{t}}\left(\sqrt{1+\lambda \frac{c\tilde{t}}{3 }}-1\right),  \label{this}
} 
with $\alpha$ satisfying the trace flow equation $\alpha^2+\frac{\alpha}{\tilde{t}}-\frac{c\lambda}{12 \, \tilde{t} }=0$.  
We picked the branch yielding the correct conformal anomaly in the undeformed $\tilde{t}\rightarrow0$ limit, i.e.~$\lim\limits_{\tilde{t}\rightarrow 0}\langle \hat{T}_{\mu\nu}\hat{g}^{\mu\nu}\rangle
=\frac{c}{6}\epsilon^2\lambda
=-\frac{c}{12}R_{\hat{g}}$. 
Plugging back 
\eqref{this} into equation  \eqref{tracefloweqT0hatTt}, we get the following trace flow equation for the stress tensor $t_{\mu\nu}$ 
\begin{align} \label{traceflowthat}
	&\left(T_0+2\tilde{t}e^{-\sigma}\alpha \right)t^{\mu}_{\mu}=\frac{c}{6}\lambda+\frac{c}{12}\Box \sigma - \frac{e^{\sigma}}{2\tilde{t}} (T_0^2-1)+\tilde{t}e^{-\sigma}\mathcal{O}_{t\bar{t}}-2\tilde{t}e^{-\sigma}\alpha^2-2\alpha T_0,
\end{align}
now written with respect to the metric $\gamma_\mn$, 
with $t^{\mu}_{\mu}=t_{\mu\nu}\gamma^{\mu\nu}$ and $\mathcal{O}_{t\bar{t}}=t_{\mu\nu}t_{\alpha\beta}\gamma^{\alpha\mu}\gamma^{\beta\nu}-(t_{\mu\nu}\gamma^{\mu\nu})^2$. 

To obtain the gravitational part $\hat{S}$ in \eqref{splitghat} in functional form, we can either solve for $W_{T\bar{T}}[\hat{g}(\gamma)]$ in functional form and subtract it from the full effective theory $W$, or we can directly solve the flow equation \eqref{traceflowthat} for $\hat{S}$. Proceeding with the latter, we solve \eqref{traceflowthat} 
following the same solving strategy of section \ref{subsec:solvingtheflow}: we start from an ansatz for the local part of $\hat{S}$ of the form 
\ali{
	\hat{S}[\gamma,\sigma]=\frac{c}{48\pi}
	\int d^2x \sqrt{-\gamma}\left(\hat D(\sigma) R_{\gamma}+\frac{1}{2}\hat E(\sigma)(\nabla\sigma)^2+ \hat V(\sigma)\right)+\text{higher}\;\text{derivatives},
	}  
and solve \eqref{traceflowthat} in a derivative expansion. At zeroth order in derivatives, equation \eqref{traceflowthat} fixes $\hat V$ to 
\ali{
	\hat V(\sigma) &=  \frac{4e^{\sigma}}{\epsilon^2} \left( \sqrt{1-\lambda \epsilon^2 e^{-\sigma}}+e^{-\sigma}\left(1-\sqrt{1-\lambda \epsilon^2}\right)-T_0 \right). \label{Vsigmasol} 
}
 Up to the non-local contributions discussed in Appendix \ref{app:nonlocal}, the next order in derivatives fixes $\hat D$ to
\begin{align} \label{Dsigmasol}
	\hat D(\sigma)&=\sigma+2\log\left(1+\sqrt{1-\lambda \epsilon^2 e^{-\sigma}}\right)+\hat{D}_0,
\end{align}
with $\hat{D}_0$ a $\sigma$-independent integration constant. As mentioned above, this procedure should be consistent with subtracting from the full $W$ the $T\bar T$-deformed sector, i.e. $\hat{S}[\gamma,\sigma]=W[\gamma,\Phi=\sigma-2\log\epsilon]-W_{T\bar{T}}[\hat{g}(\gamma)]$. We can thus use our result \eqref{WDSTTlike} with $T_0=1$, $e^{\Phi}=1/\epsilon^2$ (matched by the evaluation of the on-shell gravitational action (\ref{Wttbarghatbulk}) for the chosen bulk saddles \eqref{bulkmetricansatz}-\eqref{warpingfactors})
\begin{align}
	W_{T\bar{T}}[\hat{g}(\gamma)]=&\frac{c}{48\pi}\int d^2x \sqrt{-\gamma}\Big[ \left(2\log\left(\frac{1}{\epsilon}+\sqrt{\frac{1}{\epsilon^2}-\lambda  }\right)-2\log(2-\lambda^2)\right)R_{\gamma} \nonumber
	\\
	& \qquad \qquad \qquad \qquad \qquad \qquad \qquad  + \frac{4}{\epsilon^2}\left(\sqrt{1-\lambda \epsilon^2}-1\right)\Big], \label{Wttbarghatsplit}
\end{align} 
to find that we must fix 
\begin{equation} \label{hatD0}
	\hat{D}_0=-2\log(1+\sqrt{1-\lambda \epsilon^2}).
\end{equation}
We have thus determined the following gravitational sector
\ali{
	\hat{S}[\gamma,\sigma] 
	&=\frac{1}{32\pi G}\int d^2 x \sqrt{-\gamma} \left[ 2 \log \left( \frac{e^{\sigma/2}+\sqrt{e^{\sigma}-\lambda\epsilon^2}}{1+\sqrt{1-\lambda \epsilon^2}}\right) R_\gamma +\frac{1}{2}\hat{E}(\sigma)(\nabla \sigma)^2 \right. \nonumber \\ 
	&\qquad \left.+  \frac{4e^{\sigma}}{\epsilon^2} \left( \sqrt{1-\lambda \epsilon^2 e^{-\sigma}}+e^{-\sigma}\left(1-\sqrt{1-\lambda \epsilon^2}\right)-T_0 \right) \right] +\, \text{higher} \, \text{derivatives}.  \label{STT}
}
The leading asymptotic order within the higher derivative terms in this gravitational sector can be computed using the strategy outlined in Appendix \ref{app:higherderivatives}, and can be obtained directly from \eqref{higherderivatives} via the substitution $\Phi=\sigma-2\log\epsilon$  
\begin{align} 
	\text{higher}\;\text{derivative}\;\text{terms}\;\text{in}\; \hat{S}[\gamma,\sigma]=-\frac{c}{48\pi}\epsilon^2\int &d^2x\sqrt{-\gamma}\frac{e^{-\sigma}}{32}\Big((\nabla \sigma)^4-4\Box\sigma (\nabla \sigma)^2 \nonumber
	\\
	&-8 R_{\mu\nu}[\gamma]\nabla^{\mu}\sigma \nabla^{\nu}\sigma\Big)+\mathcal{O}(\epsilon^4). \label{higherderivativetermsS}
\end{align}

Let us briefly consider the asymptotic or undeformed limit of this decomposition. The curvature coupling \eqref{Dsigmasol} with \eqref{hatD0} reduces to
\begin{equation}
	\lim\limits_{\epsilon \rightarrow 0}\hat{D}(\sigma)=\sigma.
\end{equation}
In order to require the asymptotic limit to be free of power law divergences\footnote{This actually requires only the leading behaviour $T_0=1+\mathcal{O}(\epsilon^2)$. The first subleading behaviour, corresponding to the choice of normalization of the undeformed Liouville potential, is arbitrarily chosen to coincide with that of the NBC/CBC tension $T_0=\sqrt{1-\lambda \epsilon^2}$ of section \ref{sect:bulk}. Upon setting the theory free, this would correspond to a NBC/CBC saddle $\sigma=0$.}, we impose $T_0=1-\frac{\lambda \epsilon^2}{2}+\mathcal{O}(\epsilon^4)$.
Thus, we find that the potential \eqref{Vsigmasol} has the asymptotic limit 
\begin{equation}
	\lim\limits_{\epsilon \rightarrow 0}\hat{V}(\sigma)=2\lambda e^{\sigma}.
\end{equation} 
Further assuming the asymptotic behaviour $\hat{E}(\sigma)=1+\mathcal{O}(\epsilon^2)$, the gravitational sector reduces to the expected timelike Liouville theory describing the integrated Weyl anomaly \cite{Skenderis:1999nb,Henningson:1998gx,Callebaut:2025thw}  
\ali{
	&\lim\limits_{\epsilon\rightarrow 0}\hat{S}[\gamma,\sigma] = S_L[\gamma,\sigma], \\
	&S_L [\gamma,\sigma]=\frac{c}{48\pi}\int d^2 x \sqrt{-\gamma}\left(\sigma R_{\gamma}+\frac{1}{2}(\nabla \sigma)^2+2\lambda e^{\sigma}\right). \label{SLiou}
} 
Notice that both the non-local terms discussed in Appendix \ref{app:nonlocal} and the higher derivative terms \eqref{higherderivativetermsS} vanish asymptotically.
Clearly, the $T\bar{T}$-deformed matter sector reduces to a CFT$_2$ in the undeformed limit, i.e.  $\lim\limits_{\epsilon \rightarrow 0} W_{T\bar{T}}[\hat{g}(\gamma)]=W_{CFT}[\gamma]$, such that
\begin{equation}
	\lim\limits_{\epsilon \rightarrow 0}W[\gamma,\Phi]\equiv W_{CFT}^{(T_0)}[e^{\sigma}\gamma]
	=W_{CFT}[\gamma]+S_L[\gamma,\sigma] \label{SLiouinTTbarsplitboundary}
\end{equation}
consistent with the 
Liouville description of the integrated Weyl anomaly, whose holographic derivation we will discuss in more detail in section \ref{subsLiou} and Appendix \ref{AppLiou}. The tension
superscript on $W_{CFT}^{(T_0)}$ refers to the inclusion of a non-unit tension in the dual bulk on-shell action $S_{tot}$ and is responsible for the Liouville theory containing a non-zero cosmological constant term in \eqref{SLiou}, or the corresponding Weyl anomaly containing an extra constant contribution.
It indicates a choice of renormalization scheme where rather than using the standard counterterm $T_0=1$, one uses a brane tension term with arbitrary tension $T_0=1+\mathcal{O}(\epsilon^2)$.

It follows from the asymptotic limit \eqref{SLiouinTTbarsplitboundary} that we can interpret equation \eqref{STT} as the action for a $T\bar{T}$-like deformed Liouville theory, with the deformation defined by the trace flow \eqref{traceflowthat}. Equivalently, we could also have treated the asymptotic action as the seed theory 
\ali{
	\hat{S}^{(0)} = S_L 
}
in a perturbative expansion of $\hat{S}$ in the deformation parameter $\tilde{t}$ (i.e. an asymptotic expansion)
\ali{
	\hat{S} = \sum_{n \geq 0} \tilde t^n \hat{S}^{(n)}.  
}
Expanding the stress tensor correspondingly 
\ali{
	t_\mn = \sum_{n \geq 0} \tilde t^n t_{\mu\nu}^{(n)} = \sum_{n \geq 0} \tilde t^n \frac{4\pi}{\sqrt{-\gamma}} \frac{\delta \hat{S}^{(n)}}{\delta \gamma^\mn},  
}
we use this equivalent approach to solve in Appendix \ref{app:defliouorderbyorder} for $\hat{S}^{(1)}$, i.e.~the order $\epsilon^2$ contribution. This illustrates a systematic way in which the effective gravity can be determined order by order in the asymptotic expansion as in \cite{Callebaut:2025thw}. The result \eqref{Ssigmacorr1epsilongeneral} is consistent with the $\epsilon$-expanded action \eqref{STT}. 

In summary, the resulting decomposition $W[\gamma,\Phi] = W_{T\bar{T}}[\hat{g}(\gamma)] + \hat{S}[\gamma,\sigma]$ into a $T\bar T$ matter sector and gravitational sector given by \eqref{STT} gives the $T_0 \neq 1$ generalization of our interpretation of the induced brane theory in the $T_0 = 1$ case presented in \cite{Callebaut:2025thw}.  
In particular, we argued there that the so-called `cutoff CFT' of e.g.~\cite{Gubser:1999vj}, representing the matter sector of an effective braneworld theory, is given by a $T\bar{T}$-deformation of the CFT when the brane is non-asymptotic. The action $\hat{S}[\gamma,\sigma]$ in \eqref{STT} gives the direct generalization of the $T\bar{T}$-like deformed timelike Liouville gravity sector called $\tilde{S}_{L}$ in \cite{Callebaut:2025thw}, with a local potential and curvature coupling for the Weyl factor $\sigma$ that is known in closed form for any brane location $\epsilon$.  

\subsection{Undeformed conformal matter}  \label{subsec:undeformedmatter}

Finally, we consider extrapolating a matter sector that is simply an undeformed CFT:
\ali{
	W[\gamma,\Phi] = W_{CFT}[\gamma] + \bar{S}[\gamma,\Phi].  \label{CFTsplit}
} 
A CFT on $\gamma_\mn$ obeys the trace anomaly 
\ali{ 
	W_{mat}=W_{CFT}[\gamma]: \qquad \vev{\hat T_\mn \gamma^\mn} = -\frac{c}{12} R_\gamma, 
} 
which we can think of as the zeroth-order or undeformed limit of a $T\bar T$-deformed trace flow equation. By symmetry, we deduce that the matter stress tensor takes the form
\ali{
	\vev{\hat T_\mn} = \alpha \gamma_{\mu\nu}, \qquad \alpha = \frac{\lambda c}{12}, 
}
with $\alpha$ fixed by the anomaly. Using the result \eqref{Wttbarghatsplit} for the $T\bar{T}$-deformed generating function, we can take its undeformed limit to obtain 
\begin{align}
	W_{CFT}[\gamma]=\lim\limits_{\epsilon \rightarrow 0}W_{T\bar{T}}[\hat{g}(\gamma)]=&\frac{c}{48\pi}\int d^2x \sqrt{-\gamma}\Big[ \left(2\log2-2\log(2-\lambda^2)\right)R_{\gamma} \nonumber
	\\
	& \qquad \qquad \qquad \qquad \qquad   -2\lambda\Big]+S_{div}. \label{WCFTSdiv}
\end{align}
The divergent term $S_{div}=-\lim\limits_{\epsilon \rightarrow 0}\left(\frac{c}{48\pi}\int d^2x \sqrt{-\gamma} (2\log \epsilon )R_{\gamma}\right)$, famously associated to the Weyl anomaly of a CFT$_2$ \cite{Henningson:1998gx}, is non-covariant and may be cancelled by a convenient choice of renormalization scheme \cite{deHaro:2000vlm}. Doing so and subtracting $	\bar{S}[\gamma,\Phi]=W[\gamma,\Phi]-W_{CFT}[\gamma]$, we determine the corresponding gravitational sector
\begin{align} \label{Sbar}
	\bar{S}[\gamma,\Phi]&=\frac{1}{32\pi G}\int d^2x \sqrt{-\gamma}\Big[2 \log \left( \frac{ e^{\Phi/2} + \sqrt{e^\Phi - \lambda}}{2} \right) R_{\gamma} \nonumber
	\\
	& \qquad \qquad \qquad \qquad \qquad +4e^{\Phi}\left(\sqrt{1-\lambda e^{-\Phi}}-T_0+\frac{\lambda}{2}e^{-\Phi}\right)\Big]+\text{derivatives}.
\end{align}
As we will argue in section \ref{sectannular}, this proposed matter-gravity split is 
consistent with an annular deconstruction of the on-shell action \cite{Caputa:2020lpa}.  
It is also the split that is most useful for the application of the island rule within holographic set-ups, and will be used in section \ref{subsec:islandrule} precisely to provide a non-trivial check of its validity for non-asymptotic branes.

\section{Entanglement applications} \label{sect:EEapps}

In this section, we apply our obtained effective brane theory $W[\gamma,\Phi]$ for braneworld interpretations of holographic entanglement.

\subsection{$T\bar T$ entanglement as Wald entropy on the brane} \label{sect:TTEE}

\begin{figure}[t]
	\centering
	\includegraphics[scale=0.42]{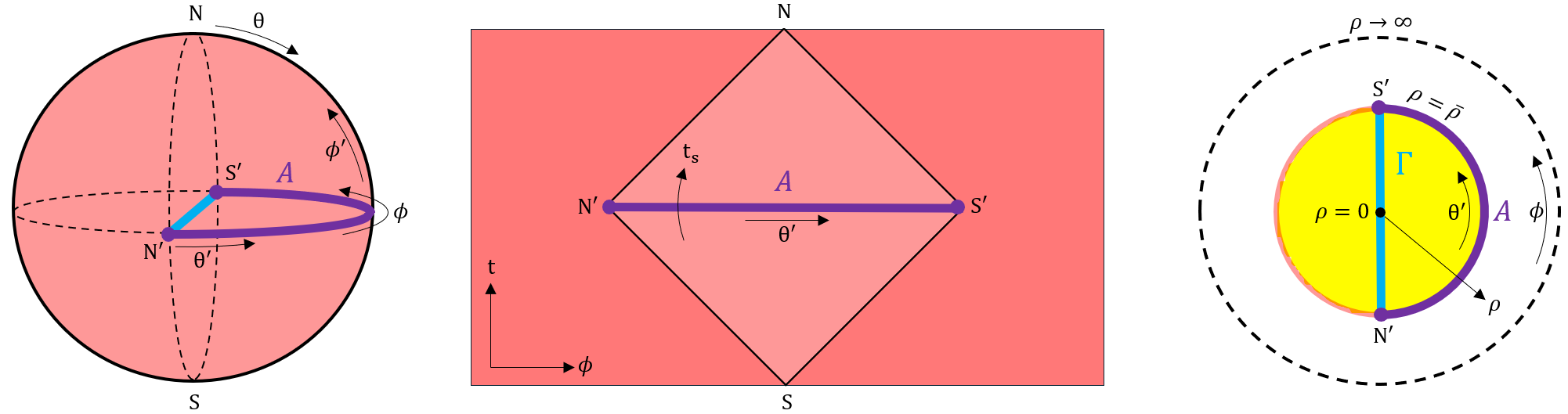} 
	\caption{
		Left: set-up relevant for the computation of the entanglement entropy $S^{ent}_{T\bar{T}}(A)$ of a $T\bar{T}$-deformed CFT$_2$ on the sphere, for an interval $A$ connecting antipodal points. Middle: Penrose diagram of global $dS_2$ (dark pink) and its static patch (light pink), which both map to the sphere. The endpoints of the interval $A$ correspond to the cosmological horizon $H$. The entanglement entropy is equal to the Wald entropy evaluated at the horizon, $S^{ent}_{T\bar{T}}(A)=S_{Wald}(H)$. Right: holographic dual set-up. The $T\bar{T}$-deformed CFT$_2$ on the sphere is dual to a subregion of AdS$_3$, shown in yellow. The sphere is embedded in the bulk geometry as a constant $\rho=\bar{\rho}$ EOW brane. The field theory result for the entanglement entropy agrees with the RT formula for the corresponding RT surface $\Gamma$, shown in blue, namely $S^{ent}_{T\bar{T}}(A)=S_{RT}(\Gamma)$. 
} \label{dSTTbarEE} 
\end{figure}

\begin{figure}[t]
	\centering
	\includegraphics[scale=0.42]{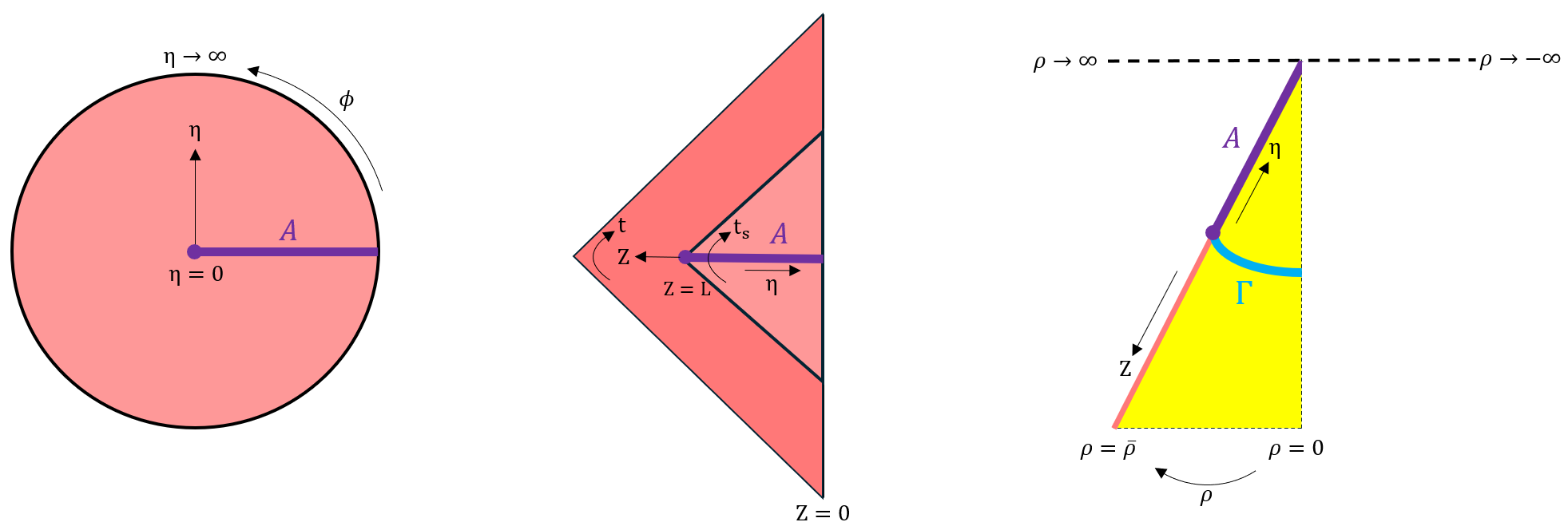} 
	\caption{ 
	     Left: set-up relevant for the computation of the entanglement entropy $S^{ent}_{T\bar{T}}(A)$ of a $T\bar{T}$-deformed CFT$_2$ on the hyperbolic plane, for an interval $A$ extending from the center to the boundary. Middle: Penrose diagram of planar $AdS_2$ (dark pink) and its Rindler $AdS_2$ patch (light pink), which both map to the hyperbolic plane. The left endpoint of the interval $A$ corresponds to the Rindler horizon $H$. The entanglement entropy is equal to the Wald entropy evaluated at the horizon, $S^{ent}_{T\bar{T}}(A)=S_{Wald}(H)$. Right: holographic dual set-up. The $T\bar{T}$-deformed CFT$_2$ on the hyperbolic plane is dual to a wedge of AdS$_3$, shown in yellow. The Poincaré patch of $AdS_2$ is embedded in the bulk geometry as a constant $\rho=\bar{\rho}$ EOW brane. The field theory result for the entanglement entropy agrees with the RT formula for the corresponding RT surface $\Gamma$, shown in blue, namely $S^{ent}_{T\bar{T}}(A)=S_{RT}(\Gamma)$. 	
}
	\label{AdSTTbarEE}
\end{figure}

We reconsider in this section the original set-up for the derivation of holographic $T\bar T$ entanglement in \cite{Donnelly:2018bef}. In that work, the length of a finite Ryu-Takayanagi (RT) geodesic is matched to the vacuum entanglement entropy of $T\bar T$ theory for an entangling surface
consisting of two antipodal points on a sphere, 
\ali{
	S^{ent}_{T\bar{T}}(A)=S_{RT}(\Gamma). \label{TTbarRT}
}
This is illustrated in the left and right pictures in Figure \ref{dSTTbarEE}. We will add to this equality the interpretation of the $T\bar T$ entropy as a Wald entropy for the effective brane theory $W[\gamma,\Phi]$ that we derived in section \ref{sec:braneeffective}: 
\ali{
	S^{ent}_{T\bar{T}}(A) = S_{Wald}[\Phi^\star] = S_{RT}(\Gamma) .  \label{STTRTWald}
}
The added perspective is illustrated in the middle of Figure \ref{dSTTbarEE}, and completes the 
set of dual boundary-brane-bulk perspectives on the entropy.  
Entirely analogous calculations hold for 
the entangling 
region across the hyperbolic plane pictured in Figure \ref{AdSTTbarEE}, with the corresponding statement \eqref{TTbarRT} worked out in \cite{Deng:2023pjs}. 
We first repeat the derivation of \eqref{TTbarRT} for the two separate cases in our notation, with details referred to Appendices. The reader familiar with this result can jump to the subsection on the brane perspective where we derive $S_{Wald}[\Phi^\star]$.

\paragraph{$dS_2$ case } 

Let us consider a bulk geometry of the form \eqref{bulkmetricansatz} with warping factor $e^{2A(\rho)}= (\sinh\rho)^2$ and with global de Sitter coordinates $x^\mu = (t,\phi)$ for the radial slices: 
\ali{
	ds^2 = d\rho^2 + (\sinh \rho)^2 \, (-dt^2 + (\cosh t)^2 d\phi^2 ) .     
} This is the $\lambda = -1$ case of the bulk geometries listed in \eqref{warpingfactors}. We place an EOW brane at a finite radial location $\rho = \bar \rho$ and consider the boundary entanglement region $A$ stretching along half of the global constant time slice at $t=0$, covered by the static patch.
To make this explicit, we introduce static patch coordinates $(t_s, \theta')$. It is instructive to first continue the $2d$ geometry to Euclidean signature using $t = i (\theta - \pi/2)$, arriving at the sphere geometry $d\theta^2 + (\sin \theta)^2 d\phi^2$ in $(\theta,\phi)$ coordinates. This sphere $S^2$ can also be covered by an alternative choice of spherical coordinates $(\theta',\phi')$ with $\phi'$ in the direction of $\theta$ and $\theta'$ in the direction of $\phi$ \cite{HartmanDeSitter}. Upon continuation of $d\theta'^2 + (\sin \theta')^2 d\phi'^2$ to Lorentzian signature by $t_s = i \phi'$ one obtains the static patch of $dS_2$,  
\ali{
	ds^2_{A} = d\theta'^2 - (\sin \theta')^2 dt_s^2 .  
    }  
It spans the entanglement region 
\ali{
	A: \qquad t_s = 0 = i \phi', \quad \theta': 0 \ra \pi. 
}
This is pictured in the center figure of Fig.~\ref{dSTTbarEE}. In Euclidean signature, the entanglement region $A$ covers the half-equator of the sphere\footnote{The half-equator refers here to the north and south poles $(N, S)$ in $(\theta,\phi)$ coordinates. Note that the north and south poles $(N',S')$ in $(\theta',\phi')$ coordinates are at different locations by an angle of 90 degrees. The latter form the endpoints $\p A$ of the entanglement region, i.e.~$N'$ at $\theta'=0$ and $S'$ at $\theta'=\pi$.}.   
This is illustrated in the left figure of Fig.~\ref{dSTTbarEE}. And finally, the right figure in Fig.~\ref{dSTTbarEE} shows the bulk perspective in $(\rho, \phi)$ 
coordinates at constant $t=0$, that is, the $d\rho^2 + (\sinh \rho)^2  d\phi^2$  geometry. 

The U(1) coordinate $\phi'$ that circles the boundary points $\p A$ of the entanglement region allows 
to use the replica manifold geometry $d\theta'^2 + n^2 (\sin \theta')^2 d\phi'^2$ (with $\Delta \phi' = 2\pi$) 
to calculate the entanglement entropy from the boundary perspective. This calculation was first performed in \cite{Donnelly:2018bef} and reviewed in Appendix \ref{appendix:B}. The result for the entanglement entropy for the $T\bar T$ theory on the sphere $S^2$ in vacuum state $|0\rangle_t$ (with respect to global time) 
is given in terms of the $T\bar T$ coupling $t=-\frac{3}{c}$ and sphere radius $r=\frac{1}{\epsilon}$ 
by 
\ali{
	S^{ent}_{T\bar{T}}(A) &=  \frac{c}{3} \sinh^{-1} \left( \sqrt{\frac{-3}{c\, t}} r 
		\right) = \frac{c}{3} \sinh^{-1} \left( \frac{1}{\epsilon} \right) \qquad (\lambda = -1) . \label{TTdS}
}
Holographically, it is obtained as the length of the RT geodesic $\Gamma$ connecting the $\p A$ points, given by 
\ali{
	S_{RT}(\Gamma) &=  \frac{\bar \rho}{2G} \qquad (\lambda = -1) . \label{RTdS}
} 
This indeed matches the boundary result \eqref{TTdS} under the bulk-boundary identifications of the $T\bar T$ coupling \eqref{TTbarcoupling} and $c = 3/(2G)$, and with $\bar \rho$ related to the induced boundary radius $\epsilon^{-1}$ by \eqref{defepsilon}.

\paragraph{$AdS_2$ case } 

The $AdS_2$ case proceeds in entirely analogous fashion. 
We consider a bulk geometry of the form \eqref{bulkmetricansatz} with warping factor $e^{2A(\rho)}= (\cosh\rho)^2$ and with planar $AdS_2$ coordinates $x^\mu = (t,Z)$ for the radial slices: 
\ali{
	ds^2 = d\rho^2 + (\cosh \rho)^2 \left( \frac{-dt^2 + dZ^2}{Z^2} \right) .     
} This is the $\lambda = 1$ case of the bulk geometries listed in \eqref{warpingfactors}. We again place an EOW brane at a finite radial location $\rho = \bar \rho$ and consider the boundary entanglement region $A$ stretching along the region 
of the planar constant time slice at $t=0$ that is covered by an AdS-Rindler  patch. 
To make this explicit, we introduce the coordinates $(t_s, \eta)$ that cover this patch. 
It is instructive to first continue the $2d$ geometry to Euclidean signature using $t = i t_E$, arriving at the $H^2$ geometry $(dt_E^2 + dZ^2)/Z^2$. 
The hyperbolic plane can alternatively be covered by $(\eta,\phi)$ coordinates as $d\eta^2 + (\sinh \eta)^2 d\phi^2$, with the relation between Euclidean coordinates given by 
\ali{
	t_E=\frac{L \sinh\eta \sin\phi}{\cosh\eta+\sinh\eta\cos\phi}, \qquad Z=\frac{L}{\cosh\eta+\sinh\eta\cos\phi} . 
}
Upon continuation to Lorentzian signature by $t_s = i \phi$ one obtains the AdS-Rindler patch   
\ali{
	ds^2_{A} = d\eta^2 - (\sinh \eta)^2  dt_s^2 .  
}  
It spans the entanglement region 
\ali{
	A: \qquad t_s = 0 = i \phi, \quad \eta: 0 \ra \infty   
}
or in planar $AdS_2$ coordinates, the region $A: \,\{ t=0$, $Z:L \ra 0 \}$.  
This is pictured in the center figure of Fig.~\ref{AdSTTbarEE}. In this section, we will follow \cite{Deng:2023pjs} and set the coordinate transformation parameter $L$ equal to one. 
In Euclidean signature, the entanglement region $A$ covers the half-space of the 
hyperbolic plane at constant polar angle.   
This is illustrated in the left figure of Fig.~\ref{AdSTTbarEE}. And finally, the right figure in Fig.~\ref{AdSTTbarEE} shows the bulk perspective in $(\rho, Z)$ 
coordinates at constant $t=0$, that is the $d\rho^2 + (\cosh \rho)^2  \frac{dZ^2}{Z^2}$ geometry. 

The U(1) coordinate $\phi$ that circles the boundary points $\p A$ of the entanglement region allows 
to use the replica manifold geometry $d\eta^2 + n^2 (\sinh \eta)^2  d\phi^2$ (with $\Delta \phi = 2\pi$) 
to calculate the entanglement entropy from the boundary perspective. This calculation appeared already in \cite{Deng:2023pjs} and is reviewed in Appendix \ref{appendix:B}. The result for the entanglement entropy for the $T\bar T$ theory on the hyperbolic plane $H^2$ in vacuum state $|0\rangle_t$ (with respect to planar time) 
is given in terms of the $T\bar T$ coupling $t=-\frac{3}{c}$ and $H^2$ radius $r=\frac{1}{\epsilon}$ 
by 
\ali{
	S^{ent}_{T\bar{T}}(A) &=  \frac{c}{6} \cosh^{-1} \left( \sqrt{\frac{-3}{c\, t}} r
	\right) = \frac{c}{6} \cosh^{-1} \left( \frac{1}{\epsilon} \right) \qquad (\lambda = 1) . \label{TTAdS}
}
From the bulk perspective in $AdS_2$ slicing, 
the dual of the entire AdS$_3$ (before the insertion of the EOW brane) consists of 
\emph{two} CFT's on an $AdS_2$ background, mirrored 
across the $\rho=0$ slice. The RT surface we need is the quotient of the RT surface connecting $\p A$ on one $T\bar T$ theory and $\p A$ on the other. It is given by the geodesic $\Gamma$ that stretches from $\rho=0$ to the $\p A: \{Z=1, \rho = \bar \rho\}$ point. Its length is  
\ali{
	S_{RT}(\Gamma) &=  \frac{\bar \rho}{4G} \qquad (\lambda = 1) .  \label{RTAdS}
} 
This indeed matches the boundary result \eqref{TTAdS} under the bulk-boundary identifications of the $T\bar T$ coupling \eqref{TTbarcoupling} and $c = 3/(2G)$, and with $\bar \rho$ related to the induced boundary radius $\epsilon^{-1}$ by \eqref{defepsilon}.

\paragraph{Brane perspective } 

Now we come to the brane perspective, given by our effective theory $W[\gamma,\Phi]$ in \eqref{Weff}. It provides us a dilaton gravity action  (with dynamical $\gamma$ and $\Phi$) by construction of the braneworld theory \eqref{Neumann} as $Z_{NBC}$ in \eqref{ZNBCsetfree}.  
For a given solution, the associated Wald entropy \cite{Iyer:1994ys} is effectively given by the factor multiplying the curvature $R_\gamma$, evaluated on the horizon $H$ of the solution\footnote{
	Since we will only consider constant solutions for the Weyl field $\Phi$, we can neglect contributions to the Wald entropy coming from the curvature-dependent parts of the higher derivative terms \eqref{higherderivatives} and non-local term \eqref{Wnonlocalappendix}. 
},  
\begin{equation}
	S_{Wald}[\Phi^\star(H)] 
	=\frac{c}{12}D(\Phi^\star(H)).
\end{equation}
On the solution $\Phi^\star$ for the Weyl field $\Phi$ given in (\ref{Phistarsol}), and $\gamma^\star = \gamma$, 
this leads to 
\ali{
	S_{Wald}[\Phi^\star(H)] &= \left\{ \begin{array}{l} \frac{c}{3} \sinh^{-1} \left( \frac{1}{\epsilon} \right) \qquad (\lambda = -1) \\ \frac{c}{6} \cosh^{-1} \left( \frac{1}{\epsilon} \right) \qquad (\lambda = 1)
	\end{array}\right.  \label{SWald}
}
for the Wald entropy associated with the cosmological horizon 
of $dS_2$ and the Rindler horizon of $AdS_2$ respectively, i.e.~the endpoints $H = \p A$ of the entanglement region $A$. The first consists of two points and the latter of one, which explains the factor two difference in the prefactor.  
The expression \eqref{SWald} matches the $T\bar T$ entanglement 
given in \eqref{TTdS} and \eqref{TTAdS},  
and the RT entropies \eqref{RTdS} and \eqref{RTAdS}. 
As discussed in section \ref{sect:bulk}, the bulk geometries \eqref{bulkmetricansatz}-\eqref{warpingfactors} used for the RT calculations provide the on-shell bulks corresponding to the saddle point $\Phi^\star$ of the brane theory, with the relation \eqref{Dtobarrho} between $\bar \rho$ and $\Phi^\star$, so that the equivalence to the RT result follows immediately from \eqref{Dtobarrho}. 
This completes the boundary-brane-bulk interpretations for the entropy, as summarized in \eqref{STTRTWald}, for branes at any constant finite distance into the bulk.

We conclude that the effective action $W[\gamma,\Phi]$ dynamically reproduces 
the result from the  boundary replica trick calculation for the $T\bar{T}$ entanglement entropy on curved space, and 
from the RT calculation within the holographic $T\bar{T}$ proposal. 
For the $\lambda=-1$ case ($dS_2$), this was accomplished by Hawking, Maldacena and Strominger in \cite{Hawking:2000da}, well before the RT and holographic $T\bar{T}$ proposals, in the asymptotic limit $\epsilon \rightarrow 0$. 
In that case,  $S_{Wald}\approx\frac{c}{3}\log\left(\frac{2}{\epsilon}\right)$, 
and the effective action is sourced entirely by the Weyl anomaly of the dual CFT$_2$. The Weyl factor across boundary points $\p A$ then provides the correct entropy for CFT on curved space \cite{Fiola:1994ir}. In this section, we showed that using our effective action we can reproduce their boundary-brane-bulk correspondence between entropies  
exactly at \emph{finite} $\epsilon$ and for both choices of curvature $\lambda=\pm1.$ 

\subsection{Island rule for non-asymptotic brane} \label{subsec:islandrule}

\begin{figure}[t]
	\centering
	\includegraphics[scale=0.42]{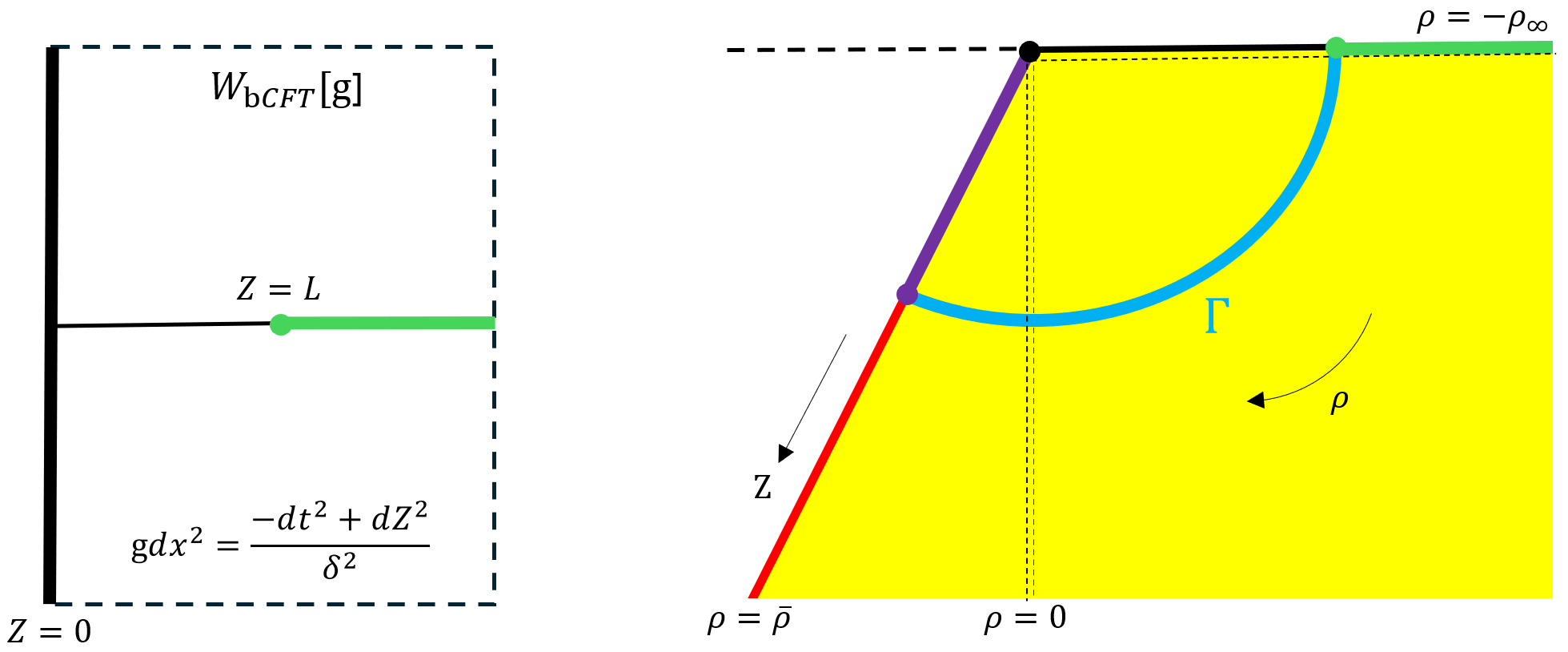}
	\caption{
	Left: set-up relevant for the computation of the entanglement entropy of a bCFT$_2$ on the half plane, for a constant time interval $Z>L$, shown in green. Right: holographic dual set-up. The bCFT$_2$ is dual to a subregion of AdS$_3$, shown in yellow, bounded on one side by half of the asymptotic boundary and on the other by an $AdS_2$ EOW brane located at constant radius $\rho=\bar{\rho}$, determined by the value of the subcritical tension. By splitting the bulk dual geometry into two regions, to the left and right of the middle $\rho=0$ slice, the set-up may equivalently be interpreted as a bCFT$_2$ with vanishing boundary entropy (dual to the region $\rho<0$) coupled to a $T\bar{T}$-deformed CFT$_2$ living on the $AdS_2$ EOW brane (dual to the region $0<\rho<\bar{\rho}$, as illustrated in Figure~\ref{AdSTTbarEE}). The entanglement entropy is reproduced holographically by the length of the corresponding RT surface $\Gamma$, shown in blue. Its contribution can similarly be separated into a bulk entanglement entropy contribution from the bCFT$_2$ region ($\rho<0$) and a contribution from the bCFT$_2$ boundary entropy, equivalently interpreted as the entanglement entropy of the $T\bar{T}$-deformed CFT$_2$ on $AdS_2$ ($0<\rho<\bar{\rho}$).} \label{AdSTTbarEEisland0} 
\end{figure} 

In this section we revisit an island rule set-up for holographic boundary CFT$_2$ \cite{Takayanagi:2011zk,Fujita:2011fp} that has previously been
applied for asymptotic branes \cite{Suzuki:2022xwv}, and extend the analysis to non-asymptotic branes by making use of our effective theory $W[\gamma,\Phi]$. 

We consider the AdS$_3$/bCFT$_2$ set-up illustrated in Figure \ref{AdSTTbarEEisland0}. The bulk geometry is empty 
AdS$_3$, \eqref{bulkmetricansatz}. Foliated by $AdS_2$ or flat slices, the bulk geometry reads
\begin{equation} 
	G_{MN}dX^MdX^N=d\rho^2+(\cosh\rho)^2\frac{-dt^2+dZ^2}{Z^2}=\frac{-dt^2+d\tilde{y}^2+d\tilde{z}^2}{\tilde{z}^2},  \label{ds3}
\end{equation}
with $\tilde{z}=Z/\cosh\rho$ and $\tilde{y}=Z\tanh\rho$. The spacetime is cut off by an EOW brane sitting at $\rho=\bar{\rho}$
\begin{equation} 
	EOW: \qquad \qquad \rho=\bar{\rho},
\end{equation}
such that the range of the radial coordinate is $\rho \in (-\rho_{\infty}, \bar{\rho}]$. Neumann boundary conditions are imposed on the EOW brane at $\rho=\bar{\rho}$, whereas Dirichlet boundary conditions fix the metric at the remaining portion of the asymptotic boundary, i.e. on the $\rho=-\rho_{\infty}$ slice.
We will calculate the entanglement entropy of the dual bCFT for the `radiation region' 
$\tilde{y}>L$ (for a given $L>0$). We start with the bulk calculation and then match to the result from the brane calculation. 

The length of a geodesic connecting the asymptotic boundary and the EOW brane is 
\begin{equation}
	\int^{\bar{\rho}}_{-\rho_{\infty}} d\rho=\bar{\rho}+\rho_{\infty}.
\end{equation}
For the RT surface being the arc of radius $L$, that is $\tilde y = L$, and an infinitesimal UV cutoff at $\tilde{z}=\delta$, we have  $\rho_{\infty}=\sinh^{-1}\frac{L}{\delta}\approx \log\left(\frac{2L}{\delta}\right)$, where we approximated for $\delta$ infinitesimal. 
On the brane, the Neumann condition yields $T_0 = \sqrt{1-\epsilon^2} = \tanh \bar \rho$ with $\epsilon = 1/\cosh \bar \rho$, as discussed in \eqref{T0epsilon} and \eqref{defepsilon}. 
It follows that, by the Ryu-Takayanagi conjecture, the entanglement entropy is given by 
\begin{equation} \label{SRT}
	S_{RT}(\Gamma_A)=\frac{\rho_{\infty}+\bar{\rho}}{4G}=\frac{c}{6}\log\left(\frac{2L}{\delta}\right)+\frac{c}{6}\cosh^{-1}\left(\frac{1}{\epsilon}\right).
\end{equation} 
From the bCFT perspective, the contribution $\frac{c}{6}\log\left(\frac{2L}{\delta}\right)$ is the standard bCFT entropy  
for the geometric region $\tilde{y}>L$, whereas $\frac{\bar{\rho}}{4G}$ is the boundary entropy contribution \cite{Takayanagi:2011zk,Fujita:2011fp}. 
As discussed in the previous section, the boundary entropy contribution can also be reinterpreted as the entanglement entropy of a $T\bar{T}$-deformed CFT on $AdS_2$. 
The boundary dual can thus be identified 
either as a bCFT with a specific boundary entropy, or as a bCFT (with vanishing boundary entropy) in the bath region that is coupled to a $T\bar{T}$-like deformed CFT on $AdS_2$ with effective action $W$, in the brane region. As discussed in section \ref{sec:identifyingmatter}, the brane theory $W$ can be split into a matter and gravity sector in different ways. We will proceed here with the 
split that separates out undeformed conformal matter, \eqref{CFTsplit} with \eqref{Sbar}.

\begin{figure}[t]
	\centering
	\includegraphics[scale=0.40]{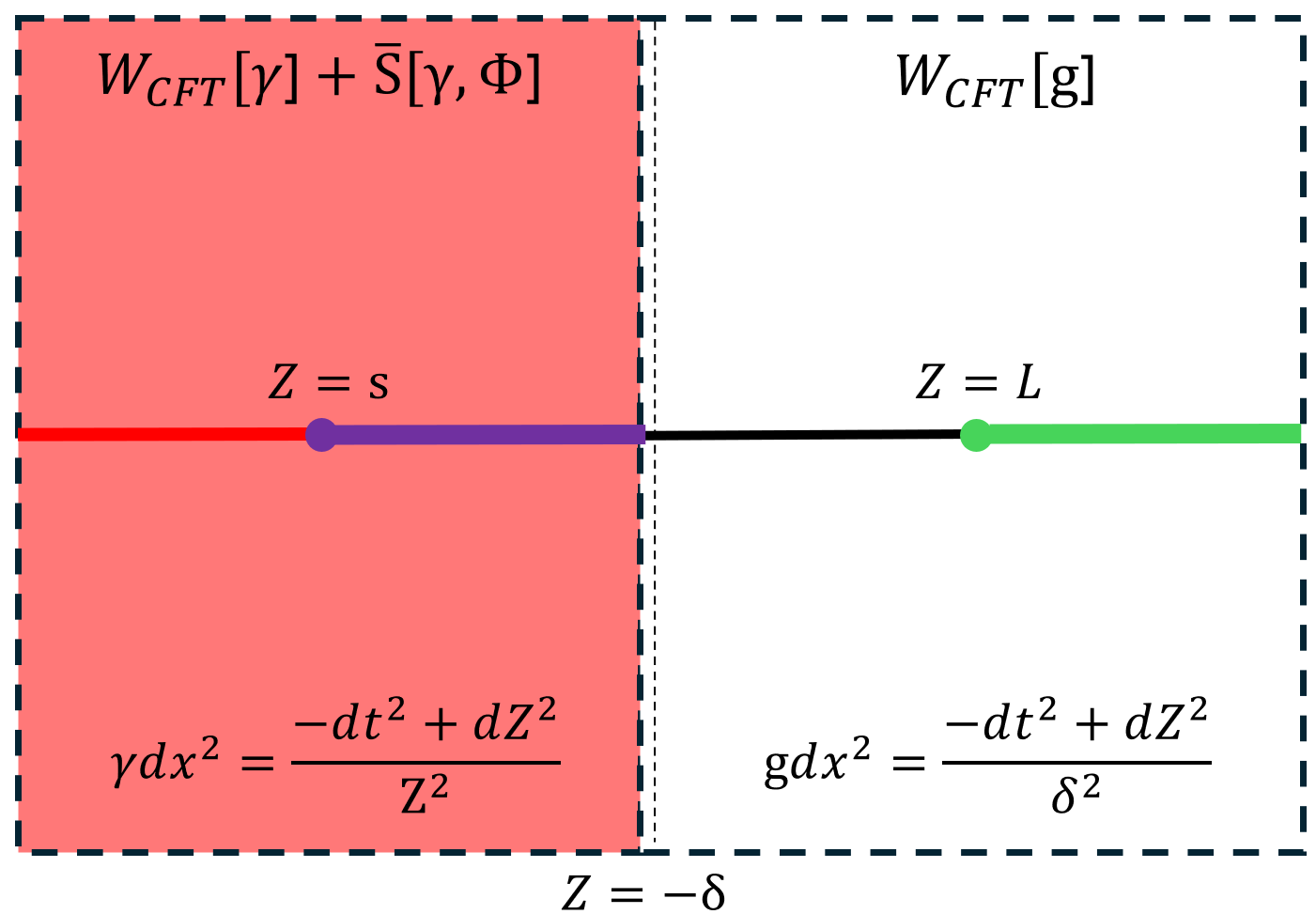}
	\caption{Set-up relevant for the island computation in braneworld holography corresponding to the AdS/bCFT configuration shown in Figure~\ref{AdSTTbarEEisland0}. According to our dictionary, the boundary and bulk descriptions are dual to a CFT$_2$ defined on a background geometry that is partly fixed to be flat (for $Z>-\delta$) and partly dynamical (for $Z<-\delta$), corresponding respectively to half of the asymptotic boundary and the EOW brane. In the $Z<-\delta$ region, the CFT$_2$ is 
	coupled to braneworld gravity with effective action $\bar{S}[\gamma,\Phi]$. The entanglement entropy of the region $Z>L$ must therefore be computed using the island rule, including a possible island on the $Z<s$ region together with a Wald entropy contribution from $\bar{S}[\gamma,\Phi]$ evaluated at the island boundary. In the semi-classical approximation, $\gamma$ for $Z<-\delta$ has an $AdS_2$ solution with unit radius.  
	The resulting island computation exactly reproduces both the bulk and boundary results.} \label{AdSTTbarEEisland1}
\end{figure}

The two-dimensional set-up from the brane perspective is presented in Figure \ref{AdSTTbarEEisland1}. In the bath region, we have a conformal field theory $W_{CFT}[g]$ on the induced metric $g(x) dx^2 = \frac{1}{\delta^2} (-dt^2 + d\tilde y^2)$ at the infinitesimal cutoff $\tilde z = \delta$ in the bulk metric \eqref{ds3}. In the brane region, we have the full effective theory $W[g]$.  It consists of conformal matter $W_{CFT}[\gamma]$
coupled to gravity $\bar{S}[\gamma,\Phi]$, with the gravitational action given explicitly in \eqref{Sbar} with $\lambda=1$
	\begin{align} 
		\bar{S}[\gamma,\Phi]=\frac{1}{32\pi G}\int d^2x \sqrt{-\gamma}&\Big[\left(2\cosh^{-1}\left(e^{\Phi/2}\right)-2\log 2\right) R_{\gamma} \nonumber
		\\
		&+ 4e^{\Phi}\left(\sqrt{1 -  e^{-\Phi}} + \frac{1}{2} e^{-\Phi} - T_0\right)\Big] +\text{derivatives}
	\end{align}
with $\Phi$ and $\gamma$ dynamical.   
In the semi-classical approximation, a solution to the  
equations of motion is provided by the $AdS_2$ metric $\gamma^{\star}=\gamma$
\ali{ 
	\gamma(x) dx^2 = \frac{-dt^2 + dZ^2}{Z^2}
} 
and by a constant conformal factor 
\begin{equation} 
	\Phi^{\star}=\log\left(\frac{1}{1-T_0^2}\right). \label{phistar}
\end{equation} 
The CFT then extends into both the bath and brane region with the background geometry given by\footnote{Note that $\tilde{y}\approx -Z$ and $d\tilde{y}^2\approx dZ^2$ on the $\rho=-\rho_{\infty}$ slice.} 
\ali{
	e^{2\hat \Phi(x)}(-dt^2 + dZ^2) 
}
with 
\ali{
	\left\{ \begin{array}{ll} e^{2\hat \Phi(x)} = \frac{1}{Z^2} & \text{ for } Z < - \delta  \\
		e^{2\hat \Phi(x)} = \frac{1}{\delta^2} & \text{ for } Z > - \delta	
	\end{array}. \right. \label{Phihatcases}
}
At the interface where the brane and bath touch, $Z = - \delta$, the background metrics on the two sides match, and transparent boundary conditions are imposed. Since the bulk geometry is empty AdS$_3$, the CFT is in the vacuum state. 

From this brane perspective, we now show how to obtain the entropy \eqref{SRT} 
by applying the island rule \cite{Almheiri:2019hni} in the holographic braneworld theory illustrated in Figure \ref{AdSTTbarEEisland1}. 

The Wald entropy corresponding to the saddle of $\bar{S}[\gamma,\Phi]$ 
and evaluated in a point is 
\begin{equation}
	S_{Wald}[\Phi^\star(P)]=\frac{1}{4G}\cosh^{-1}\left( e^{\Phi^\star(P)/2}\right)-\frac{\log 2}{4 G}. \label{SWaldisland}
\end{equation}  
We let the island be the 
geometric region $Z<s$, with $s<-\delta$. 
The `radiation' entropy for the region $Z>L$ 
consists of a contribution from the CFT 
and the Wald entropy contribution from the brane gravity evaluated at the endpoint of the island region\footnote{For the CFT contribution, we use the purity of the state to compute the entropy of the union of the radiation and island regions from that of its complementary interval.}
\ali{
	S_A = \left(\frac{c}{3}\log\left(\frac{L-s}{ \sqrt{e^{-\hat{\Phi}(s)}e^{-\hat{\Phi}(L)}}}\right)+\frac{1}{4G}\cosh^{-1}\left( e^{\Phi^{\star}(s)/2}\right)-\frac{\log 2}{4 G}\right).  \label{SAisland}
}
Its extremization $dS_A/ds=0$ imposed by the island rule is solved by 
\begin{equation}
	s=-L.
\end{equation}
Hence, the result of the island rule for the entanglement entropy is $S_{island}= S_A|_{s=-L}$ or 
\begin{equation} \label{finalislandresult}
	S_{island}=\frac{c}{6}\log\left(\frac{2L}{\delta}\right)+\frac{c}{6}\cosh^{-1}\left(\frac{1}{\epsilon}\right),
\end{equation}
with $\epsilon=\sqrt{1-T_0^2}$. 
This matches the bulk result \eqref{SRT} exactly even for finite $\epsilon$.  

We end with some comments. For the constant $\Phi^{\star}$ solution in \eqref{phistar}, the gravity contribution \eqref{SWaldisland} to the entropy is actually constant ($s$-independent), so that the island location is in fact uninfluenced by that Wald entropy contribution. It is however influenced by the Weyl factor $\hat{\Phi}(s)$ of $\gamma^\star$, given in equation \eqref{Phihatcases}. 
We further comment 
that the asymptotic brane limit $\epsilon \ra 0$ of our result, $S_{island} \approx  \frac{c}{6} \log \left( 2L/\delta \right) + \frac{c}{6} \log (2/\epsilon)$, can be alternatively obtained from a set-up as in Figure \ref{AdSTTbarEEisland1} with $W_{CFT}[g]$ being the full theory in both the left (brane) and right (bath) region. 
That is the 
perspective employed by Suzuki and Takayanagi in \cite{Suzuki:2022xwv} for their island calculation. In this interpretation, there is no Wald entropy contribution, only a contribution of the form of the first term in \eqref{SAisland}. 
In the $\epsilon\rightarrow 0$ limit, our Wald entropy contribution is 	$S_{Wald}[\Phi^{\star}(P)]\approx \frac{c}{12}\Phi^{\star}=\frac{c}{6}\log\left(\frac{1}{\epsilon}\right)$ and $W[\gamma,\Phi]$ reduces to $W^{(T_0)}_{CFT}[g=e^{\Phi}\gamma]$. Therefore, in the asymptotic limit the same Wald entropy contribution can be obtained by considering a CFT$_2$ on a fixed background geometry $g=(-dt^2+dZ^2)/\delta^2$ for $Z>-\delta/\epsilon$ and $g=e^{\Phi^{\star}}\gamma=(-dt^2+dZ^2)/(\epsilon Z)^2$ for $Z<-\delta/\epsilon$, with no gravity left \cite{Suzuki:2022xwv}.

The matching between the island and RT result is to our knowledge the first 
non-trivial check of the validity of the island rule in holographic set-ups beyond the asymptotic brane regime and with an explicit bulk-induced braneworld.

\section{Brane effective action from a fluctuating bulk cutoff} \label{sec:fluctuating}

The discussion so far has focused on the non-derivative in $\Phi$ terms in the effective brane theory $W[\gamma,\Phi]$. This is sufficient to show that the action \eqref{Weff} for the dynamical Weyl factor $\Phi$ and metric $\gamma$ reproduces as a saddle point the \emph{constant} location of the brane that one determines from the Neumann boundary condition from a bulk perspective (to be explicit, this refers to \eqref{Dtobarrho} facilitating the equality of \eqref{ZNBCT0new} and \eqref{ZNBCT0}). To go beyond this and include the `derivatives' terms in $W[\gamma,\Phi]$ one can proceed with the strategy of solving the radial Hamilton-Jacobi equation to higher order in the derivative expansion, presented in Appendix \ref{app:higherderivatives}. In section \ref{subsec:solvingtheflow}, this resulted
in equation \eqref{higherderivatives}  for the leading asymptotic contribution of the higher derivative terms that are to be added to the local sector of the effective boundary theory $W[\gamma,\Phi]$. 
From the bulk perspective, an alternative strategy to the use of the radial flow is to explicitly calculate the bulk on-shell action by integrating out the radial direction up to a \emph{non-constant} radial location. 
In this section, we employ this strategy, which is a more traditional route in recent literature on braneworld holography \cite{Geng:2022slq,Geng:2022tfc,Neuenfeld:2024gta,Deng:2022yll,Aguilar-Gutierrez:2023tic}, and examine to which extent we can match to the results \eqref{Weff}-\eqref{higherderivatives} in section \ref{holomatch}. Importantly, it is the same strategy that allows us, in the asymptotic limit, to holographically derive the timelike Liouville action describing the integrated Weyl anomaly in most general terms. We discuss this derivation in Appendix \ref{AppLiou}, providing a generalization to non-unit tension of the result of \cite{Callebaut:2025thw}.

The starting point is to consider the bulk geometry $G$ in \eqref{bulkmetricansatz} and place an EOW brane at the fluctuating radial location 
\ali{
	EOW: \qquad \rho = \tilde \rho(x) = \bar \rho + \tilde \phi(x).  \label{phitildedef}
}
The field $\tilde \phi(x)$ describes the fluctuation around constant $\bar \rho$. 
The induced metric on the brane is
\begin{equation}
	g_{\mu\nu}\big|_{\rho=\tilde{\rho}} 
	=e^{2 A(\tilde{\rho})}\gamma_{\mu\nu} + \p_\mu \tilde \rho \p_\nu \tilde \rho.   \label{induced} 
\end{equation}
It provides the background for the proposed dual theory $W[g]$ to the gravitational theory $S_{tot}[G]$ defined in \eqref{Stot}, through 
\ali{
	W[g] \equiv W[\gamma,\tilde \rho] = S_{tot}[G(\rho < \tilde \rho(x),x)]. \label{Wgbulk}  
}
Here the notation in the argument of the action refers to the upper limit of the $\rho$ integral being provided by $\tilde \rho(x)$. 
By direct calculation, we then find that the on-shell total action induces on the EOW brane the effective theory 
\begin{align} \label{bulkactionfluctuatinggamma}
	W[\gamma,\tilde{\rho}]&=\frac{1}{16\pi G}\int d^2x \sqrt{-\gamma}\Bigg(\tilde{\rho}R_{\gamma}+2\dot{A}e^{2A}+\frac{2\dot{A} (\nabla \tilde{\rho})^2}{1+e^{-2A}(\nabla\tilde{\rho})^2}+\frac{e^{-2A}\nabla^{\mu}\tilde{\rho}\nabla_{\mu}(\nabla\tilde{\rho})^2}{1+e^{-2A}(\nabla\tilde{\rho})^2} \nonumber
	\\
	&\qquad \qquad \qquad \qquad \qquad \qquad \qquad \qquad  -2T_0e^{2A}\sqrt{1+e^{-2A}(\nabla\tilde{\rho})^2}\Bigg)\Bigg|_{\rho=\tilde{\rho}},
\end{align}
with $(\nabla\tilde{\rho})^2=\nabla^{\alpha}\tilde{\rho}\nabla_{\alpha}\tilde{\rho}=\gamma^{\alpha\beta}\nabla_{\alpha}\tilde{\rho}\nabla_{\beta}\tilde{\rho}$. The details of the derivation are presented in Appendix \ref{app:fluctuatingcutoffaction}. In particular, for the three cases $\lambda=+1,0,-1$ we have 
\begin{align}
	&\lambda=+1 : \; W[\gamma,\tilde{\rho}]=\frac{1}{16\pi G}\int d^2x\sqrt{-\gamma}\left(\tilde{\rho}R_{\gamma}+2\sinh(\tilde{\rho})\cosh\tilde{\rho}-2T_0\left( \cosh(\tilde{\rho})\right)^2\right) 
	\\
	&\lambda=0: \;\;\;\; W[\gamma,\tilde{\rho}]=\frac{1}{16\pi G}\int d^2x\sqrt{-\gamma}\tilde{\rho}R_{\gamma}
	\\
	&\lambda=-1: \; W[\gamma,\tilde{\rho}]=\frac{1}{16\pi G}\int d^2x\sqrt{-\gamma}\left(\tilde{\rho}R_{\gamma}+2\sinh(\tilde{\rho})\cosh\tilde{\rho}-2T_0\left( \sinh(\tilde{\rho})\right)^2\right)
\end{align}
where we left out the explicit `$\, + \, \text{derivatives}$' referring to derivatives of $\tilde \rho$. As a reminder, we are neglecting contributions from the lower bound of integration over $\rho$.

\subsection{Holographic match} \label{holomatch}

The obtained $W[\gamma,\tilde \rho]$ 
compares directly to the result \eqref{Weff} we found for the effective brane theory $W[\gamma,\Phi]$ from integrating the radial Hamiltonian flow (bulk perspective) or $T\bar T$-like flow (boundary perspective) to leading order in a derivative expansion. Given that the induced metric on the brane takes the form
\ali{
	g_{\mu\nu} = e^\Phi \gamma_{\mu\nu} = e^{2 A(\tilde \rho)} \gamma_{\mu\nu} + \text{derivatives}, \label{gPhitilderho}
}
we can relate the Weyl field $\Phi$ of section \ref{sec:braneeffective} and the fluctuating brane location $\tilde \rho$ of the present section as 
\ali{
	\Phi = 2 A(\tilde \rho) + \text{derivatives},
}
and thus match the potential
\begin{equation}
	2e^{\Phi}\left(\sqrt{1-\lambda e^{-\Phi}}-T_0\right)=\left(2\dot{A}e^{2A}-2T_0e^{2A}\right)\Big|_{\rho=\tilde{\rho}} 
\end{equation}
and curvature coupling 
\begin{equation} \label{DPhi}
	D(\Phi)=2 A^{-1}\left(\frac{\Phi}{2}\right)=2\tilde{\rho},
\end{equation}
\emph{up to derivatives}, i.e. to leading order in the derivative expansion. This is really the same holographic match between the bulk description $W[g] \equiv W[\gamma,\tilde \rho]$ and the boundary description $W[g] \equiv W[\gamma,\Phi]$ presented in section \ref{sect:bulk}.

Let us now compare the kinetic and higher derivative terms. 
The derivative terms in the bulk description \eqref{bulkactionfluctuatinggamma} are given by 
\begin{align}
	\text{derivative}\;\text{terms}\;\text{in}\; W[\gamma,\tilde{\rho}]&=\frac{c}{24\pi}\int d^2x\sqrt{-\gamma}\Bigg(C(\tilde{\rho})(\nabla \tilde{\rho})^2+\frac{T_0}{4}e^{-2A(\tilde{\rho})}(\nabla \tilde{\rho})^4 \nonumber \\ 
	&\qquad \qquad \qquad -e^{-2A(\tilde{\rho})}\Box \tilde{\rho}(\nabla \tilde{\rho})^2 +\mathcal{O}\left(e^{-4A(\tilde{\rho})}\right)\Bigg), \label{bulkder}
\end{align}
with $C(\tilde{\rho})=2\sqrt{1-\lambda e^{-2A(\tilde{\rho})}}-T_0$, whereas the curvature-independent derivative terms in the boundary description from section \ref{subsec:solvingtheflow} are 
\begin{align}
	\text{derivative}\;\text{terms}\;\text{in}\; W[\gamma,\Phi]&=\frac{c}{48\pi}\int d^2x\sqrt{-\gamma}\Bigg(\frac{1}{2}E(\Phi)(\nabla \Phi)^2-\frac{1}{32}e^{-\Phi}(\nabla \Phi)^4 \nonumber \\ 
	& \qquad \qquad \qquad +\frac{1}{8}e^{-\Phi}\Box\Phi (\nabla \Phi)^2  +\mathcal{O}(e^{-2\Phi})\Bigg),
	\label{bdyder}
\end{align}
with $E(\Phi)=1+\mathcal{O}(e^{-\Phi})$ an undetermined scalar function. 
We observe that these derivative terms have the same structure in both expressions at order $e^{-\Phi}$. Moreover, the asymptotic expansion of \eqref{bulkactionfluctuatinggamma} clearly reveals that at
order $\mathcal{O}(e^{-(n+1)2A(\tilde{\rho})})$ or $\mathcal{O}(e^{-(n+1)\Phi})$ the local effective action requires
terms containing at most $4+2n$ derivatives, consistent with the asymptotic behaviour dictated by the deformation flow. The terms that are quadratic in derivatives will make up the kinetic terms of the asymptotic Liouville field associated with the integrated Weyl anomaly, when separating out a holographic matter sector. We will therefore first discuss the split into a matter sector for the bulk-derived theory $W[\gamma, \tilde \rho]$ in the next subsection \ref{subsLiou}, similar to the discussion for $W[\gamma, \Phi]$ in section \ref{subsec:TTbarmatter}, and then proceed to match the terms quadratic in derivatives explicitly in the asymptotic limit, using the bulk on-shell NBC/CBC relation $T_0=K/2=1+\mathcal{O}(e^{-\Phi})$, which additionally requires fixing the asymptotic behaviour of the function $E(\Phi)$ as in equation \eqref{EPhiasymptotic}.

Beyond the asymptotic limit, a detailed comparison of the derivative corrections obtained from the bulk and boundary perspectives is complicated by the presence of non-local and curvature-dependent higher derivative terms, which may be related to gauge redundancy, both in the effective action and in the field redefinitions relating the two descriptions\footnote{While the bulk description \eqref{bulkactionfluctuatinggamma} of the effective brane theory is free of non-local and curvature-dependent higher derivative terms, the boundary description \eqref{Weff} is not, as explained in Appendices \ref{app:nonlocal}-\ref{app:higherderivatives}.} . In fact, we will show in Appendix \ref{App:tildephisigmarelation} that non-localities appear also in the relation between the Weyl mode $\Phi$ and radial location $\tilde \rho$, or more specifically between the Liouville field $\sigma$ (defined in \eqref{sigmadef1}-\eqref{sigmadef2}) and the radial fluctuation $\tilde \phi$ (defined in \eqref{phitildedef}), which we work out explicitly for the $\lambda=0$ case to first subleading order in the asymptotic expansion.
With this result, focusing specifically on the $e^{-2A(\tilde{\rho})}\Box \tilde{\rho}(\nabla \tilde{\rho})^2$ term in \eqref{bulkder} and the  $e^{-\Phi}\Box\Phi (\nabla \Phi)^2$ term in \eqref{bdyder}, in Appendix \ref{App:matchingderivatives} we show that they can be matched exactly, at least for the flat slicing case ($\lambda = 0$). Thus, the finding of Appendix \ref{App:matchingderivatives} suggests that the derivative terms in the two descriptions of the effective action may be matched at higher order in perturbation theory. Since Appendices \ref{App:tildephisigmarelation} and \ref{App:matchingderivatives} are written in terms of the Liouville fields, they are better read after subsection \ref{subsLiou} 
deriving the deformed Liouville action on the brane. 

In conclusion, we will compare some of the derivative terms at the level of the expressions for the deformed Liouville action, which we now proceed to derive. 

\subsection{Bulk calculation of $T\bar T$-like deformed Liouville action} \label{subsLiou} 

The fluctuation dependent part  
of the bulk-derived brane theory $W$ in \eqref{bulkactionfluctuatinggamma} is 
defined through  
\ali{
	W[\gamma,\tilde \rho] = S_{tot}[G(\rho < \bar \rho,x)] + S_{\tilde \phi}[\gamma,\tilde \phi].  
}	
Written out with the use of \eqref{bulkactionconstantgamma}, it is given by  
\begin{align}
	S_{\tilde \phi}[\gamma,\tilde{\phi}]&=\frac{1}{16\pi G}\int d^2x \sqrt{-\gamma}\Bigg[\Bigg(\tilde{\phi}R_{\gamma}+2\dot{A}e^{2A}+\frac{2\dot{A} (\nabla \tilde{\phi})^2}{1+e^{-2A}(\nabla\tilde{\phi})^2}+\frac{e^{-2A}\nabla^{\mu}\tilde{\phi}\nabla_{\mu}(\nabla\tilde{\phi})^2}{1+e^{-2A}(\nabla\tilde{\phi})^2} \nonumber
	\\
	&\qquad -2T_0e^{2A}\sqrt{1+e^{-2A}(\nabla\tilde{\phi})^2}\Bigg)\Bigg|_{\rho=\bar{\rho}+\tilde{\phi}}-e^{2A}\left(2\dot{A}-2T_0\right)\Big|_{\rho=\bar{\rho}}\Bigg] \\ 
	&=\frac{1}{16\pi G}\int d^2x \sqrt{-\gamma}\Bigg[ \tilde{\phi} R_{\gamma} +\tilde{\phi}\left(2+2 \dot{A}^2 - 4 T_0\dot{A}\right)e^{2A} +\left(2\dot{A}-T_0\right)(\nabla\tilde{\phi})^2  \nonumber
	\\
	&\qquad + \tilde{\phi}^2 \left(2 \dot{A}\ddot{A}+2\dot{A}^3+2\dot{A}-4T_0\dot{A}^2-2T_0\ddot{A}\right)e^{2A} +\mathcal{O}(\tilde{\phi}^3)\Bigg]\Bigg|_{\rho=\bar{\rho}},   \label{Sphi}	
\end{align}
with $(\nabla \tilde{\phi})^2=\gamma^{\mu\nu}\nabla_{\mu} \tilde{\phi} \nabla_{\nu}\tilde{\phi}$ 
and where in the second line we expanded in small radial fluctuations $\tilde{\phi}$ while keeping the constant bulk radius $\bar{\rho}$ general. We perform this expansion to compare to recent literature on AdS$_3$ gravity with NBC \cite{Geng:2022slq,Geng:2022tfc,Neuenfeld:2024gta,Deng:2022yll,Aguilar-Gutierrez:2023tic,Neuenfeld:2026hen}, where one is interested in this limit. Our result is  then  consistent with theirs. In particular, to linear order in the radial fluctuation, $S_{\tilde \phi}[\gamma,\tilde{\phi}]$ reduces to the action for $AdS_2$ ($T_0<1$) or $dS_2$ ($T_0>1$) Jackiw-Teitelboim gravity. However, here we are mainly interested in keeping $\tilde{\phi}$ general. 

\begin{figure}[t]
	\centering
	\includegraphics[scale=0.45]{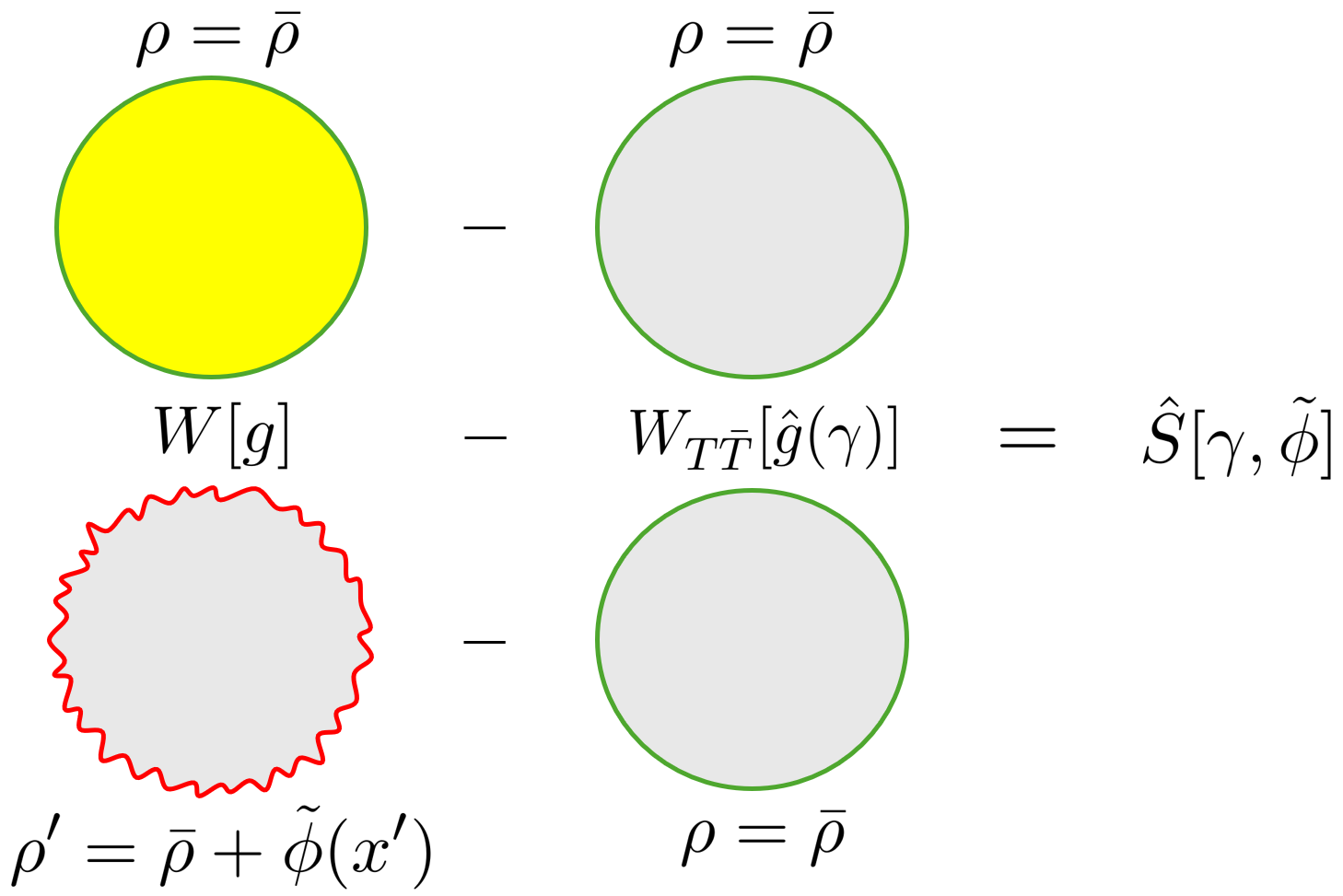}
	\caption{Illustration of the bulk geometries entering the holographic computation of the deformed Liouville action \eqref{Stotcritgrav}, which captures a bulk extension of the integrated Weyl anomaly of the dual CFT$_2$ with general brane tension $T_0$, defined through Eq.~\eqref{that}. This provides a finite-radius generalization of the asymptotic set-up shown in Figure~\ref{holoLiouville3} in Appendix \ref{AppLiou}.  
	}
	\label{holoLiouville4}
\end{figure}

In analogy with section \ref{subsec:TTbarmatter}, it is natural to extract a matter sector 
\ali{
	W_{T\bar T}[\hat g(\gamma)] &= S_{grav}[G(\rho < \bar \rho,x)] 
	= S_{tot}[G(\rho < \bar \rho,x)] - \frac{c}{12\pi} \int d^2 x \sqrt{-g} (1-T_0)\Big|_{\rho=\bar{\rho}} \label{TTbulk}
}
given explicitly in \eqref{Wttbarghatbulk}. The $(1-T_0)$ correction just comes from the difference between the bulk action $S_{tot}$ used in braneworld settings compared to the bulk action $S_{grav}$ used in absence of branes. 
One can then proceed to write the effective brane theory in the form 
\ali{
	W[g] = W_{T\bar T}[\hat g(\gamma)] + \hat{S}[\gamma,\tilde \phi]  \label{that}
}
with the effective gravity on the brane $\hat{S}[\gamma,\tilde \phi]$ calculated as the difference of bulk on-shell actions
in \eqref{Wgbulk} and \eqref{TTbulk}, to be 
\ali{
	\hat{S}[\gamma,\tilde \phi] &= \frac{1}{16\pi G}\int d^2x \sqrt{-\gamma}\Bigg[\Bigg(\tilde{\phi}R_{\gamma}+2\dot{A}e^{2A}+\frac{2\dot{A} (\nabla \tilde{\phi})^2}{1+e^{-2A}(\nabla\tilde{\phi})^2}+\frac{e^{-2A}\nabla^{\mu}\tilde{\phi}\nabla_{\mu}(\nabla\tilde{\phi})^2}{1+e^{-2A}(\nabla\tilde{\phi})^2} \nonumber
	\\
	&\qquad -2T_0e^{2A}\sqrt{1+e^{-2A}(\nabla\tilde{\phi})^2}\Bigg)\Bigg|_{\rho=\bar{\rho}+\tilde{\phi}}-e^{2A}\left(2\dot{A}-2\right)\Big|_{\rho=\bar{\rho}}\Bigg]. \label{Sbulk}
}  
The calculation is visualized in the bottom row of Figure \ref{holoLiouville4}. 
The result \eqref{Sbulk} for the effective braneworld gravity is a deformed timelike Liouville theory. This is clear from considering its asymptotic limit and 
asymptotic expansion in terms of a large cutoff radius $\bar{\rho}$ or small $\epsilon$, defined in \eqref{defepsilon}, 
\begin{align} \label{Stotcritgrav}
	\hat{S}[\gamma,\tilde{\phi}]=&\frac{1}{16\pi G}\int d^2 x \sqrt{-\gamma}\Bigg[\tilde{\phi}R_{\gamma}+(\nabla \tilde{\phi})^2+\lambda e^{2\tilde{\phi}}+\epsilon^2\Bigg(\frac{\lambda^2}{4}\left(3-e^{2\tilde{\phi}}-e^{-2\tilde{\phi}}\right) \nonumber
	\\
	&+\frac{\lambda}{2}\left(1-2e^{-2\tilde{\phi}}\right)(\nabla \tilde{\phi})^2+\frac{e^{-2\tilde{\phi}}}{4}(\nabla \tilde{\phi})^4-e^{-2\tilde{\phi}}\Box\tilde{\phi}(\nabla \tilde{\phi})^2\Bigg) +\mathcal{O}(\epsilon^4)\Bigg],   
\end{align}    
where we made use of the bulk on-shell NBC/CBC relation between the tension and the constant cutoff radius given in \eqref{T0epsilon}\footnote{More generally, the renormalization of the asymptotic theory simply requires $T_0=1+\mathcal{O}(\epsilon^2)$. However, here we make the arbitrary choice $T_0=\sqrt{1-\lambda \epsilon^2}$ before actually imposing the boundary conditions in order to show all terms at order $\mathcal{O}(\epsilon^2)$ explicitly in \eqref{Stotcritgrav}.}.  
The asymptotic timelike Liouville theory 
\begin{equation} \label{Stotcritgravasy}
	\lim\limits_{\epsilon\rightarrow 0 }\hat{S}[\gamma,\tilde{\phi}]=\frac{1}{16\pi G}\int d^2 x \sqrt{-\gamma}\left(\tilde{\phi}R_{\gamma}+(\nabla \tilde{\phi})^2+\lambda e^{2\tilde{\phi}}\right)\equiv S_L[\gamma,\tilde{\phi}] 
\end{equation}
describes the integrated Weyl anomaly of the dual CFT\footnote{We refer to the comment under equation \eqref{SLiouinTTbarsplitboundary} for the explanation of the $T_0$ superscript in $W_{CFT}^{(T_0)}$.} 
\ali{
	W_{CFT}^{(T_0)}[e^{\phi} g_{(0)}] = W_{CFT}[g_{(0)}] + S_L[g_{(0)},\phi]   \label{CFTT0}
}
with 
\ali{\lim\limits_{\epsilon \rightarrow 0}W[g]\equiv 	W_{CFT}^{(T_0)}[e^{\phi} g_{(0)}]\qquad \lim\limits_{\epsilon \rightarrow 0}W_{T\bar{T}}[\hat g]= W_{CFT}[g_{(0)}]  
}
and 
\ali{
	g_{(0)} = \frac{\gamma}{4}, \qquad \phi = \lim_{\epsilon \ra 0} 2 \tilde \phi . 
}
The non-asymptotic relation between the asymptotic Weyl mode $\phi$ and the brane fluctuation $\tilde \phi$ is given explicitly in \eqref{phitildephi}.  
Appendix \ref{AppLiou} details the bulk calculation of $S_L[g_{(0)},\phi]$, essentially following the asymptotic limit of \eqref{that}, but treating carefully the difference between the $\phi$ and $\tilde \phi$ modes and generalizing to arbitrary asymptotically-AdS$_3$ geometries in FG gauge. This type of bulk derivation of a Liouville action was first performed in \cite{Carlip:2005tz} with the standard holographic counterterm rather than a tension term\footnote{See also Appendix A of \cite{Takayanagi:2018pml} for a similar derivation.}, although we add a precise holographic interpretation in line with \cite{Callebaut:2025thw}.
The asymptotic Weyl mode $\phi$ is the one that labels the Brown-Henneaux diffeomorphism between the relevant bulk geometries. Basically, one can think of $g$ in \eqref{induced} as being the induced metric in the bulk geometry $G$ at the non-constant location $\tilde \rho(x)$, or as the induced metric at the constant location $\bar \rho$ in a bulk geometry $G'$ that is related to $G$ by a finite bulk diffeomorphism. The equivalence between these viewpoints is highlighted pictorially in Figure \ref{holoLiouville4}, as a direct extension of Figure \ref{holoLiouville3} in Appendix \ref{AppLiou}. 

As a side result, we include in Appendix \ref{app:curvedbanados} the form of the finite Brown-Henneaux diffeomorphisms between Poincar\'e AdS$_3$ (bulk metric $G$) and a Banados geometry (bulk metric $G'$) with asymptotic $(A)dS_2$ slicing. This extends the known diffeomorphisms \cite{Roberts:2012aq} from Poincar\'e AdS$_3$ to flat-sliced Banados, and brings the Banados geometry in a curved-sliced form \eqref{curvedBanados+-1}. 

\subsubsection{Holographic match for deformed Liouville action}  

Having worked out the 
split of the bulk-derived brane theory \eqref{that} into a $T\bar T$ matter sector and effective gravity action $\hat{S}[\gamma,\tilde\phi]$ in \eqref{Sbulk}, 
we can compare directly to the split of the $T\bar T$-like deformed brane theory \eqref{splitghat} into a $T\bar T$ matter sector and effective gravity action $\hat{S}[\gamma,\sigma]$ in \eqref{STT}. We compare the gravitational sectors in an asymptotic expansion\footnote{To compare the first asymptotically subleading terms between the two descriptions explicitly, we made the arbitrary choice $T_0=1-\frac{\lambda\epsilon^2}{2}-\frac{\lambda^2\epsilon^4}{8}+\mathcal{O}(\epsilon^6)$ in both descriptions.}, respectively given in \eqref{Stotcritgrav} and by
\begin{align} \label{Ssigmacorr1epsilongeneralnew}
	&\hat{S}[\gamma,\sigma]=\frac{c}{48\pi}\int  d^2x \sqrt{-\gamma}\Bigg[\sigma R_{\gamma}+\frac{1}{2}(\nabla \sigma)^2+2\lambda e^{\sigma}-\left(\frac{\epsilon}{4}\right)^2\Bigg(\frac{1}{2}e^{-\sigma}(\nabla \sigma)^4-2e^{-\sigma}(\nabla \sigma)^2\Box\sigma  \nonumber
	\\
	&-4R_{\mu\nu}[\gamma]\nabla^{\mu}\sigma\nabla^{\nu}\sigma-8\lambda^2 e^{\sigma}+8\lambda^2e^{-\sigma}-8\lambda^2+\hat{E}^{(\epsilon^2)}(\sigma)(\nabla \sigma)^2+8\lambda( e^{-\sigma}-1)R_{\gamma}\Bigg)+\mathcal{O}(\epsilon^4)\Bigg].
\end{align}  
This equation is derived in Appendix \ref{app:defliouorderbyorder}.
The derivative-independent terms  
are matched exactly using again the relation \eqref{gPhitilderho} 
with $\sigma=\Phi+2\log \epsilon$ and $\tilde{\rho}=\bar{\rho}+\tilde{\phi}$, which implies
\begin{align}
	\sigma&=2\log\left(\cosh(\tilde{\phi})+\sqrt{1-\lambda \epsilon^2}\sinh(\tilde{\phi})\right)+\text{derivatives}
	\\
	&=2\tilde{\phi}+\lambda \epsilon^2\frac{e^{-2\tilde{\phi}}-1}{2}+\text{derivatives}+\mathcal{O}(\epsilon^4) \label{sigmatildephiepslon}
\end{align}
for the asymptotic relation between the Weyl field $\sigma$ and the radial fluctuation $\tilde \phi$, or boundary versus bulk description.   
The `derivatives' terms in this relation are addressed in Appendix \ref{App:tildephisigmarelation}, where a procedure is outlined for obtaining the $\sigma$ to $\tilde \phi$ relation in an asymptotic expansion. More precisely, for the flat slicing case ($\lambda=0$), we determine the first (order $\epsilon^2$) non-trivial `derivatives' correction to \eqref{sigmatildephiepslon} explicitly and find that it is intrinsically non-local. 
The result of Appendix \ref{App:tildephisigmarelation} is then used in Appendix \ref{App:matchingderivatives} to match derivative terms in $\hat{S}[\gamma,\tilde\phi]$ and $\hat{S}[\gamma,\sigma]$ up to order $\mathcal{O}(\tilde{\phi}^3)\sim \mathcal{O}(\sigma^3)$ (meaning including cubic order in fluctuations, but excluding quartic order or higher).

\subsection{Annular deconstruction of bulk on-shell action} \label{sectannular}

In section \ref{sec:fluctuating}, we have so far discussed the bulk derivation of the brane effective theory $W[g]$, its split into a $T\bar T$ matter sector, and the matching to the brane effective theory determined from integration of the $T\bar T$-like flow in section \ref{sec:braneeffective}. In this subsection, we will discuss the split of the bulk-derived theory into a CFT matter sector, and compare to the boundary perspective results of section \ref{subsec:undeformedmatter}. 

To split off a CFT matter sector from $W[g]$, we first identify the dual CFT as 
\begin{equation}
	\lim\limits_{\bar{\rho}\rightarrow \infty}S_{grav}[G(\rho<\bar{\rho},x)]= W_{CFT}[\gamma], 
\end{equation}
consistent with taking the asymptotic limit ($\bar{\rho}\rightarrow \infty$) of Eq.~\eqref{TTbulk}. The analogue of \eqref{CFTsplit} can then be obtained by considering the 
annular deconstruction \cite{Caputa:2020lpa,Ondo:2022zgf} of Eq.~(\ref{Wgbulk}), i.e. $W[g]\equiv W[\gamma,\tilde{\rho}]=S_{tot}[G(\rho < \tilde{\rho}(x),x)]$, into 
\begin{align}
	S_{tot}[G(\rho < \tilde{\rho}(x),x)]=\left(\lim\limits_{\bar{\rho}\rightarrow \infty}S_{grav}[G(\rho < \bar{\rho},x)]\right)-S_{ann}[G(\rho>\tilde{\rho}(x),x)],
\end{align}
where we defined the annular action
\begin{align}
	S_{ann}[G(\rho>\tilde{\rho}(x),x)]=&\frac{1}{16\pi G}\int^{\bar{\rho}\rightarrow \infty}_{\tilde{\rho}}d^3X\sqrt{-G}(R_G+2)+\frac{1}{8\pi G}\int\limits_{\rho=\bar{\rho}\rightarrow \infty} d^2x\sqrt{-g}(K-1) \nonumber
	\\
	&-\frac{1}{8\pi G}\int\limits_{\rho=\tilde{\rho}} d^2x\sqrt{-g}(K-T_0). 
\end{align}
We can evaluate it using our previous results \eqref{Wttbarghatbulk}, \eqref{WCFTSdiv}, \eqref{Wgbulk} and \eqref{bulkactionfluctuatinggamma}:
\begin{align}
	S_{ann}[G(\rho>\tilde{\rho}(x),x)]=&-\frac{1}{16\pi G}\int d^2x \sqrt{-\gamma}\Big[\big(\tilde{\rho}-\log 2+\log(2-\lambda^2)\big)R_{\gamma} \nonumber
	\\
	&\qquad \qquad \qquad \qquad +2\dot{A}e^{2A} -2T_0e^{2A}+\lambda\Big]\Big|_{\rho=\tilde{\rho}}+\text{derivatives}. 
\end{align}
Just like in section \ref{subsec:undeformedmatter}, we actually eliminated the divergent contribution, 
\begin{equation}
	S_{div}=-\lim\limits_{\epsilon \rightarrow 0}\left(\frac{c}{48\pi}\int d^2x \sqrt{-\gamma}(2\log\epsilon)R_{\gamma}\right),
\end{equation}
by a specific choice of renormalization scheme for the undeformed CFT \cite{Henningson:1998gx,deHaro:2000vlm}.
The annular action $S_{ann}$ thus provides the remaining gravitational sector. Using the relation \eqref{DPhi}, i.e.~$D(\Phi)=2 A^{-1}\left(\frac{\Phi}{2}\right)=2\tilde{\rho}+\text{derivatives}$, between the Weyl mode and the brane location, we can compare it to the gravitational sector \eqref{Sbar} determined in section \ref{subsec:undeformedmatter}.  
At least up to the derivative terms, we find 
\begin{equation}
	S_{ann}[G(\rho>\tilde{\rho}(x),x)]=-\bar S[\gamma,\Phi]
\end{equation}
and thus 
\begin{equation}
	W[g]=W_{CFT}[\gamma]+\bar S[\gamma,\Phi],
\end{equation} 
consistent with the result \eqref{CFTsplit} obtained from the integration of the $T\bar{T}$-like flow.

\section{Discussion and outlook} \label{sectdiscussion}

In this paper, we have sharpened our proposal for semi-classical AdS$_3$ gravity with NBC put forward in \cite{Callebaut:2025thw}, generalizing it to  branes with non-unit tension and to CBC. The proposal for the putative holographic dual can be summarized as setting free the $T\bar T$-like deformed CFT dual to the DBC problem.

We first reinterpreted the radial Hamilton-Jacobi equation as a generalized $T\bar T$-like trace flow equation for the deformation of the asymptotic CFT. We then developed a new strategy to solve for the brane effective action dual to the DBC problem, based on a derivative expansion. This allowed us to determine the potential and curvature coupling in closed form, while systematically solving for the terms involving derivatives of the brane conformal factor order by order in the asymptotic expansion. Given the resulting effective action, we checked that our set-free proposal, for constant-conformal-factor solutions, correctly reproduces the bulk results corresponding to empty AdS$_3$ saddles in maximally-symmetric slicings. Finally, we have suggested an alternative bulk interpretation of our effective action as emerging from a position-dependent cutoff location, which becomes dynamical once the boundary theory is set free. 

Our holographic proposal provides a controlled laboratory in which to study entanglement entropy in a quantum theory coupled to dynamical gravity. 
In particular, we highlight our check of the island rule in a holographic set-up beyond the asymptotic brane regime.  It involves using a decomposition of the effective action in a matter and gravitational sector, for which we discussed several other useful options.

We end with some comments and open questions.

\paragraph{Completing the effective brane theory}
A first open question concerns the derivative  
and non-local sectors of the effective brane theory. While the derivative expansion provides a controlled strategy to determine $W[\gamma,\Phi]$, the trace-flow equation directly constrains only the part of the effective action that contributes to the trace of the stress tensor. In particular, our analysis determines only the leading contribution in an asymptotic expansion of the kinetic term and the first subleading contribution of the higher derivative terms of the local sector. Determining a closed form for the kinetic and higher-derivative terms and understanding how they are constrained by the trace-flow equation is therefore an important open problem.

As discussed in Appendix \ref{app:nonlocal}, we have determined only the non-zero-trace part of the leading non-local contribution. An additional traceless contribution at the same derivative order may be required to enforce the consistency condition $\langle T_{\mu\nu}\gamma^{\mu\nu}\rangle=
-\frac{4\pi}{\sqrt{-\gamma}}\frac{\delta W[\gamma,\Phi]}{\delta\Phi}$, 
and further non-local contributions are expected at higher derivative orders. It would therefore be important to characterize the complete non-local sector of the effective action and understand to what extent it is fixed by the radial Hamilton--Jacobi equation together with the requirement that $W[\gamma,\Phi]$ originates from a functional $W[g]$ of the induced metric.

It remains unclear to what extent the derivative corrections obtained from the radial flow in section \ref{sec:braneeffective} can be matched uniquely to those obtained from the fluctuating-cutoff calculation of section \ref{sec:fluctuating}. As illustrated in Appendices \ref{App:tildephisigmarelation} and \ref{App:matchingderivatives}, part of this difficulty originates from the non-local relation between the radial fluctuation and the Weyl factor, while physical information encoded in the derivative terms may not yet have been fully disentangled from gauge-dependent contributions. In the flat-slicing case we were able to construct this relation explicitly to first non-trivial order and verify the matching of a non-trivial derivative term. It would be interesting to extend this analysis to higher orders and to curved slicings, and to understand whether the bulk description in terms of a position-dependent cutoff can provide a systematic way of fixing the local curvature-dependent and non-local completions of the effective brane theory.

\paragraph{Beyond maximally-symmetric branes and empty AdS}  
Another important direction is to move beyond the constant-conformal-factor solutions considered in this work. These solutions were sufficient 
to reproduce maximally-symmetric branes in empty AdS$_3$, but the effective theory $W[\gamma,\Phi]$ should contain considerably more information. It would be interesting to study non-constant solutions for $\Phi$, and to identify the corresponding bulk geometries. 
The bulk constructions developed in the appendices may provide useful tools in this direction. In Appendix \ref{AppLiou}, we gave a precise interpretation to the bulk derivation of the timelike Liouville action describing the integrated Weyl anomaly. In particular, the geometric ``bulk triangle'' interpretation of Figure \ref{trianglecirclesfigure} relates a change of the cutoff surface to a boundary Weyl transformation through a bulk diffeomorphism. Moreover, Appendix \ref{app:curvedbanados} shows how curved-sliced empty AdS$_3$ geometries can be mapped to curved-sliced Banados geometries. It would be interesting to use this framework to extend the effective brane description to excited CFT states and to understand how the corresponding state-dependent data are encoded in the brane theory.

\paragraph{Matter-gravity splits and connections to other gravitational descriptions} We find it interesting 
that the split into a CFT matter sector on the brane, \eqref{CFTsplit}, suggests there is no need for a cutoff scale in the matter sector, which is typically implied by the use of the terminology `cutoff CFT' in braneworld holography. There is an overall cutoff scale present in the effective theory \eqref{CFTsplit}, but not in the matter sector.   
The structure of \eqref{CFTsplit} is also reminiscent of alternative formulations of $T\bar T$-deformed CFT's as undeformed CFT's coupled to a two-dimensional gravitational sector such as flat JT or massive gravity \cite{Dubovsky:2017cnj,Cardy:2018sdv,Dubovsky:2018bmo,Tolley:2019nmm,Callebaut:2019omt},  
as well as holographic proposals based on mixed boundary conditions and related braneworld descriptions \cite{Guica:2019nzm,Callebaut:2025uye,Hirano:2025cjg}. It would be interesting to understand the relation to these constructions better,  
for which the annular-deconstruction perspective of section \ref{sectannular} might be particularly useful.

\paragraph{Holographic entanglement and island rule}  Our example of an island rule application was deliberately simple: the bulk is empty AdS$_3$, there is no black hole, and the configuration realizes the island phase without a competing non-island saddle. A natural next step is to apply the same finite-cutoff braneworld description to geometries containing black holes, where one can study competing island and non-island saddles and the associated Page transitions \cite{Penington:2019npb,Almheiri:2019hni,Almheiri:2019yqk,Almheiri:2019qdq,Penington:2019kki,Rozali:2019day,Almheiri:2020cfm,Chen:2020uac,Hartman:2020swn,Verheijden:2021yrb}. It would be particularly interesting to determine whether the exact agreement between the island and bulk RT calculations found here at finite $\epsilon$ persists in these more general situations.
Another direction is to derive the holographic entanglement prescription directly from the gravitational replica trick at finite cutoff and for the different boundary conditions considered here, extending recent work for CBC \cite{Caceres:2026yvs}.

\paragraph{Emergent gravity from entanglement}
It would be interesting to understand whether the gravitational equations of motion of the effective braneworld theory admit an entanglement interpretation. In holography, gravitational equations are closely related to first-law-type relations for entanglement entropy \cite{Jacobson:1995ab,Savonije:2001nd,VanRaamsdonk:2010pw,Verlinde:2010hp,Casini:2011kv,Faulkner:2013ica,Czech:2015qta,deBoer:2015kda,Jacobson:2015hqa,Asplund:2016koz,Callebaut:2018nlq,Callebaut:2018xfu,Fujiki:2025rtx,Neuenfeld:2026hen}. Our construction provides a setting in which both the gravitational action and its Wald entropy arise dynamically from the same $T\bar T$-like flow. This raises the question of whether the equations of motion following from the brane effective action can themselves be derived from appropriate thermodynamic or entanglement relations. Establishing such a relation could provide a useful perspective on the emergence of braneworld gravity from the entanglement structure of the underlying deformed quantum theory.

\paragraph{Higher dimensions and well-posedness of the boundary conditions} 
While we focused on the three-dimensional bulk case in the present paper, for the higher-dimensional case we proposed a $T^2$ set free interpretation of $Z_{NBC}$ in \cite{Callebaut:2025thw}, using \cite{Hartman:2018tkw}. The higher-dimensional case raises more questions regarding boundary conditions. 
We considered in this paper different types of boundary conditions imposed at the EOW brane: DBC, NBC and CBC. There is considerable debate on which boundary conditions render the Cauchy problem well-posed in gravity. There are arguments against DBC and for CBC, but also CBC is still under debate \cite{Witten:2022xxp,Anninos:2023epi,Liu:2025xij}. We took an agnostic perspective on these issues here, since we are working with three-dimensional gravity, in which there are no propagating modes.

\paragraph{Other extensions }
Throughout this work we have  
neglected one-dimensional corner terms associated with intersections of the brane with the asymptotic boundary. Understanding the degrees of freedom associated with these corners may  
add to the connection between the double-holographic description \cite{Almheiri:2019hni,Chen:2020uac,Neuenfeld:2024gta,Neuenfeld:2026hen} and lower-dimensional holographic duals such as Schwarzian quantum mechanics \cite{Engelsoy:2016xyb,Maldacena:2016upp,Griguolo:2025kpi,Griguolo:2026eoi}.  
Finally, the strategy developed here is not intrinsically tied to a negative bulk cosmological constant. It is natural to ask whether analogous flow equations and set-free constructions can be formulated for different bulk cosmological constants, and what effective theories they induce on dynamical cutoff surfaces. Extensions to de Sitter space are naturally connected to $T\bar T+\Lambda$-type deformations, dS wedge/braneworld constructions and finite-time holography \cite{Strominger:2001pn,Gorbenko:2018oov,Chen:2020tes,Aguilar-Gutierrez:2023tic,Neuenfeld:2024gta,Godet:2024ich,Aguilar-Gutierrez:2024nst,Fumagalli:2024msi,Araujo-Regado:2025elv}. The Minkowski case could instead provide a bridge to flat-space holography, including approaches relating flat space directly to AdS constructions \cite{Ball:2019atb,Fujiki:2025yyf,Neuenfeld:2025wnl}. These extensions would provide useful tests of which aspects of our construction are specific to AdS$_3$/$T\bar T$ holography and which are more general features of gravitational theories with dynamical boundaries.

We will report on some of these directions in upcoming work.

\section*{Acknowledgments}
 
We would like to thank Dominik Neuenfeld, Simon Ross and Erik Verlinde for useful discussions. 
The research of MS is funded by the Deutsche Forschungsgemeinschaft (DFG, German Research Foundation) – Projektnummer 277101999 – TRR 183.

\appendix

\section{Non-local contributions} \label{app:nonlocal}
Consider the non-local term
\begin{equation}
	W_{\text{non-local}}[\gamma,\Phi]=\int d^2x \sqrt{-\gamma} G(\Phi)(\nabla \Phi)^2\Box^{-1}R_{\gamma}. \label{Wnonlocalappendix}
\end{equation}
It can be localized through the addition of an auxiliary field $\Psi$ obeying
\begin{equation}
	\Box \Psi= R_{\gamma} \implies \Psi= \Box^{-1}R_{\gamma}.
\end{equation}
This constraint can be enforced by adding yet another field $N$ acting as a Lagrange multiplier. The localized form of the above non-local term is thus
\begin{align}
	W_{\text{non-local}}[\gamma,\Phi]=&\int d^2x \sqrt{-\gamma}\left[X \Psi +N (\Box \Psi - R_{\gamma})\right]
	\\
	&=\int d^2x \sqrt{-\gamma}\left(X \Psi - \nabla_{\mu}N\nabla^{\mu}\Psi-N R_{\gamma}\right),
\end{align}
where we defined $X=G(\Phi)(\nabla \Phi)^2$ and performed integration by parts. The equation of motion conjugate to $\Psi$ is
\begin{equation}
	\Box N= -X.
\end{equation}
The stress-tensor contribution corresponding to the non-local term \eqref{Wnonlocalappendix} reads
\begin{align}
	\frac{1}{\sqrt{-\gamma}}\frac{	\delta W_{\text{non-local}}[\gamma,\Phi]}{\delta \gamma^{\mu\nu}}=&G(\Phi)\Psi\nabla_{\mu}\Phi\nabla_{\nu}\Phi-\frac{1}{2}\left(\nabla_{\mu}N\nabla_{\nu}\Psi+\nabla_{\nu}N\nabla_{\mu}\Psi\right) \nonumber
	\\
	&-\frac{1}{2}\gamma_{\mu\nu}(X \Psi-\nabla_{\alpha}N \nabla^{\alpha}\Psi)-\gamma_{\mu\nu}\Box N+\nabla_{\mu}\nabla_{\nu}N.
\end{align}
Its trace is
\begin{equation}
	\gamma^{\mu\nu}	\frac{1}{\sqrt{-\gamma}}\frac{\delta	W_{\text{non-local}}[\gamma,\Phi]}{\delta \gamma^{\mu\nu}}=-\Box N =X=G(\Phi)(\nabla\Phi)^2.
\end{equation}
Accounting for the contribution of the non-local piece (\ref{Wnonlocalappendix}) in the derivations of section \ref{subsec:solvingtheflow} allows us to fix
\begin{equation}
	G(\Phi)=-\frac{c}{48\pi}D''(\Phi)=\frac{c\lambda}{96\pi}\frac{e^{-\Phi}}{(1-\lambda e^{-\Phi})^{3/2}}.
\end{equation}
Note that we determined only the non-zero trace part of the leading
non-local contribution in the derivative expansion of the effective
action. An additional traceless part at the same order in derivatives and at leading order $\mathcal{O}(e^{-\Phi})$ in the asymptotic expansion is allowed, and is possibly required to enforce the consistency condition $\langle T_{\mu\nu}\gamma^{\mu\nu}\rangle=-\frac{4\pi}{\sqrt{-\gamma}}\frac{\delta W[\gamma,\Phi]}{\delta \Phi}$. Further non-local contributions are also expected
at higher derivative orders.

\section{Higher derivative terms} \label{app:higherderivatives}

We can deal with the higher derivative terms in \eqref{Weff} using an expansion in 
$e^{-\Phi}$. Requiring the asymptotic limit $\Phi\rightarrow\infty$ to
reduce to the Liouville description, all higher-derivative
corrections start at least at order $\mathcal{O}(e^{-\Phi})$. At this
order, the $T\bar T$ operator in the trace-flow equation depends only
on the leading $\mathcal{O}(1)$ stress tensor \eqref{stressCarroll}. Since this contains at
most two derivatives, its $T\bar T$ combination contains at most four
derivatives. Thus, at order $\mathcal{O}(e^{-\Phi})$, the local part of
the deformation flow requires terms with at most four derivatives. More generally, the structure of the deformation flow implies that at
order $\mathcal{O}(e^{-(n+1)\Phi})$ the local effective action requires
terms containing at most $4+2n$ derivatives.

It is important that the trace-flow equation does not by itself
distinguish local and non-local contributions to the effective action.
As discussed in Appendix \ref{app:nonlocal}, non-local terms are
already required for consistency of the derivative expansion at two derivatives, and higher derivative orders are also expected to contribute to the trace. We therefore use
the trace-flow equation here to obtain the deformation flow of the
action and determine its local four-derivative part directly, leaving
the non-local completion separate.

The same scaling argument used in \eqref{consistentflows} relates the
$t$-flow of the effective action to its integrated trace. Restricting
\eqref{TTbarlikeTraceflow} to four derivatives gives
\begin{align}
	\left.\frac{dW}{dt}\right|_{(4)}
	&=
	\frac{1}{4\pi t}
	\int d^2x\sqrt{-\gamma}\,
	\left.
	\langle \gamma^{\mu\nu} T_{\mu\nu}\rangle
	\right|_{(4)}
	\nonumber\\
	&=
	\frac{1}{4\pi T_0}
	\int d^2x\sqrt{-\gamma}\,
	e^{-\Phi}
	\big\langle T_{\mu\nu}T_{\alpha\beta}\gamma^{\mu\alpha}\gamma^{\nu\beta}-(T_{\mu\nu}\gamma^{\mu\nu})^2\big\rangle\big|_{(4)}.
	\label{higherderivativeflow}
\end{align}
The remaining terms in \eqref{TTbarlikeTraceflow} contain at most two
derivatives and therefore do not enter this equation. Moreover, the
asymptotic relation
\begin{equation}
	T_0=1+\mathcal{O}(e^{-\Phi})
\end{equation}
implies that $1/T_0$ may be replaced by one at the order considered.

The four-derivative part of the $T\bar{T}$ operator can therefore be
evaluated using only the leading two-derivative stress tensor \eqref{stressCarroll}. Using
$D'(\Phi)=1+\mathcal{O}(e^{-\Phi})$,
$D''(\Phi)=\mathcal{O}(e^{-\Phi})$, and
$E(\Phi)=1+\mathcal{O}(e^{-\Phi})$, one finds
\begin{align}
	\big\langle T_{\mu\nu}T_{\alpha\beta}\gamma^{\mu\alpha}\gamma^{\nu\beta}-(T_{\mu\nu}\gamma^{\mu\nu})^2\big\rangle\big|_{(4)}
	=&
	\left(\frac{c}{12}\right)^2
	\Bigg(
	\frac{1}{8}(\nabla\Phi)^4
	+\nabla^\mu\nabla^\nu\Phi
	\nabla_\mu\nabla_\nu\Phi
	+\frac{1}{2}(\nabla\Phi)^2\Box\Phi
	\nonumber\\
	&-\nabla_\mu\Phi\nabla_\nu\Phi
	\nabla^\mu\nabla^\nu\Phi
	-(\Box\Phi)^2
	\Bigg)+\mathcal{O}(e^{-\Phi}).
	\label{OTTbarfourderivative}
\end{align}
Integrating the deformation flow and using
$t=-3/c$, therefore gives directly
\begin{align}
	\text{higher}\;\text{derivatives}\;\text{in}\;W
	=
	&-\frac{c}{48\pi}
	\int d^2x\sqrt{-\gamma}\,
	\frac{e^{-\Phi}}{4}
	\Bigg(
	\frac{1}{8}(\nabla\Phi)^4
	+\nabla^\mu\nabla^\nu\Phi
	\nabla_\mu\nabla_\nu\Phi
	+\frac{1}{2}(\nabla\Phi)^2\Box\Phi
	\nonumber\\
	&\qquad \qquad \qquad -\nabla_\mu\Phi\nabla_\nu\Phi
	\nabla^\mu\nabla^\nu\Phi
	-(\Box\Phi)^2
	\Bigg)
	+\mathcal{O}(e^{-2\Phi}).
	\label{higherderivativepreIBP}
\end{align}

This is now directly an expression inside the effective action. We can
therefore simplify it by integrations by parts. In particular, up to
one-dimensional corner terms,
\begin{align}
	&\int d^2x\sqrt{-\gamma}\,
	e^{-\Phi}
	\nabla^\mu\nabla^\nu\Phi
	\nabla_\mu\nabla_\nu\Phi
	\nonumber\\
	&\qquad =
	\int d^2x\sqrt{-\gamma}\,
	e^{-\Phi}
	\Big(
	\nabla_\mu\Phi\nabla_\nu\Phi
	\nabla^\mu\nabla^\nu\Phi
	+(\Box\Phi)^2
	-\Box\Phi(\nabla\Phi)^2
	-R_{\mu\nu}[\gamma]\nabla^\mu\Phi\nabla^\nu\Phi
	\Big),
	\label{higherderivativeIBP}
\end{align}
with $R_{\mu\nu}[\gamma]$ the Ricci tensor of $\gamma$. Substituting \eqref{higherderivativeIBP} into
\eqref{higherderivativepreIBP}, the terms proportional to
$\nabla_\mu\Phi\nabla_\nu\Phi\nabla^\mu\nabla^\nu\Phi$ and
$(\Box\Phi)^2$ cancel, and we obtain
\begin{align}
	\text{higher}\;\text{derivatives}\;\text{in}\;W[\gamma,\Phi]
	=
	-\frac{c}{48\pi}
	\int d^2x\sqrt{-\gamma}&\,
	\frac{e^{-\Phi}}{32}
	\Big(
	(\nabla\Phi)^4
	-4\Box\Phi\,(\nabla\Phi)^2 \nonumber	
	\\
	&-8R_{\mu\nu}[\gamma]
	\nabla^\mu\Phi\nabla^\nu\Phi
	\Big)
	+\mathcal{O}(e^{-2\Phi}).
	\label{higherderivativesolutionappendix}
\end{align}
Once again, we stress that this result only refers to the leading order in the asymptotic expansion of the higher derivative terms within the local sector.

\section{Entanglement entropy of $T\bar{T}$-deformed CFT$_2$} \label{appendix:B}

\subsection{$T\bar{T}$ on the sphere}	\label{Appendix:B.1}

We review here the result of \cite{Donnelly:2018bef}, following in this appendix the conventions of that paper, with relations to the conventions in the main text pointed out in footnotes where needed. 

Under a variation of the background metric, the response of the partition function can be expressed in terms of the stress tensor as\footnote{In this appendix we work in Euclidean signature. Note also that the stress tensor convention in this appendix differs by a factor of $2\pi$ relative to the main text. The result for the entanglement entropy is of course invariant under choice of convention.}
\begin{equation} \label{eq:1.4.1}
	\delta \log Z = -\frac{1}{2} \int d^2 x \sqrt{g} \langle \mathcal T^{\mu\nu} \rangle \delta g_{\mu\nu}.
\end{equation}
A $T\bar{T}$-deformed CFT$_2$ with large central charge $c$, partition function $Z_{T\bar{T}}(g)=e^{-W_{T\bar{T}}[g]}$ and stress tensor $\vev{\mathcal T_{\mu\nu}}=-\frac{2}{\sqrt{g}}\frac{\delta W_{T\bar{T}}[g]}{\delta g^{\mu\nu}}$ obeys the following trace flow equation\footnote{In these conventions, the relation between the deformation parameter $\mu$ and the bulk parameters is $\mu=16\pi G \ell=\frac{24\pi}{c}\ell^2$, with $\ell$ the AdS$_3$ radius.}
\begin{equation} \label{eq:1.4.2}
	\langle \mathcal T_{\mu}^{\mu} \rangle =-\frac{c}{24 \pi} R - \frac{\mu}{4}\left( \langle \mathcal T_{\mu\nu}\rangle \langle \mathcal T^{\mu\nu}\rangle - \langle \mathcal T_{\mu}^{\mu} \rangle^2  \right).
\end{equation} 
Here we consider a $T\bar{T}$-deformed CFT$_2$ on a sphere with metric
\begin{equation}
	g_{\mu\nu}dx^{\mu}dx^{\nu}=r^2(d\theta^2+(\sin\theta)^2 \; d\phi^2).
\end{equation}
The stress tensor must take the form $\mathcal T_{\mu\nu}=\alpha \, g_{\mu\nu}$ due to symmetry. Combining this form of the stress tensor with the flow equation \eqref{eq:1.4.2}, leads to a quadratic equation with solution
\begin{equation}
	\alpha= \frac{2}{\mu} \left(1-\sqrt{1+\frac{c \mu}{24 \pi r^2}}\right),
\end{equation}
where we chose the branch which yields the CFT$_2$ trace anomaly in the undeformed limit $\mu \rightarrow 0 $ and we used that, on the sphere, $R=\frac{2}{r^2}$. From \eqref{eq:1.4.1} we see that varying the radius of the sphere, the partition function responds as 
\begin{equation} \label{eq:1.4.6}
	\partial_r\log Z_{T\bar{T}}(g) = - \frac{1}{r} \int d^2 x \sqrt{g} \langle \mathcal T_{\mu}^{\mu} \rangle.
\end{equation}
This implies the following differential equation
\begin{equation} \label{eq:1.4.7}
	\partial_r \log Z_{T\bar{T}}(g) = \frac{16 \pi}{\mu} \left(\sqrt{r^2+ \frac{c \mu}{24 \pi}}-r\right).
\end{equation}	
We can integrate this differential equation to obtain the partition function of the $T\bar{T}$-deformed CFT$_2$ on the sphere. For $\mu \ge 0$, the right-hand-side of \eqref{eq:1.4.7} is real-valued for any $r \ge 0$. For positive $\mu$, we obtain
\begin{align}
	\mu \ge 0: \;\; \log Z_{T\bar{T}}(g) 
	& = \frac{c}{3} \log\left(\sqrt{\frac{24 \pi}{c \mu}}r+\sqrt{1+\frac{24 \pi}{c \mu}r^2}\right) + \frac{8 \pi}{\mu} \left(r \sqrt{r^2+\frac{c \mu}{24 \pi}}-r^2\right) \\
	&= \frac{c}{3} \sinh^{-1}\left(\sqrt{\frac{24 \pi}{c \mu}}r\right) + \frac{8 \pi}{\mu} \left(r \sqrt{r^2+\frac{c \mu}{24 \pi}}-r^2\right), \label{normZS2}
\end{align}
where the integration constant is fixed by demanding the boundary condition $\log Z_{T\bar{T}}(g)=0$ at $r=0$. The result can be checked to satisfy $\mu \frac{d}{d\mu} \log Z_{T\bar T}(g) = -\frac{r}{2} \frac{d}{dr} \log Z_{T\bar T}(g) \sim \alpha$ in accordance with scaling arguments based on the dimension of $\mu$ and $r$.  

\par \noindent We are interested in calculating the von Neumann entropy for an entangling surface $A$ consisting of two antipodal points on the sphere, i.e. 
\begin{equation}
	A:\{\phi=0,\; 0\le\theta\le\pi\} . 
\end{equation}
Using the replica trick, the von Neumann entropy is given by
\begin{equation} \label{eq:1.4.10}
	S^{ent}_{T\bar{T}}(A) = \lim_{n\rightarrow1} (1-n \partial_n) \log Z_{T\bar{T}}(g^{(n)}),
\end{equation}
where $Z_{T\bar{T}}(g^{(n)})$ is the partition function of the $T\bar{T}$-deformed CFT on our replica manifold $g^{(n)}_{\mu\nu}$, i.e. $n$ copies of the sphere glued cyclically along the branch points. 
The metric of the replica manifold is
\begin{equation}
	g^{(n)}_{\mu\nu}dx^{\mu}dx^{\nu}=r^2(d\theta^2+n^2 (\sin\theta)^2d\phi^2).
\end{equation}
Using \eqref{eq:1.4.1}, we see that the variation of the replica partition function upon variation of the replica index $n$ is
\begin{equation}
	\partial_n \log Z_{T\bar{T}}(g^{(n)}) = - \frac{1}{n} \int d^2 x \sqrt{g^{(n)}} \langle \mathcal T^{\phi}_{\phi} \rangle . 
\end{equation}
Combining this with the fact that the stress tensor on the sphere must be isotropic, such that $\mathcal T^{\mu}_{\mu}=2 \mathcal T^{\phi}_{\phi}$, and comparing to \eqref{eq:1.4.6}, we see that 
\begin{equation}
	\lim_{n\rightarrow1} \partial_n \log Z_{T\bar{T}}(g^{(n)}) = \frac{r}{2} \partial_r \log Z_{T\bar{T}}(g).
\end{equation}
Therefore, from \eqref{eq:1.4.10} we infer that the von Neumann entropy of the $T\bar{T}$-deformed CFT on the sphere for an entangling region consisting of antipodal points is given by
\begin{equation}
	S^{ent}_{T\bar{T}}(A)=\left(1-\frac{r}{2}\partial_r\right) \log Z_{T\bar{T}}(g).
\end{equation}
For positive deformation parameter, one obtains
\begin{equation} \label{eq:B.15}
	\mu \ge 0: \;\; 	S^{ent}_{T\bar{T}}(A)
	= \frac{c}{3} \log\left(\sqrt{\frac{24 \pi}{c \mu}}r+\sqrt{1+\frac{24 \pi}{c \mu}r^2}\right) = \frac{c}{3} \sinh^{-1}\left(\sqrt{\frac{24 \pi}{c \mu}}r\right).	
\end{equation}
This result is matched holographically in \cite{Donnelly:2018bef} using the relation $\mu=16\pi G \ell=\frac{24\pi}{c}\ell^2$, with $\ell$ the dual AdS$_3$ radius. 

\subsection{$T\bar{T}$ on the hyperbolic plane} \label{Appendix:B.2}

We now derive the entanglement entropy of $T\bar{T}$-deformed CFT$_2$ on the hyperbolic plane for the radial interval extending from the center to the boundary. We follow the methodology of \cite{Donnelly:2018bef}. The same result is obtained, from both boundary and bulk calculations, in \cite{Deng:2023pjs}.  
\par Consider a $T\bar{T}$-deformed CFT$_2$ with large central charge $c$, partition function $Z_{T\bar{T}}(g)=e^{-W_{T\bar{T}}[g]}$ and stress tensor $\vev{\mathcal T_{\mu\nu}}=-\frac{2}{\sqrt{g}}\frac{\delta W_{T\bar{T}}[g]}{\delta g^{\mu\nu}}$ on a hyperbolic plane with metric
\begin{equation}
	g_{\mu\nu}dx^{\mu}dx^{\nu}=r^2(d\eta^2+(\sinh\eta)^2 \; d\phi^2).
\end{equation}
The stress tensor must take the form $\mathcal T_{ab}=\alpha \, g_{ab}$ due to symmetry. Combining this form of the stress tensor with the trace flow equation \eqref{eq:1.4.2}, leads to a quadratic equation with solution
\begin{equation}
	\alpha= \frac{2}{\mu} \left(1-\sqrt{1-\frac{c \mu}{24 \pi r^2}}\right),
\end{equation}
where we chose the branch which yields the CFT$_2$ trace anomaly in the undeformed limit $\mu \rightarrow 0 $ and we used that, on the hyperbolic plane, $R=-\frac{2}{r^2}$. From \eqref{eq:1.4.1} we see that varying the radius of curvature of the hyperbolic plane, the partition function responds as 
\begin{equation} \label{eq:1.34}
	\partial_r \log Z_{T\bar{T}}(g) = - \frac{1}{r} \int d^2 x \sqrt{g} \langle \mathcal T_{\mu}^{\mu} \rangle.
\end{equation}
Using the form of the stress tensor determined through the trace flow equation, and
using a renormalized volume of unit hyperbolic space equal to $(-2\pi)$  \cite{Deng:2023pjs}, one obtains the following differential equation
\begin{equation}
	\partial _r \log Z_{T\bar{T}}(g) = -\frac{8 \pi}{\mu} \left(\sqrt{r^2- \frac{c \mu}{24 \pi}}-r\right).
\end{equation}	
This differential equation can be integrated to obtain the partition function of the $T\bar{T}$-deformed CFT$_2$ on the hyperbolic plane
\begin{equation} 
	\mu \ge 0: \;\; \log Z_{T\bar{T}}(g) = \frac{c}{6} \cosh^{-1}\left(\sqrt{\frac{24 \pi}{c \mu}}r\right) - \frac{4 \pi}{\mu} \left(r \sqrt{r^2-\frac{c \mu}{24 \pi}}-r^2\right), \label{normZH2}
\end{equation}
where the choice of integration constant will be explained soon.
\par We are interested in calculating the von Neumann entropy for a radial interval $A$ extending from the center of the disk to the boundary
\begin{equation}
	A:\{\phi=0,\; 0\le\eta < \infty\}.
\end{equation}
Using the replica trick, the von Neumann entropy is given by
\begin{equation} \label{eq:1.37}
	S^{ent}_{T\bar{T}}(A) = \lim_{n\rightarrow1} (1-n \partial_n) \log Z_{T\bar{T}}(g^{(n)}),
\end{equation}
where $Z_{T\bar{T}}(g^{(n)})$ is the partition function of the $T\bar{T}$-deformed CFT$_2$ on our replica manifold with metric $g^{(n)}_{\mu\nu}$, i.e.~$n$ copies of the hyperbolic disk glued cyclically along the branch points. The metric for the replica manifold is
\begin{equation}
	g^{(n)}_{\mu\nu}dx^{\mu}dx^{\nu}=r^2(d\eta^2+n^2 (\sinh\eta)^2 \; d\phi^2).
\end{equation}
Using \eqref{eq:1.4.1}, we see that the variation of the replica partition function upon variation of the replica index $n$ is
\begin{equation}
	\partial_n \log Z_{T\bar{T}}(g^{(n)}) = - \frac{1}{n} \int d^2 x \sqrt{g^{(n)}}\langle \mathcal T^{\phi}_{\phi} \rangle .
\end{equation}
Combining this with the fact that the stress tensor on the hyperbolic plane is isotropic, such that $\mathcal T^{\mu}_{\mu}=2 \mathcal T^{\phi}_{\phi}$, and comparing to \eqref{eq:1.34}, we see that
\begin{equation}
	\lim_{n\rightarrow1} \partial_n \log Z_{T\bar{T}}(g^{(n)}) = \frac{r}{2} \partial_r \log Z_{T\bar{T}}(g).
\end{equation}
Therefore, from \eqref{eq:1.37} we infer that the von Neumann entropy of the $T\bar{T}$-deformed CFT on the hyperbolic plane for the interval $A$ extending from the center to the boundary is 
\begin{equation}
	S^{ent}_{T\bar{T}}(A)=\left(1-\frac{r}{2}\partial_r\right) \log Z_{T\bar{T}}(g).
\end{equation}
Thus, we obtain  
\begin{align} \label{eq:1.4.16}
	\mu \ge 0: \;\; S^{ent}_{T\bar{T}}(A) &= \frac{c}{6} \log\left(\sqrt{\frac{24 \pi}{c \mu}}r+\sqrt{-1+\frac{24 \pi}{c \mu}r^2}\right) \\ &= \frac{c}{6} \cosh^{-1}\left(\sqrt{\frac{24 \pi}{c \mu}}r\right). 
\end{align}
The integration constant in \eqref{eq:1.4.16} was chosen by demanding that, for $\mu \ge 0$, $S^{ent}_{T\bar{T}}(A)=0$ at the critical value of $r$ below which one has imaginary solutions, $r_{crit}=\sqrt{\frac{c \mu}{24 \pi}}$. 

In the main text we use the dimensionless $\epsilon$ defined in \eqref{defepsilon} as the physical measure for the distance into the bulk associated with the $T\bar T$ deformation. It relates to the cutoff location $\bar \rho$ as 
\ali{
	\epsilon = \left\{ \begin{array}{ll} 
		\frac{1}{\sinh \bar \rho} & \quad (\lambda = -1) \\ 
		\frac{1}{\cosh \bar \rho} & \quad (\lambda = 1)  
	\end{array} . \right. 
}
Therefore it relates to the induced radius notation of this appendix, $r = \ell \sinh \bar \rho$ for $\lambda = -1$ and $r = \ell \cosh \bar \rho$ for $\lambda = 1$, as 
\ali{
	\epsilon = \frac{\ell}{r} = \sqrt{\frac{c\mu}{24\pi}} \frac{1}{r}. 
}

\section{Solving for the deformed Liouville $\hat{S}[\gamma,\sigma]$ in an asymptotic expansion}\label{app:defliouorderbyorder}
In section \ref{subsec:TTbarmatter}, we solved for the deformed Liouville action $\hat{S}[\gamma,\sigma]$ in a derivative expansion, following the strategy of section \ref{subsec:solvingtheflow}. Here we instead show how to obtain the action by solving the $T\bar{T}$-like trace flow \eqref{traceflowthat} in an asymptotic expansion in the deformation parameter $\tilde{t}=-3\epsilon^2/c$, using the power law decompositions
\begin{equation}
	\hat{S}=\sum_{n\ge0}\tilde{t}^n \hat{S}^{(n)}=\int d^2x \sqrt{-\gamma}\sum_{n\ge 0}\tilde{t}^n\mathcal{L}^{(n)},
\end{equation}
and
\begin{equation}
	t_{\mu\nu}=\sum_{n\ge 0}\tilde{t}^n t_{\mu\nu}^{(n)}=\sum_{n\ge 0}\tilde{t}^n \frac{4\pi}{\sqrt{-\gamma}}\frac{\delta \hat{S}^{(n)}}{\delta \gamma^{\mu\nu}},
\end{equation}
while taking the timelike Liouville theory as the seed theory in the undeformed limit 
\begin{equation} \label{seedsigma}
	\hat{S}^{(0)}[\gamma,\sigma]=\lim\limits_{\epsilon\rightarrow 0}\hat{S}[\gamma,\sigma]=\frac{c}{48\pi}\int d^2 x \sqrt{-\gamma}\left(\sigma R_{\gamma}+\frac{1}{2}(\nabla \sigma)^2+2\lambda e^{\sigma}\right).
\end{equation}
The corresponding undeformed stress tensor
\begin{align} \label{undefstress}
	t^{(0)}_{\mu\nu}=&\frac{4\pi}{\sqrt{-\gamma}}\frac{\delta \hat{S}^{(0)}[\gamma,\sigma]}{\delta \gamma^{\mu\nu}} \nonumber
	\\
	=&-\frac{c}{24}\left(-\nabla_{\mu}\sigma\nabla_{\nu}\sigma+\gamma_{\mu\nu}\left(\frac{1}{2}(\nabla \sigma)^2+2\lambda e^{\sigma}\right)+2\left(\nabla_{\mu}\nabla_{\nu}\sigma-\gamma_{\mu\nu}\Box\sigma\right)\right).
\end{align}
solves the undeformed limit of the trace flow (\ref{traceflowthat}), i.e.
\begin{equation}\label{tracetasy}
	t^{(0)\mu}_{\mu}=\lim\limits_{\epsilon\rightarrow 0}t^{\mu}_{\mu}=\frac{c}{12}\Box\sigma-\frac{c}{6}\lambda e^{\sigma}.
\end{equation}
Assuming for example $T_0=1-\frac{\lambda\epsilon^2}{2}-\frac{\lambda^2\epsilon^4}{8}+\mathcal{O}(\epsilon^6)$, at first order the trace flow equation \eqref{traceflowthat} reads
\begin{equation} \label{hatt1}
	t^{(1)\mu}_{\mu}=e^{-\sigma}\mathcal{O}_{t\bar{t}}^{(0)}-\frac{c^2}{36}\lambda(1+e^{-\sigma})\left(\frac{1}{2}\Box \sigma -\lambda e^{\sigma}\right)-\frac{c^2}{72}\lambda^2e^{-\sigma}-\frac{c^2}{72}\lambda^2,
\end{equation}
with $\mathcal{O}_{t\bar{t}}^{(0)}=t^{(0)}_{\mu\nu}t^{(0)}_{\alpha\beta}\gamma^{\alpha\mu}\gamma^{\beta\nu}-(t^{(0)}_{\mu\nu}\gamma^{\mu\nu})^2$, or explicitly
\begin{align} \label{OTT0}
	\mathcal{O}_{t\bar{t}}^{(0)}=&\left(\frac{c}{24}\right)^2\Bigg(\frac{1}{2}(\nabla \sigma)^4+2(\nabla \sigma)^2\Box\sigma-4(\Box\sigma)^2+4\nabla_{\mu}\nabla_{\nu}\sigma \nabla^{\mu}\nabla^{\nu}\sigma \nonumber
	\\
	&\qquad \qquad \qquad -4\nabla_{\mu}\sigma\nabla_{\nu}\sigma \nabla^{\mu}\nabla^{\nu}\sigma+8\lambda e^{\sigma}\Box\sigma -8\lambda^2e^{2\sigma}\Bigg).
\end{align}
Using the same strategy as in \ref{app:higherderivatives} for the higher derivative terms, we find that equation \eqref{hatt1} is solved by the following first order correction
\begin{align} \label{Ssigmacorr1epsilongeneral}
	\hat{S}_{\sigma}[\gamma,\sigma]=&\hat{S}_{\sigma}^{(0)}[\gamma,\sigma]-\frac{c}{48\pi}\left(\frac{\epsilon}{4}\right)^2\int  d^2x \sqrt{-\gamma}\Bigg(\frac{1}{2}e^{-\sigma}(\nabla \sigma)^4-2e^{-\sigma}(\nabla \sigma)^2\Box\sigma -4 R_{\mu\nu}[\gamma]\nabla^{\mu}\sigma \nabla^{\nu}\sigma \nonumber
	\\
	&-8\lambda^2 e^{\sigma}+8\lambda^2e^{-\sigma}-8\lambda^2+\hat{E}^{(\epsilon^2)}(\sigma)(\nabla \sigma)^2+(\hat{D}_0^{(\epsilon^2)}+8\lambda e^{-\sigma})R_{\gamma}\Bigg)+\mathcal{O}(\epsilon^4),
\end{align} 
up to a non-local term of the form $\sim \epsilon^2 e^{-\sigma}(\nabla \sigma)^2\Box^{-1}R_{\gamma}$ which cancels unwanted contributions of the form $\sim e^{-\sigma}(\nabla \sigma)^2$ in $t^{(1)\mu}_{\mu}$, as described in Appendix \ref{app:nonlocal} with $\Phi=\sigma-2\log \epsilon$. Note that the first three terms in parentheses in equation \eqref{Ssigmacorr1epsilongeneral}, containing four derivatives each, are obtained from the $\mathcal{O}_{t\bar{t}}^{(0)}$ operator in \eqref{OTT0} after repeated integration by parts. The function $\hat{E}^{(\epsilon^2)}(\sigma)$ and constant $\hat{D}_0^{(\epsilon^2)}$ are undetermined by this solving strategy, but they are related respectively to $\hat{E}(\sigma)$ and $\hat{D}_0$ of section \ref{subsec:TTbarmatter} via $\hat{E}(\sigma)=1-\frac{\epsilon^2}{8}\hat{E}^{(\epsilon^2)}(\sigma)+\mathcal{O}(\epsilon^4)$ and $\hat{D}_0=-2\log2-\frac{\epsilon^2}{16}\hat{D}_0^{(\epsilon^2)}+\mathcal{O}(\epsilon^4)$. Thus, by comparing directly with the bulk calculation of section \ref{sect:bulk}, we find that we must fix
\begin{equation} \label{D0epsilon2}
	\hat{D}_0^{(\epsilon^2)}=-8\lambda.
\end{equation}

\section{Derivation of the brane effective action from a fluctuating bulk cutoff}
\label{app:fluctuatingcutoffaction}

In this appendix, we derive the brane effective action
\eqref{bulkactionfluctuatinggamma} by explicitly evaluating the bulk
on-shell action up to a fluctuating radial cutoff. We consider the
three maximally symmetric slicings $\lambda=0,\pm1$ together. The bulk
metric is
\begin{equation}
	ds^2
	=
	d\rho^2+e^{2A(\rho)}
	\gamma_{\mu\nu}(x)dx^\mu dx^\nu ,
	\label{bulkmetricfluctuatingappendix}
\end{equation}
with
\begin{equation}
	e^{A(\rho)}
	=
	\begin{cases}
		\cosh\rho, & \lambda=+1,\\[2mm]
		e^\rho, & \lambda=0,\\[2mm]
		\sinh\rho, & \lambda=-1,
	\end{cases}
	\qquad
	R_\gamma=-2\lambda .
	\label{Acasesfluctuatingappendix}
\end{equation}
In all three cases, the warp factor satisfies
\begin{equation}
	\dot A^2
	=
	1-\lambda e^{-2A},
	\qquad
	\ddot A
	=
	\lambda e^{-2A},
	\label{Arelationsfluctuatingappendix}
\end{equation}
where a dot denotes a derivative with respect to $\rho$.

We place the EOW brane at the fluctuating radial location
\begin{equation}
	\rho=\tilde\rho(x).
\end{equation}
The induced metric is
\begin{equation}
	g_{\mu\nu}
	=
	e^{2A(\tilde\rho)}\gamma_{\mu\nu}
	+
	\nabla_\mu\tilde\rho\nabla_\nu\tilde\rho ,
	\label{inducedmetricfluctuatingappendix}
\end{equation}
and therefore
\begin{equation}
	\sqrt{-g}
	=
	\sqrt{-\gamma}\,
	e^{2A(\tilde\rho)}
	\sqrt{
		1+e^{-2A(\tilde\rho)}
		(\nabla\tilde\rho)^2
	},
	\label{detfluctuatingappendix}
\end{equation}
where
\begin{equation}
	(\nabla\tilde\rho)^2
	\equiv
	\gamma^{\mu\nu}
	\nabla_\mu\tilde\rho
	\nabla_\nu\tilde\rho .
\end{equation}

The total bulk action is
\begin{equation}
	S_{\rm tot}
	=
	\frac{1}{16\pi G}
	\int_{\mathcal M}d^3X\sqrt{-G}\,(R_G+2)
	+
	\frac{1}{8\pi G}
	\int_{\rm EOW}d^2x\sqrt{-g}\,(K-T_0).
	\label{Stotfluctuatingappendix}
\end{equation}
We now evaluate its three contributions.

The bulk Ricci scalar for
\eqref{bulkmetricfluctuatingappendix} is
\begin{equation}
	R_G
	=
	e^{-2A}R_\gamma
	-4\ddot A
	-6\dot A^2.
\end{equation}
Thus, we find that the Einstein--Hilbert contribution evaluates to
\begin{equation}
	S_{\rm EH}
	=
	\frac{1}{16\pi G}
	\int d^2x\sqrt{-\gamma}
	\left(
	\tilde\rho R_\gamma
	-2\dot A e^{2A}
	\right)_{\rho=\tilde\rho}
	+\cdots ,
	\label{SEHfluctuatingappendix}
\end{equation}
where here and in the following the dots denote contributions from the
lower limit of the radial integration, which are independent of
$\tilde\rho$.

We next evaluate the Gibbons--Hawking term. We first define the
outward-pointing unit normal in a neighborhood of the fluctuating
surface as
\begin{equation}
	n_\rho(\rho,x)
	=
	\frac{1}{
		\sqrt{1+e^{-2A(\rho)}(\nabla\tilde\rho)^2}
	},
	\qquad
	n_\mu(\rho,x)
	=
	-\frac{\nabla_\mu\tilde\rho}{
		\sqrt{1+e^{-2A(\rho)}(\nabla\tilde\rho)^2}
	}.
	\label{normalfluctuatingappendix}
\end{equation}
The normal and its derivatives are evaluated on the surface
$\rho=\tilde\rho(x)$ only after computing the extrinsic curvature.
The non-vanishing Christoffel symbols that enter the calculation are
$\Gamma^\rho_{\mu\nu}=-\dot A e^{2A}\gamma_{\mu\nu}$,
$\Gamma^\mu_{\rho\nu}=\dot A\delta^\mu_\nu$, and
$\Gamma^\mu_{\nu\alpha}=\Gamma^{(\gamma)\mu}_{\nu\alpha}$.

The trace of the extrinsic curvature is
\begin{align}
	K
	&=
	G^{MN}\nabla_Mn_N
	=
	\nabla_\rho n_\rho
	+
	e^{-2A}\gamma^{\mu\nu}\nabla_\mu n_\nu
	\nonumber\\
	&=
	\partial_\rho n_\rho
	+
	e^{-2A}\gamma^{\mu\nu}
	\left(
	\partial_\mu n_\nu
	-\Gamma^\rho_{\mu\nu}n_\rho
	-\Gamma^\alpha_{\mu\nu}n_\alpha
	\right)
	\nonumber\\
	&=
	\frac{
		\dot A e^{-2A}(\nabla\tilde\rho)^2
	}{
		\left(
		1+e^{-2A}(\nabla\tilde\rho)^2
		\right)^{3/2}
	}
	+
	\frac{
		2\dot A
	}{
		\sqrt{
			1+e^{-2A}(\nabla\tilde\rho)^2
		}
	}
	+
	e^{-2A}
	\gamma^{\mu\nu}
	\nabla^{(\gamma)}_\mu n_\nu .
	\label{Kfluctuatingappendix}
\end{align}
Here and in the following, the right-hand side is evaluated at
$\rho=\tilde\rho(x)$ after the derivatives have been taken.

For the last term, using
\begin{equation}
	n_\nu
	=
	-\frac{\nabla_\nu\tilde\rho}{
		\sqrt{1+e^{-2A}(\nabla\tilde\rho)^2}
	},
\end{equation}
we obtain
\begin{align}
	&
	\sqrt{
		1+e^{-2A}(\nabla\tilde\rho)^2
	}
	\,
	\gamma^{\mu\nu}
	\nabla^{(\gamma)}_\mu n_\nu
	\nonumber\\
	&\qquad =
	-\Box\tilde\rho
	+
	\frac{1}{2}
	\frac{
		e^{-2A}
		\nabla^\mu\tilde\rho
		\nabla_\mu(\nabla\tilde\rho)^2
	}{
		1+e^{-2A}(\nabla\tilde\rho)^2
	}.
	\label{normalderivativefluctuatingappendix}
\end{align}
Combining \eqref{detfluctuatingappendix},
\eqref{Kfluctuatingappendix}, and
\eqref{normalderivativefluctuatingappendix}, we therefore find
\begin{align}
	S_{\rm GH}
	=
	\frac{1}{8\pi G}
	\int d^2x\sqrt{-\gamma}
	\Bigg[
	&2\dot A e^{2A}
	+
	\frac{
		\dot A(\nabla\tilde\rho)^2
	}{
		1+e^{-2A}(\nabla\tilde\rho)^2
	}
	-\Box\tilde\rho
	\nonumber\\
	&+
	\frac{1}{2}
	\frac{
		e^{-2A}
		\nabla^\mu\tilde\rho
		\nabla_\mu(\nabla\tilde\rho)^2
	}{
		1+e^{-2A}(\nabla\tilde\rho)^2
	}
	\Bigg]_{\rho=\tilde\rho}.
	\label{SGHbeforeIBPfluctuatingappendix}
\end{align}
The term containing $\Box\tilde\rho$ contributes only a
one-dimensional corner term, since
\begin{align}
	\int d^2x\sqrt{-\gamma}\,\Box\tilde\rho
	&=
	\int d^2x\sqrt{-\gamma}\,
	\nabla_\mu\nabla^\mu\tilde\rho
	\nonumber\\
	&=
	\int d^2x\,
	\partial_\mu
	\left(
	\sqrt{-\gamma}\,\nabla^\mu\tilde\rho
	\right).
\end{align}
Neglecting such corner terms, as throughout this work, the
Gibbons--Hawking contribution becomes
\begin{align}
	S_{\rm GH}
	=
	\frac{1}{8\pi G}
	\int d^2x\sqrt{-\gamma}
	\Bigg[
	&2\dot A e^{2A}
	+
	\frac{
		\dot A(\nabla\tilde\rho)^2
	}{
		1+e^{-2A}(\nabla\tilde\rho)^2
	}
	\nonumber\\
	&+
	\frac{1}{2}
	\frac{
		e^{-2A}
		\nabla^\mu\tilde\rho
		\nabla_\mu(\nabla\tilde\rho)^2
	}{
		1+e^{-2A}(\nabla\tilde\rho)^2
	}
	\Bigg]_{\rho=\tilde\rho}.
	\label{SGHfluctuatingappendix}
\end{align}
Finally, the tension contribution follows directly from
\eqref{detfluctuatingappendix},
\begin{equation}
	S_{T_0}
	=
	-\frac{1}{8\pi G}
	\int d^2x\sqrt{-\gamma}
	\left[
	T_0e^{2A}
	\sqrt{
		1+e^{-2A}(\nabla\tilde\rho)^2
	}
	\right]_{\rho=\tilde\rho}.
	\label{STfluctuatingappendix}
\end{equation}

Adding \eqref{SEHfluctuatingappendix},
\eqref{SGHfluctuatingappendix}, and
\eqref{STfluctuatingappendix}, and dropping the contribution from the
lower limit of the radial integration, gives
\begin{align}
	W[\gamma,\tilde{\rho}]
	=&
	\frac{1}{16\pi G}
	\int d^2x \sqrt{-\gamma}
	\Bigg(
	\tilde{\rho}R_{\gamma}
	+2\dot{A}e^{2A}
	+\frac{
		2\dot{A}(\nabla \tilde{\rho})^2
	}{
		1+e^{-2A}(\nabla\tilde{\rho})^2
	}
	\nonumber\\
	&\qquad\qquad
	+\frac{
		e^{-2A}\nabla^{\mu}\tilde{\rho}
		\nabla_{\mu}(\nabla\tilde{\rho})^2
	}{
		1+e^{-2A}(\nabla\tilde{\rho})^2
	}
	-2T_0e^{2A}
	\sqrt{
		1+e^{-2A}(\nabla\tilde{\rho})^2
	}
	\Bigg)\Bigg|_{\rho=\tilde{\rho}},
	\label{bulkactionfluctuatinggammaappendix}
\end{align}
which reproduces \eqref{bulkactionfluctuatinggamma}.

\section{Relating the radial fluctuation $\tilde{\phi}$ and the Weyl factor $\sigma$}\label{App:tildephisigmarelation}

We outline here a strategy to obtain a relation between the radial fluctuation $\tilde{\phi}$ and the Weyl factor $\sigma$ in an asymptotic expansion. For simplicity, we restrict to the flat-slicing, i.e. the $\lambda=0$ case. The bulk geometry is Poincaré AdS$_3$, with line element $ds^2=d\rho^2+e^{2\rho}\eta_{\mu\nu}dx^{\mu}dx^{\nu}$. The induced line element on a fluctuating EOW brane located at $\rho=\bar{\rho}+\tilde{\phi}(x)$ is
\begin{equation}
	\lambda=0: \qquad 	g_{\mu\nu}dx^{\mu}dx^{\nu}=\left(\frac{e^{2\tilde{\phi}(x)}}{\epsilon^2}\eta_{\mu\nu}+\partial_{\mu}\tilde{\phi}\partial_{\nu}\tilde{\phi}\right)dx^{\mu}dx^{\nu}=\frac{e^{\sigma(w)}}{\epsilon^2}\eta_{\mu\nu}dw^{\mu}dw^{\nu}, \label{tildephisigma0}
\end{equation}
where we have introduced coordinates $(w^\mu)$ in which the induced metric is brought to conformal gauge. We want to use the constant cutoff radius $\epsilon=e^{-\bar{\rho}}$ as the expansion parameter. We adopt the following ansatz for the diffeomorphisms
\begin{equation}
	w^{\mu}=x^{\mu}+\epsilon^2 \xi^{\mu}(x)+\mathcal{O}(\epsilon^4),
\end{equation}
and for the Weyl factor
\begin{equation} \label{sigmaK}
	\sigma(x)=2\tilde{\phi}(x)+\epsilon^2J(x)+\mathcal{O}(\epsilon^4).
\end{equation}
Plugging our ansatz into the r.h.s. of Eq.~(\ref{tildephisigma0}), expanding and matching at order $\mathcal{O}(\epsilon^2)$, we obtain 
\begin{equation}
	J\eta_{\mu\nu}+2\xi^{\alpha}\partial_{\alpha}\tilde{\phi}\eta_{\mu\nu}+\partial_{\mu}\xi_{\nu}+\partial_{\nu}\xi_{\mu}=e^{-2\tilde{\phi}}\partial_{\mu}\tilde{\phi}\partial_{\nu}\tilde{\phi}. \label{Kfull}
\end{equation}
This equation can be split into a trace part
\begin{equation}
	J=\frac{1}{2}e^{-2\tilde{\phi}}(\partial\tilde{\phi})^2-\partial_{\mu}\xi^{\mu}-2\xi^{\mu}\partial_{\mu}\tilde{\phi},
\end{equation}
and a traceless part (obtained by subtracting the trace part multiplied by the Minkowski metric from (\ref{Kfull}))
\begin{equation}
	\partial_{\mu}\xi_{\nu}+\partial_{\nu}\xi_{\mu}-\partial_{\alpha}\xi^{\alpha}\eta_{\mu\nu}=e^{-2\tilde{\phi}}\left(\partial_{\mu}\tilde{\phi}\partial_{\nu}\tilde{\phi}-\frac{1}{2}\eta_{\mu\nu}(\partial\tilde{\phi})^2\right).
\end{equation}
Taking the divergence of the traceless part we have
\begin{equation}
	\Box \xi_{\mu}=e^{-2\tilde{\phi}}\left(\Box \tilde{\phi}-(\partial \tilde{\phi})^2\right)\partial_{\mu}\tilde{\phi}.
\end{equation}
The traceless part determines the following solution for the diffeomorphism
\begin{equation}
	\xi_{\mu}=\frac{1}{\Box}\left(e^{-2\tilde{\phi}}\left(\Box \tilde{\phi}-(\partial \tilde{\phi})^2\right)\partial_{\mu}\tilde{\phi}\right)+\xi_{\mu}^{(0)},
\end{equation}
with $\xi_{\mu}^{(0)}$ a conformal Killing vector satisfying $\partial_{\mu}\xi^{(0)}_{\nu}+\partial_{\nu}\xi^{(0)}_{\mu}-\partial_{\alpha}\xi^{\alpha}_{(0)}\eta_{\mu\nu}=0$. The given solution for the diffeomorphisms is manifestly non-local.
Plugging it back into the trace part, we obtain
\begin{equation}
	J=\frac{1}{2}e^{-2\tilde{\phi}}(\partial \tilde{\phi})^2-\left(\partial_{\mu}+2\partial_{\mu}\tilde{\phi}\right)\left[\frac{1}{\Box}\left(e^{-2\tilde{\phi}}\left(\Box \tilde{\phi}-(\partial \tilde{\phi})^2\right)\partial^{\mu}\tilde{\phi}\right)+\xi^{\mu}_{(0)}\right]. \label{Ktildephi}
\end{equation}
Thus, we have determined the following relation between the Weyl factor $\sigma$ and the radial fluctuation $\tilde{\phi}$ explicitly to first non-trivial order in the asymptotic expansion
\begin{align} \label{sigmatildephilambda0}
	&\lambda=0: 
	\\
	& \sigma=2\tilde{\phi}+\epsilon^2\Bigg\{\frac{1}{2}e^{-2\tilde{\phi}}(\partial \tilde{\phi})^2-\left(\partial_{\mu}+2\partial_{\mu}\tilde{\phi}\right)\left[\frac{1}{\Box}\left(e^{-2\tilde{\phi}}\left(\Box \tilde{\phi}-(\partial \tilde{\phi})^2\right)\partial^{\mu}\tilde{\phi}\right)+\xi^{\mu}_{(0)}\right]\Bigg\}+\mathcal{O}(\epsilon^4). \nonumber
\end{align}
The above relation is manifestly non-local already at order $\mathcal{O}(\epsilon^2)$.

\section{Matching derivative terms} \label{App:matchingderivatives}

It remains unclear to what extent the derivative corrections in the effective action can be matched uniquely between the bulk and boundary descriptions. This is partly due to the fact that the physical information encoded in these terms has not yet been fully disentangled from gauge-dependent contributions, and partly because of the role played by the non-local terms appearing in the effective action and in the field redefinitions. Nevertheless, it is instructive to outline a possible strategy for relating the derivative corrections in the two approaches. For simplicity, we focus on the flat-slicing case ($\lambda=0$), where we can make use of the relation (\ref{sigmatildephilambda0}) between the Weyl factor $\sigma$ and the radial fluctuation $\tilde{\phi}$ derived in the appendix. 
Consider the $\lambda=0$ case for the boundary result (\ref{Ssigmacorr1epsilongeneral}):
\begin{equation}
	\lambda=0: \;\;\; \hat{S}_{\sigma}[\gamma,\sigma]=\frac{c}{48\pi}\int d^2x \left[\frac{1}{2}(\partial \sigma)^2+\frac{\epsilon^2}{8}\left(e^{-\sigma}(\partial \sigma)^2\Box \sigma + \mathcal{O}(\sigma^4)\right)+\mathcal{O}(\epsilon^4)\right].
\end{equation}
We dropped the curvature terms and metric determinant because in this case $\gamma=\eta$, thus $R_{\gamma}=R_{\mu\nu}[\gamma]=0$ and $\sqrt{-\gamma}=1$. We can now use the relation (\ref{sigmaK}) with the solution (\ref{Ktildephi}) to write
\begin{align}
	\frac{1}{2}(\partial \sigma)^2&=2(\partial \tilde{\phi})^2+2\epsilon^2\partial^{\mu}\tilde{\phi}\partial_{\mu}J+\mathcal{O}(\epsilon^4)
	\\
	&=2(\partial \tilde{\phi})^2+2\epsilon^2\partial^{\mu}\tilde{\phi}\partial_{\mu}\Bigg\{\frac{1}{2}e^{-2\tilde{\phi}}(\partial \tilde{\phi})^2-\partial_{\alpha}\left[\frac{1}{\Box}\left(e^{-2\tilde{\phi}}\Box \tilde{\phi}\partial^{\alpha}\tilde{\phi}\right)\right] +\mathcal{O}(\tilde{\phi}^3)\Bigg\}+\mathcal{O}(\epsilon^4)
\end{align}
and
\begin{equation}
	\frac{\epsilon^2}{8}\left(e^{-\sigma}(\partial \sigma)^2\Box \sigma + \mathcal{O}(\sigma^4)\right)=\epsilon^2\left(e^{-2\tilde{\phi}}(\partial \tilde{\phi})^2\Box \tilde{\phi} + \mathcal{O}(\tilde{\phi}^4)\right)+\mathcal{O}(\epsilon^4).
\end{equation}
Here we set the conformal Killing vector to zero ($\xi_{\mu}^{(0)}=0$). Plugging this back into the action we have
\begin{align}
	\hat{S}_{\sigma}[\gamma,\sigma(\tilde{\phi})]=&\frac{c}{24\pi}\int d^2x \Bigg\{(\partial \tilde{\phi})^2+\frac{\epsilon^2}{2}\left(e^{-2\tilde{\phi}}(\partial \tilde{\phi})^2\Box \tilde{\phi} + \mathcal{O}(\tilde{\phi}^4)\right) \nonumber
	\\
	&+\epsilon^2\partial^{\mu}\tilde{\phi}\partial_{\mu}\Bigg[\frac{1}{2}e^{-2\tilde{\phi}}(\partial \tilde{\phi})^2-\partial_{\alpha}\left[\frac{1}{\Box}\left(e^{-2\tilde{\phi}}\Box \tilde{\phi}\partial^{\alpha}\tilde{\phi}\right)\right] \Bigg]+\mathcal{O}(\epsilon^4)\Bigg\}.
\end{align}
Integrating by parts the first term in square brackets one obtains
\begin{align}
	\hat{S}_{\sigma}[\gamma,\sigma(\tilde{\phi})]=\frac{c}{24\pi}\int d^2x \Bigg\{(\partial \tilde{\phi})^2-\epsilon^2\partial^{\mu}\tilde{\phi}\partial_{\mu}\Bigg[\partial_{\alpha}\left[\frac{1}{\Box}\left(e^{-2\tilde{\phi}}\Box \tilde{\phi}\partial^{\alpha}\tilde{\phi}\right)\right]\Bigg]+\epsilon^2\mathcal{O}(\tilde{\phi}^4) +\mathcal{O}(\epsilon^4)\Bigg\}.
\end{align}
We now integrate by parts the second term twice, to arrive at
\begin{align}
	\lambda=0: \;\;\; \hat{S}_{\sigma}[\gamma,\sigma(\tilde{\phi})]=\frac{c}{24\pi}\int d^2x \Big[(\partial \tilde{\phi})^2-\epsilon^2e^{-2\tilde{\phi}}(\partial \tilde{\phi})^2\Box \tilde{\phi}+\epsilon^2\mathcal{O}(\tilde{\phi}^4) +\mathcal{O}(\epsilon^4)\Big]. \label{hatSsigmatildephi}
\end{align}
Comparing to the $\lambda=0$ case for the result (\ref{Stotcritgrav}) of the bulk calculation, we see that the first                       two terms in Eq.~(\ref{hatSsigmatildephi}) are matched exactly.

\section{Bulk derivation of the timelike Liouville action} \label{AppLiou}
We present here a bulk derivation of the timelike Liouville description of the integrated Weyl anomaly. In sections \ref{subsec:TTbarmatter} and \ref{subsLiou}, we showed, from the boundary and bulk perspectives respectively, that our brane effective action reduces to the integrated Weyl anomaly. From the boundary perspective, this follows directly from integrating the trace-anomaly equation, whereas from the bulk perspective the origin of this result is less immediate. The purpose of this appendix is to clarify the bulk interpretation, generalizing the perspective developed in \cite{Callebaut:2025thw} to arbitrary tension. 

\begin{figure}[t]
	\centering
	\includegraphics[scale=0.45]{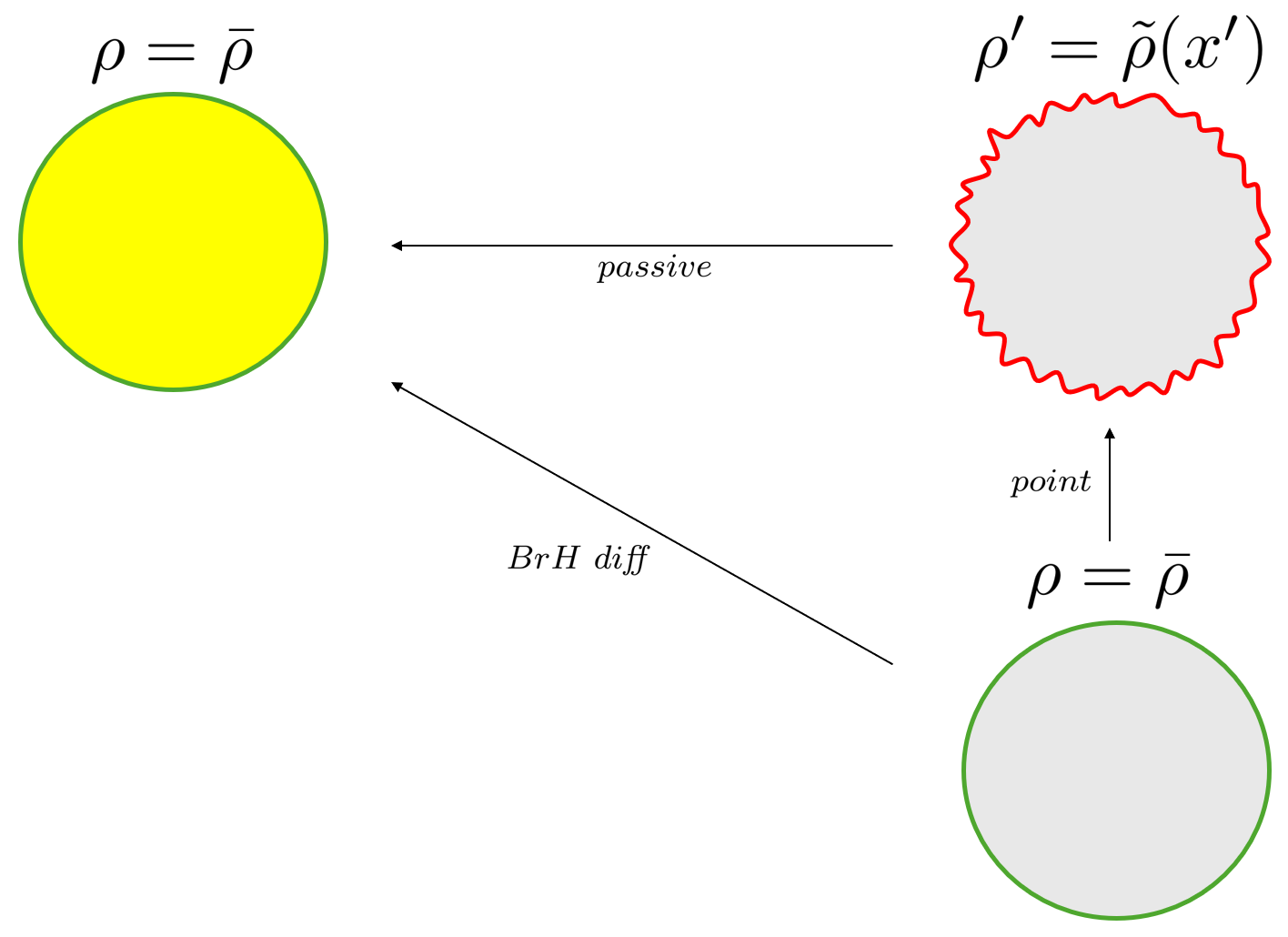}
	\caption{Pictorial `bulk triangle' representation of a Brown-Henneaux diffeomorphism taking us from a bulk metric $ G(X) dX^2$ in the lower right corner to a bulk metric $G'(X) dX^2$ in the upper left corner along the diagonal arrow, via a point transformation (vertical arrow) to $ G(X') dX'^2$ and passive coordinate transformation (left-pointing arrow). The grey color indicates the metric $ G$ and yellow the metric $G'$. Following the point transformation, a fixed boundary at the constant radial location $\rho=\bar{\rho}$ maps to a wiggly boundary at the position-dependent radius $\rho^{\prime}=\tilde{\rho}(x')$.
	} 
	\label{trianglecirclesfigure}
\end{figure}

We consider asymptotically AdS$_3$ geometries $\mathcal{M}$ with a timelike end of the world brane which constitutes (a portion or the entirety of) its outer boundary,
\begin{align}
	ds^2&=G_{MN}(X)dX^MdX^N=d\rho^2+G_{\mu\nu}(\rho,x)dx^{\mu}dx^{\nu}
	\\
	&=d\rho^2+e^{2\rho}\left(g^{(0)}_{\mu\nu}(x)+e^{-2\rho}g^{(2)}_{\mu\nu}(x)+e^{-4\rho}g^{(4)}_{\mu\nu}(x)+\mathcal{O}(e^{-6\rho})\right)dx^{\mu}dx^{\nu}. \label{ds2}
\end{align}
The asymptotic boundary is reached as $\rho \rightarrow \infty$. When the bulk equations of motion are imposed, the expansion truncates after $g^{(4)}_{\mu\nu}$. This geometry is related to the new geometry
\begin{align}
	(ds^{\prime})^2&=G_{MN}(X^{\prime})dX^{\prime M}dX^{\prime N}
	\\
	&=G^{\prime}_{MN}(X)dX^MdX^N=d\rho^2+G^{\prime}_{\mu\nu}(\rho,x)dx^{\mu}dx^{\nu}
	\\
	&=d\rho^2+e^{2\rho}\left(g^{\prime(0)}_{\mu\nu}(x)+e^{-2\rho}g^{\prime(2)}_{\mu\nu}(x)+e^{-4\rho}g^{\prime(4)}_{\mu\nu}(x)+\mathcal{O}(e^{-6\rho})\right)dx^{\mu}dx^{\nu}, \label{Gprimetransformed}
\end{align}
with
\begin{equation} \label{g0transformed}
	g_{\mu\nu}^{\prime(0)}(x)=e^{\phi(x)}	g_{\mu\nu}^{(0)}(x),
\end{equation}
by the Brown-Henneaux diffeomorphisms, which can be written in an asymptotic expansion as
\ali{
	\begin{split}
		\rho'(\rho,x) &= \rho + \frac{1}{2}\phi(x) + \sum_{j=1} e^{-2\rho j} a_\rho^{(2j)}(x), \\ 
		x'^\mu(\rho,x) &= x^\mu + \sum_{j=1} e^{-2\rho j} a_x^{(2j) \mu}(x).
	\end{split} \label{BrHexp}
}
All terms in the expansion can be found order by order by demanding preservation of the FG gauge, allowing us to determine the higher order $g_{\mu\nu}^{\prime(n>0)}$ in terms of $\phi$ and $g_{\mu\nu}^{(n)}$. The first subleading coefficients in the above expansion of the Brown-Henneaux diffeomorphisms are
\ali{
	a_\rho^{(2)} = \frac{1}{16} e^{-\phi}  g^{\mn}_{(0)} \p_\mu \phi \p_\nu \phi, \qquad a_x^{(2)\mu} = \frac{1}{4} e^{-\phi}  g^{\mn}_{(0)} \p_\nu \phi. 
}
Using the above Brown-Henneaux diffeomorphisms, we find that a constant radial location $\rho=\bar{\rho}$ maps to the fluctuating location 
\ali{
	&\rho' = \rho'\left(\bar \rho, x(\bar \rho, x')\right) = \tilde{\rho}(x^{\prime})= \bar{\rho}+\tilde{\phi}(x^{\prime}), \label{curlycutoff}   
}	
with the fluctuation $\tilde{\phi}$ related to the asymptotic Weyl mode $\phi$ as
\ali{
	\tilde{\phi}(x)=\frac{1}{2} \phi(x) - \frac{1}{16} e^{-2\bar \rho} e^{-\phi(x)} g_{(0)}^\mn(x) \p_\mu \phi(x) \p_\nu \phi(x) + \mathcal O(e^{-4\bar \rho}).  \label{phitildephi}
}
Clearly, the fluctuation and the asymptotic Weyl factor coincide asymptotically up to a factor of $2$
\begin{equation}
	\lim\limits_{\bar{\rho}\rightarrow \infty}\tilde{\phi}=\frac{\phi}{2}.
\end{equation} 
The transformation of the bulk geometry under Brown--Henneaux diffeomorphisms, decomposed into a point transformation followed by a passive coordinate transformation, together with the corresponding mapping of the boundary radial location, is illustrated in Figure~\ref{trianglecirclesfigure}. 
To holographically compute the Liouville theory 
\ali{ 
	S_L[g_{(0)},\phi]  &= W_{CFT}^{(T_0)}[e^{\phi} g_{(0)}] - W_{CFT}[g_{(0)}]  \label{W0-WCFT}
} 
in \eqref{CFTT0}, we need the subtraction of bulk on-shell actions 
\ali{ 
	S_L[g_{(0)},\phi]=\lim_{\bar \rho \ra \infty} \left( S_{tot}[G^{\prime}(\rho<\bar{\rho},x)] 
	- S_{grav}[G(\rho < \bar \rho,x)]  \right).  \label{S0phi0}
} 
It is the difference between the on-shell total action evaluated on the new geometry $G^{\prime}(X)$ and the gravitational action evaluated on the original geometry $G(X)$ in the limit in which the EOW brane at the same constant location $\rho=\bar{\rho}$ is pushed to the asymptotic boundary. We will consider bulk geometries with leading FG metric coefficient of constant positive ($\lambda=-1$), zero ($\lambda=0$) or negative ($\lambda=1$) curvature, i.e. $R_{g_{(0)}}=-8\lambda$, where we fixed the arbitrary constant for convenience. On-shell, the corresponding trace of extrinsic curvature on constant-$\bar{\rho}$ slices takes the asymptotic expansion $K=2+\frac{1}{2}R_{g_{(0)}}e^{-2\bar{\rho}}+\mathcal{O}(e^{-4\bar{\rho}})$. The trace of the Neumann condition $K=2T_0$ will thus lead to the following asymptotic expansion for the tension
\begin{equation}
	T_0=1-2\lambda e^{-2\bar{\rho}}+\mathcal{O}(e^{-4\bar{\rho}}). \label{tensionKarg}
\end{equation}
This ensures that the asymptotic theory is free of power law divergences. Having understood the equivalence between the fixed boundary at $\rho=\bar{\rho}$ in $G^{\prime}(\rho,x)$ and the curly boundary at $\rho^{\prime}=\tilde{\rho}(x^{\prime})$ in $G(\rho^{\prime},x^{\prime})$, we can use the invariance of the action under the passive change of coordinates
\ali{
	S_{tot}[G^{\prime}(\rho < \bar \rho,x)] = S_{tot}[G(\rho'< \tilde{\rho}(x^{\prime}),x^{\prime})]=S_{tot}[G(\rho<\tilde{\rho}(x),x)] \label{equalityS}, 
}
where in the last equality we renamed the dummy integration variables. Thus, to obtain the desired result we may equivalently compute
\ali{ 
	S_L[g_{(0)},\phi]=\lim_{\bar \rho \ra \infty} \left( S_{tot}[G(\rho< \tilde{\rho}(x),x)]
	- S_{grav}[G(\rho < \bar \rho,x)]  \right).  \label{hatFGmhatFG}
} 
The latter is in fact the more straightforward way, because it allows us to neglect the contribution from the inner boundaries: whereas the relation between the radial locations of the inner boundaries in $G^{\prime}(X)$ and $G(X)$ is dependent on the particular choice of solutions, we may assume the same contribution from the inner boundaries in $S_{tot}[G(\rho< \tilde{\rho}(x),x)] $ and
$S_{grav}[G(\rho < \bar \rho,x)] $, hence cancelling out in the difference. 
To proceed with the evaluation of $S_L$, we first calculate $S_{tot}[G(\rho<\tilde{\rho}(x),x)]$.
\begin{figure}[t]
	\centering
	\includegraphics[scale=0.45]{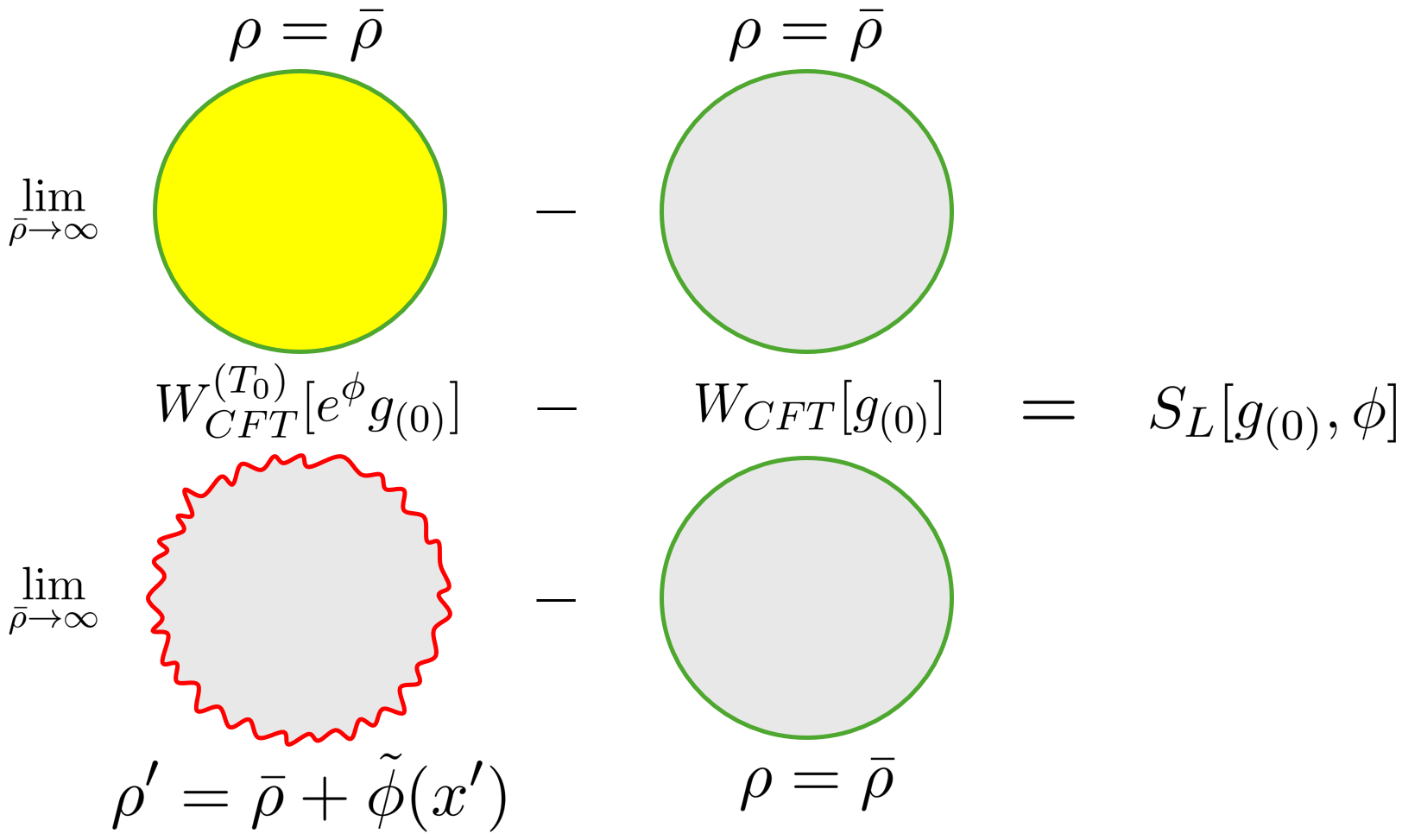}
	\caption{Depiction of the bulk geometries entering the holographic computation of the Liouville action \eqref{Sphi0}, which describes the integrated Weyl anomaly of the dual CFT$_2$ with general brane tension $T_0$, defined through Eq.~\eqref{W0-WCFT}. Computing the difference of on-shell actions along either the top or bottom rows (equivalently, along the diagonal or vertical arrows in Figure \ref{trianglecirclesfigure}) corresponds respectively to the calculations in Eqs. \eqref{S0phi0} and \eqref{hatFGmhatFG}. Their equivalence follows from the invariance \eqref{equalityS} of the action under the passive coordinate transformation represented by the horizontal arrow in Figure \ref{trianglecirclesfigure}.
	} 
	\label{holoLiouville3}
\end{figure}

The induced metric at the fluctuating EOW brane is
\begin{equation}
	g_{\mu\nu}(x)dx^{\mu}dx^{\nu}\big|_{\rho=\tilde{\rho}(x)}=ds^2\big|_{\rho=\tilde{\rho}(x)}=\big(G_{\mu\nu}(X)+\partial_{\mu} \tilde{\rho}(x)\partial_{\nu} \tilde{\rho}(x)\big)\big|_{\rho=\tilde{\rho}(x)}dx^{\mu}dx^{\nu}.
\end{equation}
Using that for $2 \times 2$ matrices $A,B$ we have the following properties of determinants $det(A+B)=(det A) ( 1+Tr(A^{-1}B))+det B$ and $det(\lambda A)=\lambda^2 det A$, we find
\begin{equation}
	\sqrt{|G|}=e^{2\rho}\sqrt{|g_{(0)}|}+\frac{1}{2}\sqrt{|g_{(0)}|} \, g_{(0)}^{\mu\nu}g^{(2)}_{\mu\nu}+\mathcal{O}(e^{-2\rho})
\end{equation}
and
\begin{align}
	\sqrt{|g|}\Big|_{\rho=\tilde{\rho}(x)}&=\left(\sqrt{| G|}\sqrt{1+G^{\mu\nu}\partial_{\mu} \tilde{\rho}\partial_{\nu} \tilde{\rho}}\right)\Big|_{\rho=\tilde{\rho}}
	\\
	&=\sqrt{|g_{(0)}|}\left(e^{2\tilde{\rho}}+\frac{1}{2}g_{(0)}^{\mu\nu}g^{(2)}_{\mu\nu}+\frac{1}{2}g^{\mu\nu}_{(0)}\partial_{\mu} \tilde{\rho}\partial_{\nu} \tilde{\rho}\right)+\mathcal{O}(e^{-2\tilde{\rho}}) . 
\end{align}
The components of the normal $n_{M}=\frac{\partial_{M}\left(\rho-\tilde{\rho}\right)}{\sqrt{|G^{AB}\partial_A (\rho-\tilde{\rho})\partial_B (\rho-\tilde{\rho})|}}$ are given by
\begin{equation}
	n_{\rho}=\frac{1}{\sqrt{1+G^{\alpha\beta}\partial_{\alpha} \tilde{\rho}\partial_{\beta} \tilde{\rho}}}, \;\;\;\; n_{\mu}=-\frac{\partial_{\mu} \tilde{\rho}}{\sqrt{1+G^{\alpha\beta}\partial_{\alpha} \tilde{\rho}\partial_{\beta} \tilde{\rho}}}.
\end{equation}
The trace of the extrinsic curvature on the boundary is thus
\begin{equation}
	K\big|_{\rho=\tilde{\rho}(x)}=\left(G^{MN}\nabla^{G}_{M}n_{N}\right)\Big|_{\rho=\tilde{\rho}}=2-e^{-2\tilde{\rho}}g_{(0)}^{\mu\nu}g^{(2)}_{\mu\nu}-e^{-2\tilde{\rho}}g^{\mu\nu}_{(0)}\nabla_{\mu}^{(0)}\partial_{\nu} \tilde{\rho}+\mathcal{O}(e^{-4\tilde{\rho}}),
\end{equation}
where we used $\Gamma^{\rho}_{\rho\rho}=\Gamma^{\mu}_{\rho\rho}=0$, $\Gamma^{\rho}_{\mu\nu}=-\frac{1}{2}\partial_{\rho}G_{\mu\nu}$ and $G^{\mu\nu}=e^{-2\rho}\left(g_{(0)}^{\mu\nu}-e^{-2\rho}g^{\mu\nu}_{(2)}+\mathcal{O}(e^{-4\rho})\right)$. Therefore, we have
\begin{equation}
	\sqrt{|g|}K\big|_{\rho=\tilde{\rho}(x)}=\sqrt{|g_{(0)}|}\left(2 e^{2\tilde{\rho}}+g^{\mu\nu}_{(0)}\partial_{\mu} \tilde{\rho}\partial_{\nu} \tilde{\rho}\right)-\partial_{\mu}\left(\sqrt{|g_{(0)}|}g^{\mu\nu}_{(0)}\partial_{\nu} \tilde{\rho}\right)+\mathcal{O}(e^{-2\tilde{\rho}}). \label{gammaKCarlip}
\end{equation}
Using that $R_G=-6$ when the Einstein equations are satisfied, we have that the bulk and boundary terms evaluate respectively to
\begin{align}
	\frac{1}{2 \kappa}\int_{\mathcal{M}} d^3X \sqrt{|G|}\left(R_G+2\right)&=-\frac{2}{\kappa}\int d^2 x \sqrt{|g_{(0)}|} \int\limits^{\rho=\tilde{\rho}}d\rho\left(e^{2\rho}+\frac{1}{2}g_{(0)}^{\mu\nu}g^{(2)}_{\mu\nu}+\mathcal{O}(e^{-2\rho})\right) 
	\\
	&=-\frac{1}{\kappa}\int d^2 x \sqrt{|g_{(0)}|}\left(e^{2\tilde{\rho}}+\tilde{\rho}g_{(0)}^{\mu\nu}g^{(2)}_{\mu\nu}+\mathcal{O}(e^{-2\tilde{\rho}})\right)+\cdots,
\end{align} 
and
\begin{align}
	\frac{1}{\kappa}\int_{EOW} d^2x \sqrt{|g|}\left(K-T_0\right)\big|_{\rho=\tilde{\rho}}=\frac{1}{\kappa}\int d^2x& \sqrt{|g_{(0)}|}\Bigg[(2-T_0)e^{2\tilde{\rho}}+\left(1-\frac{T_0}{2}\right)g_{(0)}^{\mu\nu}\partial_{\mu} \tilde{\rho}\partial_{\nu} \tilde{\rho} \nonumber
	\\
	&-\frac{T_0}{2}g_{(0)}^{\mu\nu}g^{(2)}_{\mu\nu}+\mathcal{O}(e^{-2\tilde{\rho}})\Bigg],
\end{align}
where we denoted with the dots the contributions from the lower bound of the radial integral in the bulk term and we assumed the vanishing of the total derivative term in \eqref{gammaKCarlip}. Therefore, we obtain
\ali{
	& S_{tot}[G(\rho < \tilde{\rho}(x), x)] \nonumber \\
	&\quad =  \frac{1}{2\kappa}\int d^2x \sqrt{|g_{(0)}|}\left((2-2T_0)e^{2\tilde{\rho}}+(2-T_0)g_{(0)}^{\mn}\partial_\mu \tilde{\rho}\partial_\nu \tilde{\rho} - g_{(0)}^{\mn} g^{(2)}_{\mn}\left(T_0+2 \tilde{\rho}\right)+\mathcal{O}(e^{-2\tilde{\rho}})\right)+\cdots . 
}   
\par We can now go ahead and evaluate \eqref{hatFGmhatFG}.  Requiring that the starting metric $G(X)$ is asymptotically bulk on-shell, and therefore obeys the relation
\ali{
	g_{(0)}^{\mn}  g^{(2)}_{\mn} = -\frac{1}{2}  R_{g_{(0)}} \label{hatFGonshell},  
} 
we indeed obtain the timelike Liouville action 
\begin{equation} \label{Sphi0}
	S_L[g_{(0)},\phi]=\frac{c}{48\pi}\int d^2x \sqrt{-g_{(0)}}\left(\phi R_{g_{(0)}}+\frac{1}{2}g_{(0)}^{\mu\nu}\partial_{\mu}\phi \partial_{\nu}\phi+8\lambda e^{\phi}\right).
\end{equation}
We used the Brown-Henneaux central charge $c=3/2G$. 
Figure \ref{holoLiouville3} provides a schematic illustration of the gravitational interpretation of the computation leading to Eq.~(\ref{Sphi0}).

\section{From empty AdS$_3$ to curved-sliced Banados} \label{app:curvedbanados}
We derive here a Banados solution with asymptotic $(A)dS_2$ slicing and the finite Brown--Henneaux diffeomorphisms relating it to Poincar\'e AdS$_3$. These generalize the known diffeomorphisms \cite{Roberts:2012aq} between Poincar\'e AdS$_3$ and the flat-sliced Banados geometry.

The integrated Weyl anomaly for general tension $T_0$ is expressed via the relation  $W_{CFT}^{(T_0)}[e^{\phi}g_{(0)}]-W_{CFT}[g_{(0)}]=S_L[g_{(0)},\phi]$, with the bulk-derived action (\ref{Sphi0}) 
\begin{equation} \label{Sphi0new}
	S_L[g_{(0)},\phi]=\frac{c}{48\pi}\int d^2x \sqrt{-g_{(0)}}\left(\phi R_{g_{(0)}}+\frac{1}{2}g_{(0)}^{\mu\nu}\partial_{\mu}\phi \partial_{\nu}\phi+8\lambda e^{\phi}\right),
\end{equation}
with $\lambda=+1,0,-1$ for $T_0<1$, $T_0=1$ and $T_0>1$ respectively. Consider fixing our starting geometry (\ref{ds2}) in the bulk derivation of the Liouville action to a very simple solution ($R_G=-6$) such as Poincaré AdS$_3$
\begin{equation} \label{ds2poinc}
	ds^2=d\rho^2+e^{2\rho}(-2dfd\bar{f}), \qquad \left(g^{(0)}_{\mu\nu}dx^{\mu}dx^{\nu}=-2dfd\bar{f}, g_{(2)}=g_{(4)}=0\right).
\end{equation}
Hence, the equation of motion of the action (\ref{Sphi0new}) with respect to $\delta\phi$ variations reads
\begin{equation} \label{eom0}
	\Box^{(0)}\phi=-2\partial_f \partial_{\bar{f}}\phi=8\lambda e^{\phi},
\end{equation}
with $\Box^{(0)}=g^{\mu\nu}_{(0)}\nabla^{(0)}_{\mu}\nabla^{(0)}_{\nu}$ and $\nabla^{(0)}_{\mu}$ denoting covariant differentiation w.r.t. $g_{(0)}$. A solution to the equation of motion (\ref{eom0}) is provided by
\begin{align} \label{phisol0}
	\phi=\begin{cases}
		&-\log\left(\partial f\bar{\partial}\bar{f}\right), \qquad \qquad \qquad \qquad \lambda=0,
		\\
		&-\log\left((-2)(z+\lambda \bar{z})^{2}\partial f\bar{\partial}\bar{f}\right), \qquad \lambda=\pm1,
	\end{cases}
\end{align}
in terms of the holomorphic and anti-holomorphic functions and the derivatives
\begin{equation}
	f(z), \qquad \bar{f}(\bar{z}), \qquad \partial=\frac{d}{dz}, \qquad \bar{\partial}=\frac{d}{d\bar{z}}.
\end{equation}
We will take
\begin{equation} \label{zzbarZt}
	z=\frac{Z+t}{\sqrt{2}i}, \qquad \bar{z}=\frac{Z-t}{\sqrt{2}i},
\end{equation}
with $Z$ and $t$ a spacelike and timelike coordinate respectively. In terms of a general $\phi$, the transformed solution (\ref{Gprimetransformed}) with respect to the initial solution (\ref{ds2poinc}) reads \cite{Skenderis:2000in}
\begin{align}
	(ds^{\prime})^2=&d\rho^2+e^{2\rho}\Bigg\{ e^{\phi}g_{\mu\nu}^{(0)}+e^{-2\rho}\left(-\frac{1}{4}\partial_{\mu}\phi\partial_{\nu}\phi+\frac{1}{2}\nabla^{(0)}_{\mu}\partial_{\nu}\phi+\frac{1}{8}g_{\mu\nu}^{(0)}g_{(0)}^{\alpha\beta}\partial_{\alpha}\phi\partial_{\beta}\phi\right) \nonumber
	\\
	&+e^{-4\rho}e^{-\phi}\Bigg[ \frac{1}{32}\nabla^{(0)}_{\mu}\partial_{\nu}\phi g^{\alpha\beta}_{(0)}\partial_{\alpha}\phi \partial_{\beta}\phi+\frac{1}{256}g_{\mu\nu}^{(0)}(g^{\alpha\beta}_{(0)}\partial_{\alpha}\phi \partial_{\beta}\phi)^2 +\frac{1}{16}g_{(0)}^{\alpha\beta}\nabla^{(0)}_{\mu}\partial_{\alpha}\phi \nabla_{\nu}^{(0)}\partial_{\beta}\phi \nonumber
	\\
	&-\frac{1}{64}\left(\nabla_{\mu}^{(0)}(g^{\alpha\beta}_{(0)}\partial_{\alpha}\phi \partial_{\beta}\phi)\partial_{\nu}\phi+\nabla_{\nu}^{(0)}(g^{\alpha\beta}_{(0)}\partial_{\alpha}\phi \partial_{\beta}\phi)\partial_{\mu}\phi\right)\Bigg]\Bigg\}dx^{\mu}dx^{\nu}. \label{curvedbanadosphi}
\end{align}
The $\phi$ solution (\ref{phisol0}) for the asymptotic Weyl factor is such that the leading term in the FG expansion of $(ds^{\prime})^2$ is respectively $AdS_2$, $Mink_2$ or $dS_2$ for $\lambda=+1,0,-1$, i.e.
\begin{equation}
	g^{\prime (0)}_{\mu\nu}dx^{\mu}dx^{\nu}=e^{\phi}g^{(0)}_{\mu\nu}dx^{\mu}dx^{\nu}=
	\begin{cases}
		&\frac{-dt^2+dZ^2}{4Z^2} \qquad \;\;\;\;  (AdS_2) \qquad \;\; \, \lambda=+1,
		\\
		&-dt^2+dZ^2 \qquad (Mink_2) \qquad \lambda=0,
		\\
		&\frac{-dt^2+dZ^2}{4t^2} \qquad \;\;\;\; (dS_2) \qquad \;\;\;\; \; \lambda=-1.
	\end{cases}
\end{equation}
Substituting the precise form  (\ref{phisol0}) of the $\phi$ solution into Eq.~(\ref{curvedbanadosphi}) we obtain the standard Banados solution for $\lambda=0$
\begin{equation}
	\lambda=0: \qquad  (ds^{\prime})^2=d\rho^2-2e^{2\rho}dzd\bar{z}-L(z)dz^2-\bar{L}(\bar{z})d\bar{z}^2-\frac{1}{2}e^{-2\rho}L(z)\bar{L}(\bar{z})dzd\bar{z}, \label{banadoslambda0}
\end{equation}
and a curved-sliced version of it for $\lambda=\pm1$
\begin{align}
	\lambda=\pm1: \qquad &(ds^{\prime})^2=d\rho^2+e^{2\rho}\Bigg\{\frac{dzd\bar{z}}{(z+\lambda \bar{z})^{2}}+e^{-2\rho}\left(\frac{2\lambda}{(z+\lambda \bar{z})^2}dzd\bar{z}-L(z)dz^2-\bar{L}(\bar{z})d\bar{z}^2\right) \nonumber
	\\
	&+e^{-4\rho}\Bigg[\frac{1+L(z)\bar{L}(\bar{z})(z+\lambda \bar{z})^{4}}{(z+\lambda \bar{z})^{2}}dzd\bar{z}-\lambda L(z)dz^2-\lambda \bar{L}(\bar{z})d\bar{z}^2\Bigg]\Bigg\}. \label{curvedBanados+-1}
\end{align}
We defined the functions
\begin{equation}
	L(z)= \frac{1}{2}\{f,z\}, \qquad    \qquad \bar{L}(\bar{z})=\frac{1}{2}\{\bar{f},\bar{z}\}, \label{Lschwarzian}
\end{equation}
in terms of the Schwarzian derivatives $\{f,z\}=\frac{\partial^3 f}{\partial f}-\frac{3}{2}\frac{(\partial^2 f)^2}{(\partial f)^2}$ and $\{\bar{f},\bar{z}\}=\frac{\bar{\partial}^3 \bar{f}}{\bar{\partial} \bar{f}}-\frac{3}{2}\frac{(\bar{\partial}^2 \bar{f})^2}{(\bar{\partial} \bar{f})^2}$. It is straightforward to check that the above geometries are indeed solutions.
For vanishing $L(z)=\bar{L}(\bar{z})=0$, the curved-sliced Banados (\ref{curvedBanados+-1}) reduces to curved-sliced empty AdS$_3$
\begin{align} 
	(ds^{\prime})^2&=d\rho^2+\frac{e^{2\rho}}{4}\left(\gamma_{\mu\nu}(x)+ e^{-2\rho}2\lambda\gamma_{\mu\nu}(x)+e^{-4\rho}\gamma_{\mu\nu}(x)\right)dx^{\mu}dx^{\nu}
	\\
	&=d\rho^2+e^{2 A(\rho)}\gamma_{\mu\nu}(x)dx^{\mu}dx^{\nu}, \label{dsprimecurvedempty}
\end{align}
with
\begin{align} 
	&AdS_2 \; (\lambda=1): \; \; \; \; \;\; \; \; \; \; e^{2A(\rho)}=(\cosh\rho)^2, \; \; \; \; \;\; \; \; \; \; \gamma_{\mu\nu}(x)dx^{\mu}dx^{\nu}=\frac{-dt^2+dZ^2}{Z^2}, \nonumber
	\\
	&dS_2 \; (\lambda=-1): \; \; \; \; \;\; \; \; \; \;  e^{2A(\rho)}= (\sinh\rho)^2, \; \; \; \; \;\; \; \; \; \;  \, \gamma_{\mu\nu}(x)dx^{\mu}dx^{\nu}=\frac{-dt^2+dZ^2}{t^2} .
\end{align}
For constant $L(z)=L$ and $\bar{L}(\bar{z})=\bar{L}$, the curved-sliced Banados (\ref{curvedBanados+-1}) presumably reduces to a curved-sliced version of the BTZ black hole.

\par The finite version of the infinitesimal Brown-Henneaux diffs \eqref{BrHexp} taking us from the initial Poincaré AdS$_3$ solution \eqref{ds2poinc} to the standard or curved-sliced Banados solutions  \eqref{banadoslambda0}-\eqref{curvedBanados+-1} can be expressed in terms of the asymptotic Weyl factor \eqref{phisol0} as
\ali{
	\begin{split}
		\rho'(\rho,x) &= \rho + \frac{1}{2}\phi(x) + \log\left( 1 + \frac{1}{16} e^{-2\rho} e^{-\phi(x)} g_{(0)}^\mn \p_\mu \phi \p_\nu \phi \right),  
		\\ 
		x'^\mu(\rho,x) &= x^\mu + \frac{e^{-2\rho} e^{-\phi(x)} g_{(0)}^\mn \p_\nu \phi}{4+\frac{1}{4}e^{-2\rho} e^{-\phi(x)} g_{(0)}^\mn \p_\mu \phi \p_\nu \phi},
	\end{split} , \label{BrHRobertsnew}	
}
or in terms of the old coordinates $(\rho,x):(\rho,f,\bar{f})$ and the new coordinates $(\rho^{\prime},x^{\prime}):(\rho^{\prime},y,\bar{y})$ as 
\begin{align}
	\rho^{\prime}
	&=
	\rho-\frac{1}{2}\log\left(\partial f\bar{\partial}\bar{f}\right)
	+\log\left(
	1-\frac{e^{-2\rho}}{8}
	\frac{\partial^2 f\,\bar{\partial}^2\bar{f}}
	{\partial f\,\bar{\partial}\bar{f}}
	\right), \nonumber
	\\
	y
	&=
	f+\frac{e^{-2\rho}}{4}
	\frac{
		\partial f\,\bar{\partial}\bar{f}
		\frac{\bar{\partial}^2\bar{f}}
		{(\bar{\partial}\bar{f})^2}
	}{
		1-\frac{e^{-2\rho}}{8}
		\frac{\partial^2 f\,\bar{\partial}^2\bar{f}}
		{\partial f\,\bar{\partial}\bar{f}}
	}, \label{Robertsdiffflat}
	\\
	\bar{y}
	&=
	\bar{f}+\frac{e^{-2\rho}}{4}
	\frac{
		\partial f\,\bar{\partial}\bar{f}
		\frac{\partial^2 f}
		{(\partial f)^2}
	}{
		1-\frac{e^{-2\rho}}{8}
		\frac{\partial^2 f\,\bar{\partial}^2\bar{f}}
		{\partial f\,\bar{\partial}\bar{f}} \nonumber
	},
\end{align}
for $\lambda=0$ (reducing to the diffeomorphisms of Roberts \cite{Roberts:2012aq} sending Poincaré AdS$_3$ to the standard Banados geometry), and
\begin{align} 
	\rho^{\prime}
	&=
	\rho-\frac{1}{2}\log\left(
	-2(z+\lambda\bar{z})^2
	\partial f\bar{\partial}\bar{f}
	\right)
	\nonumber\\
	&\qquad
	+\log\left[
	1+\frac{e^{-2\rho}}{4}
	\frac{
		\left(2\partial f+(z+\lambda\bar{z})\partial^2 f\right)
		\left(2\lambda\bar{\partial}\bar{f}
		+(z+\lambda\bar{z})\bar{\partial}^2\bar{f}\right)
	}{
		\partial f\,\bar{\partial}\bar{f}
	}
	\right], \nonumber
	\\
	y
	&=
	f-\frac{e^{-2\rho}}{2}
	\frac{
		\partial f\,\bar{\partial}\bar{f}
		(z+\lambda\bar{z})^2
		\left(
		\frac{\bar{\partial}^2\bar{f}}
		{(\bar{\partial}\bar{f})^2}
		+\frac{2\lambda}
		{(z+\lambda\bar{z})\bar{\partial}\bar{f}}
		\right)
	}{
		1+\frac{e^{-2\rho}}{4}
		\frac{
			\left(2\partial f+(z+\lambda\bar{z})\partial^2 f\right)
			\left(2\lambda\bar{\partial}\bar{f}
			+(z+\lambda\bar{z})\bar{\partial}^2\bar{f}\right)
		}{
			\partial f\,\bar{\partial}\bar{f}
		} 
	}, 	\label{Robertsdiffcurved}
	\\
	\bar{y}
	&=
	\bar{f}-\frac{e^{-2\rho}}{2}
	\frac{
		\partial f\,\bar{\partial}\bar{f}
		(z+\lambda\bar{z})^2
		\left(
		\frac{\partial^2 f}{(\partial f)^2}
		+\frac{2}
		{(z+\lambda\bar{z})\partial f}
		\right)
	}{
		1+\frac{e^{-2\rho}}{4}
		\frac{
			\left(2\partial f+(z+\lambda\bar{z})\partial^2 f\right)
			\left(2\lambda\bar{\partial}\bar{f}
			+(z+\lambda\bar{z})\bar{\partial}^2\bar{f}\right)
		}{
			\partial f\,\bar{\partial}\bar{f}
		} \nonumber
	}, 
\end{align}
for $\lambda=\pm1$. More generally, we can check that $(ds^{\prime})^2=d\rho^{\prime 2}+e^{2\rho^{\prime}}(-2dyd\bar{y})$ together with the transformations (\ref{Robertsdiffflat})-(\ref{Robertsdiffcurved}) yield the standard or curved-sliced Banados solutions \eqref{banadoslambda0}-\eqref{curvedBanados+-1}. As a simple example, we can consider the $\lambda=1$ case with the choice of functions $f(z)=z$ and $\bar{f}(\bar{z})=\bar{z}$, for which $L(z)=\bar{L}(\bar{z})=0$, such that the finite Brown-Henneaux diffeomorphisms (\ref{Robertsdiffcurved}) taking us from the initial Poincaré AdS$_3$ geometry (\ref{ds2poinc}), i.e. $ds^2=d\rho^2+e^{2\rho}(-2dfd\bar{f})$, to the $AdS_2$-sliced empty AdS$_3$ (\ref{dsprimecurvedempty}), i.e. $(ds^{\prime})^2=d\rho^{\prime 2}+e^{2\rho^{\prime}}(-2dyd\bar{y})=d\rho^{\prime 2}+e^{2\rho^{\prime}}(-d\tilde{t}^2+d\tilde{y}^2)=\frac{-d\tilde{t}^2+d\tilde{y}^2+d\tilde{z}^2}{\tilde{z}^2}=d\rho^2+(\cosh\rho)^2\frac{-dt^2+dZ^2}{Z^2}$ take the form \cite{Suzuki:2022xwv}
\begin{align}
	\tilde{t}=t, \qquad \qquad \tilde{z}=Z/\cosh\rho, \qquad \qquad \tilde{y}=Z\tanh\rho,
\end{align}
where we used $y=\frac{\tilde{y}+\tilde{t}}{\sqrt{2}i}$, $\bar{y}=\frac{\tilde{y}-\tilde{t}}{\sqrt{2}i}$ and $e^{-\rho^{\prime}}=\tilde{z}$, together with Eq.~(\ref{zzbarZt}).

\par We can be even more general and take as starting point empty AdS$_3$ in constant-positive-curvature ($\lambda_0=-1$) or constant-negative-curvature slicing ($\lambda_0=+1$), i.e.
\begin{align} \label{ds2emptycurved}
	ds^2=&d\rho^2+e^{2\rho}\Bigg(\frac{1}{(f+\lambda_0 \bar{f})^2}+e^{-2\rho}\frac{2\lambda_0}{(f+\lambda_0 \bar{f})^2} +e^{-4\rho}\frac{1}{(f+\lambda_0 \bar{f})^{2}}\Bigg)dfd\bar{f}, 
\end{align}
or in other words
\begin{align}
	&g^{(0)}_{\mu\nu}dx^{\mu}dx^{\nu}=\frac{dfd\bar{f}}{(f+\lambda_0 \bar{f})^{2}},
	\\
	& 	g^{(2)}_{\mu\nu}dx^{\mu}dx^{\nu}=\frac{2\lambda_0}{(f+\lambda_0 \bar{f})^2}dfd\bar{f}
	\\
	&	g^{(4)}_{\mu\nu}dx^{\mu}dx^{\nu}=\frac{dfd\bar{f}}{(f+\lambda_0 \bar{f})^{2}},
\end{align}
such that
\begin{equation}
	R_{g_{(0)}}=-8\lambda_0.
\end{equation}
Then, the equation of motion of the action (\ref{Sphi0new}) with respect to $\delta\phi$ variations reads
\begin{equation} \label{eom0new}
	\Box^{(0)}\phi=-2(-2)^{\lambda_0^2}(f+\lambda_0 \bar{f})^{2\lambda_0^2}\partial_f \partial_{\bar{f}}\phi=8\lambda e^{\phi}-8\lambda_0,
\end{equation}
with again $\Box^{(0)}=g^{\mu\nu}_{(0)}\nabla^{(0)}_{\mu}\nabla^{(0)}_{\nu}$ and $\nabla^{(0)}_{\mu}$ denoting covariant differentiation w.r.t. $g_{(0)}$. A solution to the equation of motion (\ref{eom0new}) is provided by
\begin{align} \label{phisol0new}
	\phi=\begin{cases}
		&-\log\left(\frac{\partial f\bar{\partial}\bar{f}}{-2(f+\lambda_0 \bar{f})^{2}}\right), \qquad \qquad \qquad \lambda=0,
		\\
		&-\log\left(\frac{(z+\lambda \bar{z})^{2}}{(f+\lambda_0 \bar{f})^{2}}\partial f\bar{\partial}\bar{f}\right), \qquad \qquad \;\; \lambda=\pm1,
	\end{cases}
\end{align}
In terms of a general $\phi$, the transformed solution (\ref{Gprimetransformed}) with respect to the initial solution (\ref{ds2emptycurved}) reads\footnote{Note the change of sign on the term proportional to the Riemann tensor compared to \cite{Skenderis:2000in}, due to our use of the opposite convention.} \cite{Skenderis:2000in}
\begin{align}
	(ds^{\prime})^2=&d\rho^2+e^{2\rho}\Bigg\{ e^{\phi}g_{\mu\nu}^{(0)}+e^{-2\rho}\left(g^{(2)}_{\mu\nu}-\frac{1}{4}\partial_{\mu}\phi\partial_{\nu}\phi+\frac{1}{2}\nabla^{(0)}_{\mu}\partial_{\nu}\phi+\frac{1}{8}g_{\mu\nu}^{(0)}g_{(0)}^{\alpha\beta}\partial_{\alpha}\phi\partial_{\beta}\phi\right) \nonumber
	\\
	&+e^{-4\rho}e^{-\phi}\Bigg[g^{(4)}_{\mu\nu}-\frac{1}{8}\nabla^{\alpha}_{(0)}\phi\left(\nabla^{(0)}_{\mu}g_{\nu\alpha}^{(2)}+\nabla^{(0)}_{\nu}g^{(2)}_{\mu\alpha}-2\nabla^{(0)}_{\alpha}g^{(2)}_{\mu\nu}\right)+\frac{1}{8}\left(g_{\nu\alpha}^{(2)}\nabla^{(0)}_{\mu}\nabla^{\alpha}_{(0)}\phi +g_{\mu\alpha}^{(2)}\nabla^{(0)}_{\nu}\nabla^{\alpha}_{(0)}\phi \right) \nonumber
	\\
	&-\frac{1}{16}R^{(0)}_{\alpha \mu \beta \nu}\nabla^{\alpha}_{(0)}\phi \nabla^{\beta}_{(0)}\phi+\frac{1}{256}g_{\mu\nu}^{(0)}\left((g^{\alpha\beta}_{(0)}\partial_{\alpha}\phi \partial_{\beta}\phi)^2-16 g^{(2)}_{\alpha\beta}\nabla^{\alpha}_{(0)}\phi \nabla^{\beta}_{(0)}\phi \right)\nonumber
	\\
	&+ \frac{1}{32}\nabla^{(0)}_{\mu}\partial_{\nu}\phi g^{\alpha\beta}_{(0)}\partial_{\alpha}\phi \partial_{\beta}\phi +\frac{1}{16}g_{(0)}^{\alpha\beta}\nabla^{(0)}_{\mu}\partial_{\alpha}\phi \nabla_{\nu}^{(0)}\partial_{\beta}\phi \nonumber
	\\
	&-\frac{1}{64}\left(\nabla_{\mu}^{(0)}(g^{\alpha\beta}_{(0)}\partial_{\alpha}\phi \partial_{\beta}\phi)\partial_{\nu}\phi+\nabla_{\nu}^{(0)}(g^{\alpha\beta}_{(0)}\partial_{\alpha}\phi \partial_{\beta}\phi)\partial_{\mu}\phi\right)\Bigg]\Bigg\}dx^{\mu}dx^{\nu}. \label{curvedbanadosphinew}
\end{align}
The $\phi$ solution (\ref{phisol0new}) for the asymptotic Weyl factor is such that the leading term in the FG expansion of $(ds^{\prime})^2$ is respectively $AdS_2$, $Mink_2$ or $dS_2$ for $\lambda=+1,0,-1$, i.e.
\begin{equation}
	g^{\prime (0)}_{\mu\nu}dx^{\mu}dx^{\nu}=e^{\phi}g^{(0)}_{\mu\nu}dx^{\mu}dx^{\nu}=
	\begin{cases}
		&\frac{-dt^2+dZ^2}{4Z^2} \qquad \;\;\;\;  (AdS_2) \qquad \;\; \, \lambda=+1,
		\\
		&-dt^2+dZ^2 \qquad (Mink_2) \qquad \lambda=0,
		\\
		&\frac{-dt^2+dZ^2}{4t^2} \qquad \;\;\;\; (dS_2) \qquad \;\;\;\; \; \lambda=-1.
	\end{cases}
\end{equation}
Substituting the precise form  (\ref{phisol0new}) of the $\phi$ solution into Eq.~(\ref{curvedbanadosphinew}) we recover the standard or curved-sliced Banados solutions  \eqref{banadoslambda0}-\eqref{curvedBanados+-1}.

\par In summary, we showed how to go from an initial empty AdS$_3$ geometry in general constant-curvature slicing ($\lambda_0=0,\pm1$) to the Banados solutions \eqref{banadoslambda0}-\eqref{curvedBanados+-1} in curved slicing ($\lambda=0,\pm1$) through the Brown-Henneaux diffeomorphisms in terms of an asymptotic Weyl factor $\phi$ which is a solution to the $\delta \phi$ equations of motion of the Liouville action (\ref{Sphi0new}). Note that we may take the slicing of the initial empty AdS$_3$ geometry to be arbitrary ($\lambda_0=-1,0,+1$ is a free-to-choose parameter corresponding to the choice of reference background geometry for the CFT, i.e. $g_{(0)}$, being respectively $dS_2$, $Mink_2$ or $AdS_2$), while the choice of empty AdS$_3$ fixes the initial CFT state to be the
vacuum. Then, whether the background geometry of the CFT dual to the transformed bulk geometry, i.e. $g^{\prime}_{(0)}$, will be that of $dS_2$, $Mink_2$ or $AdS_2$ (hence whether asymptotic constant-radius slices will have positive, flat or negative curvature in the transformed bulk geometry) depends entirely on the parameter $\lambda$ that appears in the potential term of the Liouville action (\ref{Sphi0new}), independently of $\lambda_0$ choice. The state of the CFT dual to the transformed geometry will instead depend on the arbitrary choice of the $f(z)$ and $\bar{f}(\bar{z})$ functions.

\bibliographystyle{JHEP}
\bibliography{referencesTTbarBWDraft}

\providecommand{\href}[2]{#2}\begingroup\raggedright\begin{thebibliography}{100}

\bibitem{Almheiri:2019hni}
A.~Almheiri, R.~Mahajan, J.~Maldacena and Y.~Zhao, \emph{{The Page curve of
  Hawking radiation from semiclassical geometry}},
  \href{https://doi.org/10.1007/JHEP03(2020)149}{\emph{JHEP} {\bfseries 03}
  (2020) 149} [\href{https://arxiv.org/abs/1908.10996}{{\ttfamily
  1908.10996}}].

\bibitem{Penington:2019npb}
G.~Penington, \emph{{Entanglement Wedge Reconstruction and the Information
  Paradox}}, \href{https://doi.org/10.1007/JHEP09(2020)002}{\emph{JHEP}
  {\bfseries 09} (2020) 002}
  [\href{https://arxiv.org/abs/1905.08255}{{\ttfamily 1905.08255}}].

\bibitem{Almheiri:2019qdq}
A.~Almheiri, T.~Hartman, J.~Maldacena, E.~Shaghoulian and A.~Tajdini,
  \emph{{Replica Wormholes and the Entropy of Hawking Radiation}},
  \href{https://doi.org/10.1007/JHEP05(2020)013}{\emph{JHEP} {\bfseries 05}
  (2020) 013} [\href{https://arxiv.org/abs/1911.12333}{{\ttfamily
  1911.12333}}].

\bibitem{Penington:2019kki}
G.~Penington, S.H.~Shenker, D.~Stanford and Z.~Yang, \emph{{Replica wormholes
  and the black hole interior}},
  \href{https://doi.org/10.1007/JHEP03(2022)205}{\emph{JHEP} {\bfseries 03}
  (2022) 205} [\href{https://arxiv.org/abs/1911.11977}{{\ttfamily
  1911.11977}}].

\bibitem{Chen:2020uac}
H.Z.~Chen, R.C.~Myers, D.~Neuenfeld, I.A.~Reyes and J.~Sandor, \emph{{Quantum
  Extremal Islands Made Easy, Part I: Entanglement on the Brane}},
  \href{https://doi.org/10.1007/JHEP10(2020)166}{\emph{JHEP} {\bfseries 10}
  (2020) 166} [\href{https://arxiv.org/abs/2006.04851}{{\ttfamily
  2006.04851}}].

\bibitem{Chen:2020hmv}
H.Z.~Chen, R.C.~Myers, D.~Neuenfeld, I.A.~Reyes and J.~Sandor, \emph{{Quantum
  Extremal Islands Made Easy, Part II: Black Holes on the Brane}},
  \href{https://doi.org/10.1007/JHEP12(2020)025}{\emph{JHEP} {\bfseries 12}
  (2020) 025} [\href{https://arxiv.org/abs/2010.00018}{{\ttfamily
  2010.00018}}].

\bibitem{Geng:2022slq}
H.~Geng, A.~Karch, C.~Perez-Pardavila, S.~Raju, L.~Randall, M.~Riojas et~al.,
  \emph{{Jackiw-Teitelboim Gravity from the Karch-Randall Braneworld}},
  \href{https://doi.org/10.1103/PhysRevLett.129.231601}{\emph{Phys. Rev. Lett.}
  {\bfseries 129} (2022) 231601}
  [\href{https://arxiv.org/abs/2206.04695}{{\ttfamily 2206.04695}}].

\bibitem{Geng:2022tfc}
H.~Geng, \emph{{Aspects of AdS$_{2}$ quantum gravity and the Karch-Randall
  braneworld}}, \href{https://doi.org/10.1007/JHEP09(2022)024}{\emph{JHEP}
  {\bfseries 09} (2022) 024}
  [\href{https://arxiv.org/abs/2206.11277}{{\ttfamily 2206.11277}}].

\bibitem{Neuenfeld:2024gta}
D.~Neuenfeld, A.~Svesko and W.~Sybesma, \emph{{Liouville gravity at the end of
  the world: deformed defects in AdS/BCFT}},
  \href{https://doi.org/10.1007/JHEP07(2024)215}{\emph{JHEP} {\bfseries 07}
  (2024) 215} [\href{https://arxiv.org/abs/2404.07260}{{\ttfamily
  2404.07260}}].

\bibitem{Almheiri:2019yqk}
A.~Almheiri, R.~Mahajan and J.~Maldacena, \emph{{Islands outside the horizon}},
   \href{https://arxiv.org/abs/1910.11077}{{\ttfamily 1910.11077}}.

\bibitem{Almheiri:2019psf}
A.~Almheiri, N.~Engelhardt, D.~Marolf and H.~Maxfield, \emph{{The entropy of
  bulk quantum fields and the entanglement wedge of an evaporating black
  hole}}, \href{https://doi.org/10.1007/JHEP12(2019)063}{\emph{JHEP} {\bfseries
  12} (2019) 063} [\href{https://arxiv.org/abs/1905.08762}{{\ttfamily
  1905.08762}}].

\bibitem{Suzuki:2022xwv}
K.~Suzuki and T.~Takayanagi, \emph{{BCFT and Islands in two dimensions}},
  \href{https://doi.org/10.1007/JHEP06(2022)095}{\emph{JHEP} {\bfseries 06}
  (2022) 095} [\href{https://arxiv.org/abs/2202.08462}{{\ttfamily
  2202.08462}}].

\bibitem{Wang:2025bcx}
D.~Wang, Z.~Wang and Z.~Wei, \emph{{Wormholes with ends of the world}},
  \href{https://doi.org/10.1007/JHEP09(2025)166}{\emph{JHEP} {\bfseries 09}
  (2025) 166} [\href{https://arxiv.org/abs/2504.12278}{{\ttfamily
  2504.12278}}].

\bibitem{Callebaut:2025thw}
N.~Callebaut and M.~Selle, \emph{{Setting $T^2$ free for braneworld
  holography}}, {\emph{JHEP} (2025) }
  [\href{https://arxiv.org/abs/2510.01099}{{\ttfamily 2510.01099}}].

\bibitem{Gubser:1999vj}
S.S.~Gubser, \emph{{AdS / CFT and gravity}},
  \href{https://doi.org/10.1103/PhysRevD.63.084017}{\emph{Phys. Rev. D}
  {\bfseries 63} (2001) 084017}
  [\href{https://arxiv.org/abs/hep-th/9912001}{{\ttfamily hep-th/9912001}}].

\bibitem{Geng:2026asi}
H.~Geng, A.~Karch, C.~Perez-Pardavila, S.~Raju, L.~Randall and M.~Riojas,
  \emph{{Seeing Page Curves and Islands with Blinders On}},
  \href{https://arxiv.org/abs/2602.06543}{{\ttfamily 2602.06543}}.

\bibitem{DeVuyst:2022bua}
J.~De~Vuyst and T.G.~Mertens, \emph{{Operational islands and black hole
  dissipation in JT gravity}},
  \href{https://doi.org/10.1007/JHEP01(2023)027}{\emph{JHEP} {\bfseries 01}
  (2023) 027} [\href{https://arxiv.org/abs/2207.03351}{{\ttfamily
  2207.03351}}].

\bibitem{Geng:2020qvw}
H.~Geng and A.~Karch, \emph{{Massive islands}},
  \href{https://doi.org/10.1007/JHEP09(2020)121}{\emph{JHEP} {\bfseries 09}
  (2020) 121} [\href{https://arxiv.org/abs/2006.02438}{{\ttfamily
  2006.02438}}].

\bibitem{Randall:1999ee}
L.~Randall and R.~Sundrum, \emph{{A Large mass hierarchy from a small extra
  dimension}}, \href{https://doi.org/10.1103/PhysRevLett.83.3370}{\emph{Phys.
  Rev. Lett.} {\bfseries 83} (1999) 3370}
  [\href{https://arxiv.org/abs/hep-ph/9905221}{{\ttfamily hep-ph/9905221}}].

\bibitem{Randall:1999vf}
L.~Randall and R.~Sundrum, \emph{{An Alternative to compactification}},
  \href{https://doi.org/10.1103/PhysRevLett.83.4690}{\emph{Phys. Rev. Lett.}
  {\bfseries 83} (1999) 4690}
  [\href{https://arxiv.org/abs/hep-th/9906064}{{\ttfamily hep-th/9906064}}].

\bibitem{Karch:2000ct}
A.~Karch and L.~Randall, \emph{{Locally localized gravity}},
  \href{https://doi.org/10.1088/1126-6708/2001/05/008}{\emph{JHEP} {\bfseries
  05} (2001) 008} [\href{https://arxiv.org/abs/hep-th/0011156}{{\ttfamily
  hep-th/0011156}}].

\bibitem{Antonini:2025sur}
S.~Antonini, C.-H.~Chen, H.~Maxfield and G.~Penington, \emph{{An apologia for
  islands}}, \href{https://doi.org/10.1007/JHEP10(2025)034}{\emph{JHEP}
  {\bfseries 10} (2025) 034}
  [\href{https://arxiv.org/abs/2506.04311}{{\ttfamily 2506.04311}}].

\bibitem{McGough:2016lol}
L.~McGough, M.~Mezei and H.~Verlinde, \emph{{Moving the CFT into the bulk with
  $ T\overline{T} $}},
  \href{https://doi.org/10.1007/JHEP04(2018)010}{\emph{JHEP} {\bfseries 04}
  (2018) 010} [\href{https://arxiv.org/abs/1611.03470}{{\ttfamily
  1611.03470}}].

\bibitem{Compere:2008us}
G.~Compere and D.~Marolf, \emph{{Setting the boundary free in AdS/CFT}},
  \href{https://doi.org/10.1088/0264-9381/25/19/195014}{\emph{Class. Quant.
  Grav.} {\bfseries 25} (2008) 195014}
  [\href{https://arxiv.org/abs/0805.1902}{{\ttfamily 0805.1902}}].

\bibitem{deBoer:1999tgo}
J.~de~Boer, E.P.~Verlinde and H.L.~Verlinde, \emph{{On the holographic
  renormalization group}},
  \href{https://doi.org/10.1088/1126-6708/2000/08/003}{\emph{JHEP} {\bfseries
  08} (2000) 003} [\href{https://arxiv.org/abs/hep-th/9912012}{{\ttfamily
  hep-th/9912012}}].

\bibitem{Allameh:2025gsa}
K.~Allameh and E.~Shaghoulian, \emph{{Timelike Liouville theory and AdS$_3$
  gravity at finite cutoff}},
  \href{https://arxiv.org/abs/2508.03236}{{\ttfamily 2508.03236}}.

\bibitem{Hawking:2000da}
S.~Hawking, J.M.~Maldacena and A.~Strominger, \emph{{de Sitter entropy, quantum
  entanglement and AdS / CFT}},
  \href{https://doi.org/10.1088/1126-6708/2001/05/001}{\emph{JHEP} {\bfseries
  05} (2001) 001} [\href{https://arxiv.org/abs/hep-th/0002145}{{\ttfamily
  hep-th/0002145}}].

\bibitem{Roberts:2012aq}
M.M.~Roberts, \emph{{Time evolution of entanglement entropy from a pulse}},
  \href{https://doi.org/10.1007/JHEP12(2012)027}{\emph{JHEP} {\bfseries 12}
  (2012) 027} [\href{https://arxiv.org/abs/1204.1982}{{\ttfamily 1204.1982}}].

\bibitem{Balasubramanian:1999re}
V.~Balasubramanian and P.~Kraus, \emph{{A Stress tensor for Anti-de Sitter
  gravity}}, \href{https://doi.org/10.1007/s002200050764}{\emph{Commun. Math.
  Phys.} {\bfseries 208} (1999) 413}
  [\href{https://arxiv.org/abs/hep-th/9902121}{{\ttfamily hep-th/9902121}}].

\bibitem{Zamolodchikov:2004ce}
A.B.~Zamolodchikov, \emph{{Expectation value of composite field T anti-T in
  two-dimensional quantum field theory}},
  \href{https://arxiv.org/abs/hep-th/0401146}{{\ttfamily hep-th/0401146}}.

\bibitem{Marolf18}
P.~Kraus, J.~Liu and D.~Marolf, \emph{{Cutoff AdS$_{3}$ versus the $
  T\overline{T} $ deformation}},
  \href{https://doi.org/10.1007/JHEP07(2018)027}{\emph{JHEP} {\bfseries 07}
  (2018) 027} [\href{https://arxiv.org/abs/1801.02714}{{\ttfamily
  1801.02714}}].

\bibitem{Witten:2022xxp}
E.~Witten, \emph{{A Note On The Canonical Formalism for Gravity}},
  \href{https://arxiv.org/abs/2212.08270}{{\ttfamily 2212.08270}}.

\bibitem{Anninos:2023epi}
D.~Anninos, D.A.~Galante and C.~Maneerat, \emph{{Gravitational observatories}},
  \href{https://doi.org/10.1007/JHEP12(2023)024}{\emph{JHEP} {\bfseries 12}
  (2023) 024} [\href{https://arxiv.org/abs/2310.08648}{{\ttfamily
  2310.08648}}].

\bibitem{Coleman:2020jte}
E.~Coleman and V.~Shyam, \emph{{Conformal boundary conditions from cutoff
  AdS$_{3}$}}, \href{https://doi.org/10.1007/JHEP09(2021)079}{\emph{JHEP}
  {\bfseries 09} (2021) 079}
  [\href{https://arxiv.org/abs/2010.08504}{{\ttfamily 2010.08504}}].

\bibitem{Chu:2021mvq}
C.-S.~Chu and R.-X.~Miao, \emph{{Conformal boundary condition and massive
  gravitons in AdS/BCFT}},
  \href{https://doi.org/10.1007/JHEP01(2022)084}{\emph{JHEP} {\bfseries 01}
  (2022) 084} [\href{https://arxiv.org/abs/2110.03159}{{\ttfamily
  2110.03159}}].

\bibitem{Anninos:2024wpy}
D.~Anninos, D.A.~Galante and C.~Maneerat, \emph{{Cosmological observatories}},
  \href{https://doi.org/10.1088/1361-6382/ad5824}{\emph{Class. Quant. Grav.}
  {\bfseries 41} (2024) 165009}
  [\href{https://arxiv.org/abs/2402.04305}{{\ttfamily 2402.04305}}].

\bibitem{Callebaut:2025uye}
N.~Callebaut, B.~Hergueta, R.~Monten and M.~Selle, \emph{{Deforming and
  dissecting AdS$_3$ with matter}},
  \href{https://arxiv.org/abs/2512.21255}{{\ttfamily 2512.21255}}.

\bibitem{Gorbenko:2018oov}
V.~Gorbenko, E.~Silverstein and G.~Torroba, \emph{{dS/dS and $ T\overline{T}
  $}}, \href{https://doi.org/10.1007/JHEP03(2019)085}{\emph{JHEP} {\bfseries
  03} (2019) 085} [\href{https://arxiv.org/abs/1811.07965}{{\ttfamily
  1811.07965}}].

\bibitem{Verlinde:1999xm}
E.P.~Verlinde and H.L.~Verlinde, \emph{{RG flow, gravity and the cosmological
  constant}}, \href{https://doi.org/10.1088/1126-6708/2000/05/034}{\emph{JHEP}
  {\bfseries 05} (2000) 034}
  [\href{https://arxiv.org/abs/hep-th/9912018}{{\ttfamily hep-th/9912018}}].

\bibitem{Porrati:2001gx}
M.~Porrati, \emph{{Mass and gauge invariance 4. Holography for the
  Karch-Randall model}},
  \href{https://doi.org/10.1103/PhysRevD.65.044015}{\emph{Phys. Rev. D}
  {\bfseries 65} (2002) 044015}
  [\href{https://arxiv.org/abs/hep-th/0109017}{{\ttfamily hep-th/0109017}}].

\bibitem{Perez-Victoria:2001lex}
M.~Perez-Victoria, \emph{{Randall-Sundrum models and the regularized AdS / CFT
  correspondence}},
  \href{https://doi.org/10.1088/1126-6708/2001/05/064}{\emph{JHEP} {\bfseries
  05} (2001) 064} [\href{https://arxiv.org/abs/hep-th/0105048}{{\ttfamily
  hep-th/0105048}}].

\bibitem{Arkani-Hamed:2000ijo}
N.~Arkani-Hamed, M.~Porrati and L.~Randall, \emph{{Holography and
  phenomenology}},
  \href{https://doi.org/10.1088/1126-6708/2001/08/017}{\emph{JHEP} {\bfseries
  08} (2001) 017} [\href{https://arxiv.org/abs/hep-th/0012148}{{\ttfamily
  hep-th/0012148}}].

\bibitem{Giddings:2000mu}
S.B.~Giddings, E.~Katz and L.~Randall, \emph{{Linearized gravity in brane
  backgrounds}},
  \href{https://doi.org/10.1088/1126-6708/2000/03/023}{\emph{JHEP} {\bfseries
  03} (2000) 023} [\href{https://arxiv.org/abs/hep-th/0002091}{{\ttfamily
  hep-th/0002091}}].

\bibitem{Verlinde:1999fy}
H.L.~Verlinde, \emph{{Holography and compactification}},
  \href{https://doi.org/10.1016/S0550-3213(00)00224-8}{\emph{Nucl. Phys. B}
  {\bfseries 580} (2000) 264}
  [\href{https://arxiv.org/abs/hep-th/9906182}{{\ttfamily hep-th/9906182}}].

\bibitem{ZZ_Liouville_gravity}
A.~Zamolodchikov and A.~Zamolodchikov, ``Lectures on liouville theory and
  matrix models.'' Online lecture notes, 2007.

\bibitem{Donnelly:2018bef}
W.~Donnelly and V.~Shyam, \emph{{Entanglement entropy and $T \overline{T}$
  deformation}},
  \href{https://doi.org/10.1103/PhysRevLett.121.131602}{\emph{Phys. Rev. Lett.}
  {\bfseries 121} (2018) 131602}
  [\href{https://arxiv.org/abs/1806.07444}{{\ttfamily 1806.07444}}].

\bibitem{Deng:2023pjs}
F.~Deng, Z.~Wang and Y.~Zhou, \emph{{End of the world brane meets $
  T\overline{T} $}}, \href{https://doi.org/10.1007/JHEP07(2024)036}{\emph{JHEP}
  {\bfseries 07} (2024) 036}
  [\href{https://arxiv.org/abs/2310.15031}{{\ttfamily 2310.15031}}].

\bibitem{Skenderis:1999nb}
K.~Skenderis and S.N.~Solodukhin, \emph{{Quantum effective action from the AdS
  / CFT correspondence}},
  \href{https://doi.org/10.1016/S0370-2693(99)01467-7}{\emph{Phys. Lett. B}
  {\bfseries 472} (2000) 316}
  [\href{https://arxiv.org/abs/hep-th/9910023}{{\ttfamily hep-th/9910023}}].

\bibitem{Henningson:1998gx}
M.~Henningson and K.~Skenderis, \emph{{The Holographic Weyl anomaly}},
  \href{https://doi.org/10.1088/1126-6708/1998/07/023}{\emph{JHEP} {\bfseries
  07} (1998) 023} [\href{https://arxiv.org/abs/hep-th/9806087}{{\ttfamily
  hep-th/9806087}}].

\bibitem{deHaro:2000vlm}
S.~de~Haro, S.N.~Solodukhin and K.~Skenderis, \emph{{Holographic reconstruction
  of space-time and renormalization in the AdS / CFT correspondence}},
  \href{https://doi.org/10.1007/s002200100381}{\emph{Commun. Math. Phys.}
  {\bfseries 217} (2001) 595}
  [\href{https://arxiv.org/abs/hep-th/0002230}{{\ttfamily hep-th/0002230}}].

\bibitem{Caputa:2020lpa}
P.~Caputa, S.~Datta, Y.~Jiang and P.~Kraus, \emph{{Geometrizing $ T\overline{T}
  $}}, \href{https://doi.org/10.1007/JHEP03(2021)140}{\emph{JHEP} {\bfseries
  03} (2021) 140} [\href{https://arxiv.org/abs/2011.04664}{{\ttfamily
  2011.04664}}].

\bibitem{HartmanDeSitter}
T.~Hartman, ``Lecture notes on classical de sitter space.''
  \url{http://www.hartmanhep.net/GR2017/desitter-lectures-v2.pdf}.

\bibitem{Iyer:1994ys}
V.~Iyer and R.M.~Wald, \emph{{Some properties of Noether charge and a proposal
  for dynamical black hole entropy}},
  \href{https://doi.org/10.1103/PhysRevD.50.846}{\emph{Phys. Rev. D} {\bfseries
  50} (1994) 846} [\href{https://arxiv.org/abs/gr-qc/9403028}{{\ttfamily
  gr-qc/9403028}}].

\bibitem{Fiola:1994ir}
T.M.~Fiola, J.~Preskill, A.~Strominger and S.P.~Trivedi, \emph{{Black hole
  thermodynamics and information loss in two-dimensions}},
  \href{https://doi.org/10.1103/PhysRevD.50.3987}{\emph{Phys. Rev. D}
  {\bfseries 50} (1994) 3987}
  [\href{https://arxiv.org/abs/hep-th/9403137}{{\ttfamily hep-th/9403137}}].

\bibitem{Takayanagi:2011zk}
T.~Takayanagi, \emph{{Holographic Dual of BCFT}},
  \href{https://doi.org/10.1103/PhysRevLett.107.101602}{\emph{Phys. Rev. Lett.}
  {\bfseries 107} (2011) 101602}
  [\href{https://arxiv.org/abs/1105.5165}{{\ttfamily 1105.5165}}].

\bibitem{Fujita:2011fp}
M.~Fujita, T.~Takayanagi and E.~Tonni, \emph{{Aspects of AdS/BCFT}},
  \href{https://doi.org/10.1007/JHEP11(2011)043}{\emph{JHEP} {\bfseries 11}
  (2011) 043} [\href{https://arxiv.org/abs/1108.5152}{{\ttfamily 1108.5152}}].

\bibitem{Deng:2022yll}
F.~Deng, Y.-S.~An and Y.~Zhou, \emph{{JT gravity from partial reduction and
  defect extremal surface}},
  \href{https://doi.org/10.1007/JHEP02(2023)219}{\emph{JHEP} {\bfseries 02}
  (2023) 219} [\href{https://arxiv.org/abs/2206.09609}{{\ttfamily
  2206.09609}}].

\bibitem{Aguilar-Gutierrez:2023tic}
S.E.~Aguilar-Gutierrez, A.K.~Patra and J.F.~Pedraza, \emph{{Entangled universes
  in dS wedge holography}},
  \href{https://doi.org/10.1007/JHEP10(2023)156}{\emph{JHEP} {\bfseries 10}
  (2023) 156} [\href{https://arxiv.org/abs/2308.05666}{{\ttfamily
  2308.05666}}].

\bibitem{Neuenfeld:2026hen}
D.~Neuenfeld and C.~Tellinger, \emph{{Large-c BCFT Entanglement Entropy with
  Deformed Boundaries from Emergent JT Gravity}},
  \href{https://arxiv.org/abs/2604.18405}{{\ttfamily 2604.18405}}.

\bibitem{Carlip:2005tz}
S.~Carlip, \emph{{Dynamics of asymptotic diffeomorphisms in (2+1)-dimensional
  gravity}}, \href{https://doi.org/10.1088/0264-9381/22/14/014}{\emph{Class.
  Quant. Grav.} {\bfseries 22} (2005) 3055}
  [\href{https://arxiv.org/abs/gr-qc/0501033}{{\ttfamily gr-qc/0501033}}].

\bibitem{Takayanagi:2018pml}
T.~Takayanagi, \emph{{Holographic Spacetimes as Quantum Circuits of
  Path-Integrations}},
  \href{https://doi.org/10.1007/JHEP12(2018)048}{\emph{JHEP} {\bfseries 12}
  (2018) 048} [\href{https://arxiv.org/abs/1808.09072}{{\ttfamily
  1808.09072}}].

\bibitem{Ondo:2022zgf}
N.~Ondo and V.~Shyam, \emph{{The role of dRGT mass terms in cutoff holography
  and the Randall--Sundrum II scenario}},
  \href{https://arxiv.org/abs/2206.04005}{{\ttfamily 2206.04005}}.

\bibitem{Dubovsky:2017cnj}
S.~Dubovsky, V.~Gorbenko and M.~Mirbabayi, \emph{{Asymptotic fragility, near
  AdS$_{2}$ holography and $ T\overline{T} $}},
  \href{https://doi.org/10.1007/JHEP09(2017)136}{\emph{JHEP} {\bfseries 09}
  (2017) 136} [\href{https://arxiv.org/abs/1706.06604}{{\ttfamily
  1706.06604}}].

\bibitem{Cardy:2018sdv}
J.~Cardy, \emph{{The $ T\overline{T} $ deformation of quantum field theory as
  random geometry}}, \href{https://doi.org/10.1007/JHEP10(2018)186}{\emph{JHEP}
  {\bfseries 10} (2018) 186}
  [\href{https://arxiv.org/abs/1801.06895}{{\ttfamily 1801.06895}}].

\bibitem{Dubovsky:2018bmo}
S.~Dubovsky, V.~Gorbenko and G.~Hern\'andez-Chifflet, \emph{{$ T\overline{T} $
  partition function from topological gravity}},
  \href{https://doi.org/10.1007/JHEP09(2018)158}{\emph{JHEP} {\bfseries 09}
  (2018) 158} [\href{https://arxiv.org/abs/1805.07386}{{\ttfamily
  1805.07386}}].

\bibitem{Tolley:2019nmm}
A.J.~Tolley, \emph{{$ T\overline{T} $ deformations, massive gravity and
  non-critical strings}},
  \href{https://doi.org/10.1007/JHEP06(2020)050}{\emph{JHEP} {\bfseries 06}
  (2020) 050} [\href{https://arxiv.org/abs/1911.06142}{{\ttfamily
  1911.06142}}].

\bibitem{Callebaut:2019omt}
N.~Callebaut, J.~Kruthoff and H.~Verlinde, \emph{{$ T\overline{T} $ deformed
  CFT as a non-critical string}},
  \href{https://doi.org/10.1007/JHEP04(2020)084}{\emph{JHEP} {\bfseries 04}
  (2020) 084} [\href{https://arxiv.org/abs/1910.13578}{{\ttfamily
  1910.13578}}].

\bibitem{Guica:2019nzm}
M.~Guica and R.~Monten, \emph{{$T\bar T$ and the mirage of a bulk cutoff}},
  \href{https://doi.org/10.21468/SciPostPhys.10.2.024}{\emph{SciPost Phys.}
  {\bfseries 10} (2021) 024}
  [\href{https://arxiv.org/abs/1906.11251}{{\ttfamily 1906.11251}}].

\bibitem{Hirano:2025cjg}
S.~Hirano and V.~Raj, \emph{{$T\bar{T}$ braneworld holography}},
  \href{https://arxiv.org/abs/2508.11471}{{\ttfamily 2508.11471}}.

\bibitem{Rozali:2019day}
M.~Rozali, J.~Sully, M.~Van~Raamsdonk, C.~Waddell and D.~Wakeham,
  \emph{{Information radiation in BCFT models of black holes}},
  \href{https://doi.org/10.1007/JHEP05(2020)004}{\emph{JHEP} {\bfseries 05}
  (2020) 004} [\href{https://arxiv.org/abs/1910.12836}{{\ttfamily
  1910.12836}}].

\bibitem{Almheiri:2020cfm}
A.~Almheiri, T.~Hartman, J.~Maldacena, E.~Shaghoulian and A.~Tajdini,
  \emph{{The entropy of Hawking radiation}},
  \href{https://doi.org/10.1103/RevModPhys.93.035002}{\emph{Rev. Mod. Phys.}
  {\bfseries 93} (2021) 035002}
  [\href{https://arxiv.org/abs/2006.06872}{{\ttfamily 2006.06872}}].

\bibitem{Hartman:2020swn}
T.~Hartman, E.~Shaghoulian and A.~Strominger, \emph{{Islands in Asymptotically
  Flat 2D Gravity}}, \href{https://doi.org/10.1007/JHEP07(2020)022}{\emph{JHEP}
  {\bfseries 07} (2020) 022}
  [\href{https://arxiv.org/abs/2004.13857}{{\ttfamily 2004.13857}}].

\bibitem{Verheijden:2021yrb}
E.~Verheijden and E.~Verlinde, \emph{{From the BTZ black hole to JT gravity:
  geometrizing the island}},
  \href{https://doi.org/10.1007/JHEP11(2021)092}{\emph{JHEP} {\bfseries 11}
  (2021) 092} [\href{https://arxiv.org/abs/2102.00922}{{\ttfamily
  2102.00922}}].

\bibitem{Caceres:2026yvs}
E.~C{\'a}ceres, H.~Krishna, H.P.~Balaji and V.~Patil, \emph{{Holographic
  entanglement entropy with conformal boundary conditions}},
  \href{https://arxiv.org/abs/2608.06277}{{\ttfamily 2608.06277}}.

\bibitem{Jacobson:1995ab}
T.~Jacobson, \emph{{Thermodynamics of space-time: The Einstein equation of
  state}}, \href{https://doi.org/10.1103/PhysRevLett.75.1260}{\emph{Phys. Rev.
  Lett.} {\bfseries 75} (1995) 1260}
  [\href{https://arxiv.org/abs/gr-qc/9504004}{{\ttfamily gr-qc/9504004}}].

\bibitem{Savonije:2001nd}
I.~Savonije and E.P.~Verlinde, \emph{{CFT and entropy on the brane}},
  \href{https://doi.org/10.1016/S0370-2693(01)00467-1}{\emph{Phys. Lett. B}
  {\bfseries 507} (2001) 305}
  [\href{https://arxiv.org/abs/hep-th/0102042}{{\ttfamily hep-th/0102042}}].

\bibitem{VanRaamsdonk:2010pw}
M.~Van~Raamsdonk, \emph{{Building up spacetime with quantum entanglement}},
  \href{https://doi.org/10.1142/S0218271810018529}{\emph{Gen. Rel. Grav.}
  {\bfseries 42} (2010) 2323}
  [\href{https://arxiv.org/abs/1005.3035}{{\ttfamily 1005.3035}}].

\bibitem{Verlinde:2010hp}
E.P.~Verlinde, \emph{{On the Origin of Gravity and the Laws of Newton}},
  \href{https://doi.org/10.1007/JHEP04(2011)029}{\emph{JHEP} {\bfseries 04}
  (2011) 029} [\href{https://arxiv.org/abs/1001.0785}{{\ttfamily 1001.0785}}].

\bibitem{Casini:2011kv}
H.~Casini, M.~Huerta and R.C.~Myers, \emph{{Towards a derivation of holographic
  entanglement entropy}},
  \href{https://doi.org/10.1007/JHEP05(2011)036}{\emph{JHEP} {\bfseries 05}
  (2011) 036} [\href{https://arxiv.org/abs/1102.0440}{{\ttfamily 1102.0440}}].

\bibitem{Faulkner:2013ica}
T.~Faulkner, M.~Guica, T.~Hartman, R.C.~Myers and M.~Van~Raamsdonk,
  \emph{{Gravitation from Entanglement in Holographic CFTs}},
  \href{https://doi.org/10.1007/JHEP03(2014)051}{\emph{JHEP} {\bfseries 03}
  (2014) 051} [\href{https://arxiv.org/abs/1312.7856}{{\ttfamily 1312.7856}}].

\bibitem{Czech:2015qta}
B.~Czech, L.~Lamprou, S.~McCandlish and J.~Sully, \emph{{Integral Geometry and
  Holography}}, \href{https://doi.org/10.1007/JHEP10(2015)175}{\emph{JHEP}
  {\bfseries 10} (2015) 175}
  [\href{https://arxiv.org/abs/1505.05515}{{\ttfamily 1505.05515}}].

\bibitem{deBoer:2015kda}
J.~de~Boer, M.P.~Heller, R.C.~Myers and Y.~Neiman, \emph{{Holographic de Sitter
  Geometry from Entanglement in Conformal Field Theory}},
  \href{https://doi.org/10.1103/PhysRevLett.116.061602}{\emph{Phys. Rev. Lett.}
  {\bfseries 116} (2016) 061602}
  [\href{https://arxiv.org/abs/1509.00113}{{\ttfamily 1509.00113}}].

\bibitem{Jacobson:2015hqa}
T.~Jacobson, \emph{{Entanglement Equilibrium and the Einstein Equation}},
  \href{https://doi.org/10.1103/PhysRevLett.116.201101}{\emph{Phys. Rev. Lett.}
  {\bfseries 116} (2016) 201101}
  [\href{https://arxiv.org/abs/1505.04753}{{\ttfamily 1505.04753}}].

\bibitem{Asplund:2016koz}
C.T.~Asplund, N.~Callebaut and C.~Zukowski, \emph{{Equivalence of Emergent de
  Sitter Spaces from Conformal Field Theory}},
  \href{https://doi.org/10.1007/JHEP09(2016)154}{\emph{JHEP} {\bfseries 09}
  (2016) 154} [\href{https://arxiv.org/abs/1604.02687}{{\ttfamily
  1604.02687}}].

\bibitem{Callebaut:2018nlq}
N.~Callebaut and H.~Verlinde, \emph{{Entanglement Dynamics in 2D CFT with
  Boundary: Entropic origin of JT gravity and Schwarzian QM}},
  \href{https://doi.org/10.1007/JHEP05(2019)045}{\emph{JHEP} {\bfseries 05}
  (2019) 045} [\href{https://arxiv.org/abs/1808.05583}{{\ttfamily
  1808.05583}}].

\bibitem{Callebaut:2018xfu}
N.~Callebaut, \emph{{The gravitational dynamics of kinematic space}},
  \href{https://doi.org/10.1007/JHEP02(2019)153}{\emph{JHEP} {\bfseries 02}
  (2019) 153} [\href{https://arxiv.org/abs/1808.10431}{{\ttfamily
  1808.10431}}].

\bibitem{Fujiki:2025rtx}
K.~Fujiki, M.~Kohara, K.~Shinmyo, Y.-k.~Suzuki and T.~Takayanagi,
  \emph{{Entropic interpretation of Einstein equation in dS/CFT}},
  \href{https://doi.org/10.1007/JHEP04(2026)072}{\emph{JHEP} {\bfseries 04}
  (2026) 072} [\href{https://arxiv.org/abs/2511.07915}{{\ttfamily
  2511.07915}}].

\bibitem{Hartman:2018tkw}
T.~Hartman, J.~Kruthoff, E.~Shaghoulian and A.~Tajdini, \emph{{Holography at
  finite cutoff with a $T^2$ deformation}},
  \href{https://doi.org/10.1007/JHEP03(2019)004}{\emph{JHEP} {\bfseries 03}
  (2019) 004} [\href{https://arxiv.org/abs/1807.11401}{{\ttfamily
  1807.11401}}].

\bibitem{Liu:2025xij}
X.~Liu, H.S.~Reall, J.E.~Santos and T.~Wiseman, \emph{{Ill-posedness of the
  Cauchy problem for linearized gravity in a cavity with conformal boundary
  conditions}}, \href{https://doi.org/10.1088/1361-6382/ae1b60}{\emph{Class.
  Quant. Grav.} {\bfseries 42} (2025) 235003}
  [\href{https://arxiv.org/abs/2505.20410}{{\ttfamily 2505.20410}}].

\bibitem{Engelsoy:2016xyb}
J.~Engels{\"o}y, T.G.~Mertens and H.~Verlinde, \emph{{An investigation of
  AdS$_{2}$ backreaction and holography}},
  \href{https://doi.org/10.1007/JHEP07(2016)139}{\emph{JHEP} {\bfseries 07}
  (2016) 139} [\href{https://arxiv.org/abs/1606.03438}{{\ttfamily
  1606.03438}}].

\bibitem{Maldacena:2016upp}
J.~Maldacena, D.~Stanford and Z.~Yang, \emph{{Conformal symmetry and its
  breaking in two dimensional Nearly Anti-de-Sitter space}},
  \href{https://doi.org/10.1093/ptep/ptw124}{\emph{PTEP} {\bfseries 2016}
  (2016) 12C104} [\href{https://arxiv.org/abs/1606.01857}{{\ttfamily
  1606.01857}}].

\bibitem{Griguolo:2025kpi}
L.~Griguolo, J.~Papalini, L.~Russo and D.~Seminara, \emph{{A new perspective on
  dilaton gravity at finite cutoff}},
  \href{https://arxiv.org/abs/2512.21774}{{\ttfamily 2512.21774}}.

\bibitem{Griguolo:2026eoi}
L.~Griguolo, J.~Papalini, L.~Russo, D.~Seminara and A.~Tarana, \emph{{Quantum
  JT Gravity in a box as a P{\"o}schl-Teller Scattering Problem}},
  \href{https://arxiv.org/abs/2607.01385}{{\ttfamily 2607.01385}}.

\bibitem{Strominger:2001pn}
A.~Strominger, \emph{{The dS / CFT correspondence}},
  \href{https://doi.org/10.1088/1126-6708/2001/10/034}{\emph{JHEP} {\bfseries
  10} (2001) 034} [\href{https://arxiv.org/abs/hep-th/0106113}{{\ttfamily
  hep-th/0106113}}].

\bibitem{Chen:2020tes}
Y.~Chen, V.~Gorbenko and J.~Maldacena, \emph{{Bra-ket wormholes in
  gravitationally prepared states}},
  \href{https://doi.org/10.1007/JHEP02(2021)009}{\emph{JHEP} {\bfseries 02}
  (2021) 009} [\href{https://arxiv.org/abs/2007.16091}{{\ttfamily
  2007.16091}}].

\bibitem{Godet:2024ich}
V.~Godet, \emph{{Quantum cosmology as automorphic dynamics}},
  \href{https://arxiv.org/abs/2405.09833}{{\ttfamily 2405.09833}}.

\bibitem{Aguilar-Gutierrez:2024nst}
S.E.~Aguilar-Gutierrez, A.~Svesko and M.R.~Visser, \emph{{$
  \textrm{T}\overline{\textrm{T}} $ deformations from AdS$_{2}$ to dS$_{2}$}},
  \href{https://doi.org/10.1007/JHEP01(2025)120}{\emph{JHEP} {\bfseries 01}
  (2025) 120} [\href{https://arxiv.org/abs/2410.18257}{{\ttfamily
  2410.18257}}].

\bibitem{Fumagalli:2024msi}
A.~Fumagalli, V.~Gorbenko and J.~Kames-King, \emph{{De Sitter Bra-Ket
  wormholes}}, \href{https://doi.org/10.1007/JHEP05(2025)074}{\emph{JHEP}
  {\bfseries 05} (2025) 074}
  [\href{https://arxiv.org/abs/2408.08351}{{\ttfamily 2408.08351}}].

\bibitem{Araujo-Regado:2025elv}
G.~Araujo-Regado, A.~Thavanesan and A.C.~Wall, \emph{{Holographic cosmology at
  finite time}}, \href{https://doi.org/10.1007/JHEP04(2026)121}{\emph{JHEP}
  {\bfseries 04} (2026) 121}
  [\href{https://arxiv.org/abs/2511.04511}{{\ttfamily 2511.04511}}].

\bibitem{Ball:2019atb}
A.~Ball, E.~Himwich, S.A.~Narayanan, S.~Pasterski and A.~Strominger,
  \emph{{Uplifting AdS$_{3}$/CFT$_{2}$ to flat space holography}},
  \href{https://doi.org/10.1007/JHEP08(2019)168}{\emph{JHEP} {\bfseries 08}
  (2019) 168} [\href{https://arxiv.org/abs/1905.09809}{{\ttfamily
  1905.09809}}].

\bibitem{Fujiki:2025yyf}
K.~Fujiki, H.~Kanda, M.~Kohara and T.~Takayanagi, \emph{{Brane cosmology from
  AdS/BCFT}}, \href{https://doi.org/10.1007/JHEP03(2025)135}{\emph{JHEP}
  {\bfseries 03} (2025) 135}
  [\href{https://arxiv.org/abs/2501.05036}{{\ttfamily 2501.05036}}].

\bibitem{Neuenfeld:2025wnl}
D.~Neuenfeld, \emph{{The Flat-Space Limit of AdS Coupled to a Bath}},
  \href{https://arxiv.org/abs/2508.03798}{{\ttfamily 2508.03798}}.

\bibitem{Skenderis:2000in}
K.~Skenderis, \emph{{Asymptotically Anti-de Sitter space-times and their stress
  energy tensor}}, \href{https://doi.org/10.1142/S0217751X0100386X}{\emph{Int.
  J. Mod. Phys. A} {\bfseries 16} (2001) 740}
  [\href{https://arxiv.org/abs/hep-th/0010138}{{\ttfamily hep-th/0010138}}].

\end{thebibliography}\endgroup

\end{document}